\documentclass[a4paper,fleqn,usenatbib]{mnras}
\pdfoutput=1
\usepackage{newtxtext,newtxmath}
\usepackage{array}

\usepackage[T1]{fontenc}
\usepackage{ae,aecompl}

\usepackage{subfig}
\usepackage{graphicx}	% Including figure files
\usepackage{amsmath}	% Advanced maths commands
\usepackage{amssymb}	% Extra maths symbols
\let\oldAA\AA
\renewcommand{\AA}{\text{\normalfont\oldAA}}
\let\footnote\thanks
\title[Extended absorbers in the high-z radio galaxies]{Detection of large scale Ly$\alpha$ absorbers at large angles to the radio axis of high-redshift radio galaxies using SOAR\footnote{Based on observations obtained at the Southern Astrophysical Research (SOAR) telescope, which is a joint project of the Minist\'erio da Ci\^encia, Tecnologia, Inova\c{c}\~ao e Comunica\c{c}\~oes (MCTIC) da Rep\'ublica Federativa do Brasil, the U.S. National Optical Astronomy Observatory (NOAO), the University of North Carolina at Chapel Hill (UNC), and Michigan State University (MSU).}}

\author[Silva, M. et al.]{
M. Silva$^{1,2}$\thanks{E-mail: marckelson.silva@astro.up.pt},
A. Humphrey$^{1}$,
P. Lagos$^{1,3}$,
R. Guimar\~aes$^{4}$,
T. Scott$^{1}$,
P. Papaderos$^{1}$ 
\newauthor
and S.G. Morais$^{1,2}$
\\
$^{1}$Institute of Astrophysics and Space Sciences, Universidade do Porto, CAUP, Rua das Estrelas, 4150-762 Porto, Portugal.\\
$^{2}$Departamento de F\'isica e Astronomia, Faculdade de Ci\^encias, Universidade do Porto, R. Campo Alegre 687, 4169-007 Porto, Portugal\\
$^{3}$Centre for Space Research, North West University, Potchefstroom 2520, South Africa\\
$^{4}$Faculty of Medicine, Universidade Federal de Minas Gerais, Belo Horizonte, 30130-100 Minas Gerais, Brazil
}

\date{Accepted 2018 August 24. Received 2018 August 1; in original form 2018 June 25}

\pubyear{2018}

\begin{document}
\label{firstpage}
\pagerange{\pageref{firstpage}--\pageref{lastpage}}
\maketitle

\begin{abstract}
We present an investigation of the properties of the extended Ly$\alpha$ halo and the large-scale \ion{H}{I} absorbing structures associated with 5 high-redshift radio galaxies at z $>$ 2, using the Goodman long-slit spectrograph on the SOAR telescope, with the slit placed at large angles ($>$45$^{\circ}$) to the radio axis, to study regions that are unlikely to be illuminated by the active nucleus.
Spatially extended Ly$\alpha$ emission is detected with large line widths (FWHM = 1000 -- 2500 km s$^{-1}$), which although impacted by resonant scattering, is suggestive of turbulent motion. We find a correlation between higher blueshifts and higher FWHM, which is an indication that radial motion dominates the bulk gas dynamics perpendicular to the radio axis, although we are unable to distinguish between outflow and infall scenarios due to the resonant nature of the Ly$\alpha$ line.
Extended, blueshifted Ly$\alpha$ absorption is detected in the direction perpendicular to the radio axis in three radio galaxies with minimum spatial extents ranging from $\ga$27 kpc to $\ga$35 kpc, supporting the idea that the absorbing structure covers the entire Ly$\alpha$ halo, consistent with being part of a giant, expanding shell of gas enveloping the galaxy and its (detected) gaseous halo.
\end{abstract}


%%%%%%%%%%%%%%%%%%%%%%%%%%%%%%%%%%%%%%%%%%%%%%%%%%

%%%%%%%%%%%%%%%%% BODY OF PAPER %%%%%%%%%%%%%%%%%%

\section{Introduction}
\label{intro}

High-redshift radio galaxies (hereafter HzRGs; z $>$ 2) are among the most spectacular objects in the sky. Producing prodigious luminosities (e.g., L$_{500 MHz}$ $>$ 10$^{27}$ W Hz$^{-1}$; \citealt{breuck2010} and references therein) from tiny volumes, with radiation spread over a broad range of frequencies, HzRGs are noteworthy in providing us with a unique opportunity to investigate the formation and evolution of massive galaxies and active galactic nuclei (hereafter AGN). 

These galaxies are among the most massive galaxies, with stellar masses up to $\sim$ 10$^{12}$ M$_{\odot}$ \citep[e.g.][]{Bre2001,seymour2007,hatch2009,breuck2010,Hatch2013,nesvadba2017}, often with evidence for significant star formation rates (up to $\sim$ 1400 M$_{\odot}$ yr$^{-1}$; \citealt{ogle2012,seymour2012,rocca2013,Hatch2013,drouart}). 
HzRGs are also commonly associated with spatially extended Ly$\alpha$ halos, with luminosities up to $\sim$ 10$^{45}$ erg s$^{-1}$, which are gas rich (M$_{HI}$ $\sim$ 10$^{9-11}$ M$_{\odot}$) and extend over tens or even a few hundred kpc \citep{fosbury1982,serego1988,Mc90,Mc93,Vo,Pe98,Fr2001,Re,VM1,Sa,Hu1,Cantalupo2014,swinbank,borisova2016,Cai2017,arrigoni2018}. These large scale Ly$\alpha$ structures usually show a clumpy and irregular morphology \citep{Re} often aligned with the radio jets \citep{Mc95} that exert significant feedback onto the surrounding intergalactic medium (hereafter IGM; \citealt{VM1,nesvadba2006,Hu2,ogle2012}).
Generally, the halos can be divided into two distinct kinematic components: quiescent and perturbed.
The quiescent component shows kinematics with full width at half maximum (hereafter FWHM) $<$ 1000 km s$^{-1}$ and no clear relationship with the radio jets \citep{Vo96,VM4,VM1,Sa}. 
This component has been reported to be a common feature of the Ly$\alpha$ halos being detected across the full spatial extent of the halos \citep[e.g.][]{VM1}.
In some HzRGs, the quiescent component seems to be infalling towards the central regions of the host galaxy \citep[e.g.][]{Hu3,VM2,Hu1,nathan2014,marckelson2018}, which may be explained by a scenario whereby cold gas streams fall onto the dark matter halos along the cosmic web \citep[e.g.][]{tobias2010} or by an alternative scenario in which gaseous debris falls back into the host galaxy after a feedback 'blowout' \citep[e.g][]{Hu1}.
The perturbed component usually shows an irregular gas kinematics with FWHM $>$ 1000 km s$^{-1}$  \citep{Vo96,VM1,Hu2,marckelson2018}. With clear spatial association with the radio structure in some HzRGs, the perturbed component provides evidence of gas that has been disturbed by the passage of the radio jets, and which may be outflowing \citep[see][]{nesvadba2006,nesvadba2008a,sandy2016}.
As such, the Ly$\alpha$ halos allow us to probe the evolution of massive galaxies during a phase of significant feedback, black hole growth and, in many cases, star formation. The study of the properties of these halos provides keys to understand how hosts of powerful radio galaxies form and evolve.

In addition, studies have shown that some HzRGs show spatially resolved \ion{H}{I} absorption features in their Ly$\alpha$ emission line profiles \citep[e.g.][]{Ro95,Vo,Bi1,Ja,Bi3,Hu6,Hu5,moyano2015,swinbank,Gu,marckelson2018}, which are thought to be produced by a giant shell of \ion{H}{I} gas enveloping the Ly$\alpha$ emitting region, and which appears to be expanding or outflowing due to feedback activity \citep[see][]{Bi3,Hu6,swinbank,marckelson2018}. 
For example, in the case of the main extended absorber associated with MRC 0943--242 (z = 2.92), \citet{marckelson2018} found a significant radial evolution in the \ion{H}{I} absorber's line of sight velocity which they argued is consistent with it being an expanding shell with a radius of at least several tens of kpc.
With \ion{H}{I} column densities in the range $\sim$ 10$^{14-20}$ cm$^{-2}$, observations with the slit placed along the radio axis have shown that strong absorbers (N(\ion{H}{I}) $>$ 10$^{18}$ cm$^{-2}$) extend over the full spatial extent of the Ly$\alpha$ emission, although the properties of the absorption does not always remain constant over the full spatial extent \citep{Vo,Bi3}. Although the precise nature and origins of these large scale absorbing structures are not well understood, they are clearly relevant for understanding issues such as feedback, the dispersion of metals through the interstellar medium (hereafter ISM) of massive galaxies and into the surrounding IGM, as well as the escape of Ly$\alpha$ and ionizing photons from HzRGs. 

Previous studies of the Ly$\alpha$ halos and \ion{H}{I} absorbers associated with HzRGs have focused mainly on the relatively high surface brightness emission regions aligned with the radio jet axis, where the jet-gas interactions and the ionizing radiation of the AGN are expected to have their greatest impact \citep[e.g.][]{rush1997,Bre2000b,Tani2001,Bre2001,VM1,nesvadba2006,Hu2,Hu4,humphrey2009,nesvadba17,nesvadba2017}.
By studying the extended Ly$\alpha$ emission regions that are located significantly away from the radio jet axis, it may be possible to obtain a more complete picture of the extended gaseous environment of HzRGs \citep[e.g.][]{Gu,sandy2016,vernet2017,marckelson2018}, allowing us to investigate questions such as what produces the Ly$\alpha$ emission when it is not illuminated by the AGN, and whether the impact of radio mode feedback is global or instead confined to the radio axis. Likewise, the 2-dimensional spatial distribution of the \ion{H}{I} absorbers is also poorly known, with spatial information predominantly coming from long slit spectra where the slit was placed along the radio axis \citep[e.g.][]{Vo}. A handful of \ion{H}{I} absorbers have now been studied using IFU spectroscopy \citep[e.g.][]{Hu6,swinbank,marckelson2018}, in each case showing that the absorber is also extended perpendicularly to the radio axis, consistent with the idea that the absorbing gas is part of a giant shell enveloping the Ly$\alpha$ emitting region. However, similar observations of a larger number of HzRGs are needed to confirm that this is a general property of this class of absorbers. 

This paper aims to characterize the properties of the extended Ly$\alpha$ halos and the large-scale \ion{H}{I} absorbing structures in the direction perpendicular to the radio axis of HzRGs, adding new information about the global properties of the extended ionized gas and extended absorbers associated with HzRGs. The paper is organized as follows. In $\S$~\ref{obs}, we describe the sample selection, observations and data reduction. In $\S$~\ref{analysis}, we discuss our data analysis methods. In $\S$~\ref{prev},we present the results of our study.
In $\S$~\ref{discussion}, we discuss the gas dynamics of the extended Ly$\alpha$ halo and the nature of the extended \ion{H}{I} absorbers. In $\S$~\ref{conclusions}, we give a brief summary concluding our results.
Throughout this paper we assume $\Omega _{\Lambda}$ = 0.713, $\Omega _{m}$ = 0.287 and $H_{\circ}$ = 69.3 km s$^{-1}$ Mpc$^{-1}$ \citep{wmap}.

\section{Sample selection and SOAR observations}
\label{obs}

We selected 5 Ly$\alpha$-bright, steep spectrum HzRGs from the list of \cite{Bre2000} within a redshift range of 2.16 $<$ z $<$ 2.76. This redshift range ensured that Ly$\alpha$, \ion{C}{IV} $\lambda$1549 and \ion{He}{II} $\lambda$1640 fell within the wavelength coverage of the spectrograph. Our sample (see Table \ref{observations}) is characterized by galaxies with relatively bright Ly$\alpha$ emission (1.95$\times$10$^{43}$ to 2.91$\times$10$^{44}$ erg s$^{-1}$, measured through a long slit aligned along the radio axis) of powerful radio sources (log($P_{325}$) ranging from 35.48 to 36.46 erg s$^{-1}$ Hz$^{-1}$), and covers a large range in radio source diameter ($<$ 2.5 to 196 kpc: \citealt{Bre2000} and references therein.). 

The observations were performed on 2014 September and on 2015 April, using the Goodman High Throughput Spectrograph (GTHS; \citealt{clemens2004}) on the Southern Astrophysical Research (SOAR) 4.1 m telescope during the commissioning of the instrument under the program SO2014B-013 (PI: R. N. Guimar\~aes) in classical mode, and SO2015A-024 (PI: R. N. Guimar\~aes) in remote mode. The observations were set to use the blue camera combined with the 600l/mm grating which allowed us to obtain a wavelength range of $\sim$ 3500 -- 6200 \AA. This wavelength range allows, for some objects, the detection of important diagnostic lines such as \ion{N}{V} $\lambda \lambda$1239,1243, \ion{C}{IV} $\lambda \lambda$1548,1551 and \ion{He}{II} $\lambda$1640 along with Ly$\alpha$ $\lambda$1216. For all observations, on-chip binning of 2 $\times$ 2 (spatial x spectral) was used, resulting in a pixel size of 0.3\arcsec. In addition, the slit widths were set to be 1.03\arcsec$\,$ and 1.68\arcsec$\,$ (see Table \ref{observations}). The instrumental profile (FWHM) was estimated using the mean of the Gaussian FWHMs measured for a number of unblended arc-lamp and the night-sky lines over the whole spectral range of a wavelength-calibrated spectrum, which gave us an instrumental profile FWHM of 3.9 \AA$\,$ (or 266 km s$^{-1}$) and 8.0 \AA$\,$ (or 560 km s$^{-1}$), respectively. The seeing was obtained by reconstructing the spatial profile of the seeing disk along the slit using a non-saturated star in images taken before or after the observation of the science target.

The exposure time for each of the five radio galaxies was set in intervals of 1800s, totaling at the end of the program in 20 hours of observation. The slit alignment was chosen to be perpendicular to the radio jets (see Table \ref{observations}) as determined from radio images in the literature \cite[e.g.][]{carilli1997,pentericci1997}, for 3 out of 5 targets. The position angle of the radio axis was determined by simply measuring the angle made by the brightest radio hotspot on either side of the nucleus. This assumes that the hotspots are well aligned with the radio jets, which is generally the case in HzRGs where jets have been detected in HzRGs \citep[e.g.][]{carilli1997,pentericci1997}.

In the case of MRC 0030--219, whose radio source is unresolved ($<$0.3\arcsec or $<$2.5 kpc), we have used the parallactic angle for the slit position. Although it is not clear whether the extended emission in our spectrum of this target is influenced by the ionizing radiation field of the AGN, we can be fairly confident that the extended gas is not affected by interaction with the radio source.

In the case of 4C--00.54 we have used an angle perpendicular to the UV-optical emission in Hubble Space Telescope image of the galaxy \citep[e.g.][]{pentericci1997,Pe2001}, rather than placing the slit perpendicular to the large scale radio axis. While most HzRGs show a close correlation between the position angles of the UV-optical emission and the radio source \citep[e.g.][]{chambers1987,Mc1987}, 4C--00.54 is somewhat unusual in showing a large misalignment between the UV-optical emission and the large-scale radio emission. This substantial $\sim$45$^{\circ}$ misalignment has been suggested to be the result of precession of the radio jets, where the radio axis defined by positions of the radio hotspots would represent the axis of the AGN $\ga$1 Myr earlier than the currently observed nuclear activity \citep{pentericci1997}. Unfortunately, existing radio observations do not allow us to definitely test this scenario due to the lack of detection of extended nuclear radio jets. Nevertheless, we have adopted the position angle of the UV-optical emission as the most probable current orientation of the AGN and radio jets. We do not expect this decision to have a significant effect on our data analysis or interpretation, because even in the extreme case where the inner radio source has the same position angle as the outer radio hotpsots, our slit position angle would still be at a large angle ($\sim$45$^{\circ}$) to the radio jets.

The data reduction was performed with {\scshape iraf}. The raw spectra were bias subtracted, flat-fielded, wavelength calibrated using Hg-Ar arc lamp spectra, and then sky subtracted. In all cases, the 1$\sigma$ uncertainty in the wavelength calibration is $\le$ 2.1 A ($\le$ 140 km s$^{-1}$). The uncertainty in the wavelength calibration was estimated by calculating the median of the difference between the theoretical and the observed wavelength of the emission lines of the arc lamp spectra and also of the night-sky emission lines.
In addition, flux calibration was carried out with the spectrophotometric standard stars LTT 1020, LTT 377, LTT 1778, EG 274 and LTT 6248. Finally, the spectra were corrected for Galactic extinction using {\scshape iraf's deredden} task, assuming the extinction curve of \citet{cardelli89}.

\begin{table*}
	\centering
	\caption{Details of the sample and log of the observations. (1) Objects. (2) Redshift. (3) Maximum angular size of the radio source in arcseconds. (4) Maximum angular size of the radio source in kpc using the cosmology adopted in this paper. (5) Ly$\alpha$ luminosity. (6) Date of the observation. (7) On target integration time. (8) Projected linear scale per angular scale. (9) Position Angle (East of North). (10) Slit width. (11) Spectral resolution.}
	\label{observations}
	\begin{tabular}{lcccccccccc}
		\hline
		\hline
		\multicolumn{1}{c}{Object}        & Redshift & Size      & Size    & L(Ly$\alpha$) & Date             & T$_{exp}$      & Scale/\arcsec & P.A.     & Slit width & FWHM          \\
		\hline
		\multicolumn{1}{c}{}              &          & (\arcsec) & (kpc)   & (erg s$^{-1}$) &                  & (sec)          & (kpc/\arcsec) & ($^{o}$) & (\arcsec)  & (km s$^{-1}$) \\
		\hline
		\multicolumn{1}{c}{(1)}           & (2)      & (3)       & (4)     & (5)           & (6)              & (7)            & (8)           & (9)      & (10)       & (11)          \\
		\hline
		\multicolumn{1}{c}{MRC 0030--219} & 2.17     & $<$ 0.3   & $<$ 2.5 &      1.95$\times$10$^{43}$         & 2014-09-(25--27) & 6$\times$1800  & 8.46          & 175      & 1.03       & 266           \\
		\hline
		MRC 0406--244                     & 2.44     & 7.3       & 61      &      2.31$\times$10$^{44}$         & 2014-09-(25--27) & 6$\times$1800  & 8.30          & 30       & 1.03       & 266           \\
		\hline
		4C--00.54                         & 2.37     & 23.5      & 196     &      1.28$\times$10$^{44}$         & 2015-04-(20--21) & 10$\times$1800 & 8.34          & 90       & 1.68       & 560      \\
		\hline
		PKS 1138--262                     & 2.16     & 11.4      & 97      &      5.06$\times$10$^{43}$         & 2015-04-20       & 9$\times$1800  & 8.47          & 360      & 1.68       & 560           \\
		\hline
		TN J0920--0712                    & 2.76     & 1.4       & 11      &      2.91$\times$10$^{44}$         & 2015-04-21       & 9$\times$1800  & 8.06          & 210      & 1.68       & 560          \\
		\hline
		\hline
	\end{tabular}
\end{table*}
\begin{table*}
	\centering
	\caption{Rest-frame UV emission lines properties. (1) Object name. (2) Detected line species. (3) Rest wavelength. (4) Observed wavelength. (5) Line flux. (6) Line width. (7) Velocity shift with respect to fiducial systemic velocity.}
	\label{instru02}
	\begin{tabular}{lcccccr} % four columns, alignment for each
		\hline
		\hline
		Object & Line & $\lambda _{rest}$ & $\lambda _{obs}$ & Line Flux & FWHM  &  $\Delta$v \\
		& & \AA		      &		\AA		     & ($\times10^{-16}$ erg cm$^{-2}$ s$^{-1}$) & (km s$^{-1}$)  &  (km s$^{-1}$) \\
		\hline
		(1)  &    (2)    &   (3)  &  (4) & (5)  & (6)  & (7) \\
		\hline	
		MRC 0030--219 & Ly$\alpha$ & 1215.7 & 3854.5 $\pm$ 0.3 & 7.06 $\pm$ 0.52  &  1102 $\pm$ 62  &  -43 $\pm$ 25$^{*}$  \\
		\hline	
		MRC 0406--244 & Ly$\alpha$ & 1215.7 & 4161.1 $\pm$ 0.2 & 21.07 $\pm$ 0.85  &  1827 $\pm$ 44  & -124 $\pm$ 13 \\
		& \ion{C}{IV} & 1548.2,1550.8 & 5278.1 $\pm$ 2.0, 5296.8 $\pm$ 2.0 & 1.54 $\pm$ 0.42 & 1445 $\pm$ 332  &  -763 $\pm$ 115 \\
		& \ion{He}{II} & 1640.4 & 5610.3 $\pm$ 1.1 & 1.93 $\pm$ 0.49 & 1359 $\pm$ 313 & -311 $\pm$ 59 \\
		\hline	
		4C--00.54 & Ly$\alpha$ & 1215.7 & 4095.4 $\pm$ 0.1 & 19.84 $\pm$ 0.54  &  1037 $\pm$ 22  & 152 $\pm$ 9 \\
		& \ion{N}{V} & 1238.8,1242.8 & 4168.2 $\pm$ 2.4, 4181.6 $\pm$ 2.4 & 1.44 $\pm$ 0.38 & 2279 $\pm$ 545  &  -219 $\pm$ 173 \\
		& \ion{C}{IV} & 1548.2,1550.8 & 5210.7 $\pm$ 3.2, 5218.9 $\pm$ 0.9 & 3.22 $\pm$ 0.52 & 1059 $\pm$ 139  &  -132 $\pm$ 52 \\
		& \ion{He}{II} & 1640.4 & 5522.2 $\pm$ 0.4 & 2.80 $\pm$ 0.40 & 999 $\pm$ 131 & 7 $\pm$ 22 \\
		\hline	
		PKS 1138--262 & Ly$\alpha$ & 1215.7 & 3846.7 $\pm$ 0.4 & 14.07 $\pm$ 0.78  &  2289 $\pm$ 86  & -95 $\pm$ 31 \\
		& \ion{C}{IV} & 1548.2,1550.8 & 4890.5 $\pm$ 2.9, 4907.8 $\pm$ 2.9 & 0.96 $\pm$ 0.28 & 2081 $\pm$ 520  &  -614 $\pm$ 178 \\
		& \ion{He}{II} & 1640.4 & 5187.4 $\pm$ 1.8 & 0.85 $\pm$ 0.20 & 1451 $\pm$ 257 & -215 $\pm$ 103 \\
		\hline	
		TN J0920--0712 & \ion{O}{VI} & 1031.9,1037.6 & 3874.3 $\pm$ 0.6,3894.8 $\pm$ 0.7 & 1.58 $\pm$ 0.24  &  738 $\pm$ 87  & -179 $\pm$ 46 \\
		& Ly$\alpha$ & 1215.7 & 4564.2 $\pm$ 0.1 & 46.78 $\pm$ 0.89  &  1775 $\pm$ 22  & -180 $\pm$ 7 \\
		& \ion{N}{V} & 1238.8,1242.8 & 4650.3 $\pm$ 2.0, 4665 $\pm$ 2.0 & 1.02 $\pm$ 0.27 & 1934 $\pm$ 467  &  -237 $\pm$ 127 \\
		& \ion{C}{IV} & 1548.2,1550.8 & 5807.7 $\pm$ 1.6, 5820.7 $\pm$ 0.5 & 3.70 $\pm$ 0.40 & 958 $\pm$ 84  &  -438 $\pm$ 83 \\
		& \ion{He}{II} & 1640.4 & 6161.0 $\pm$ 0.7 & 2.16 $\pm$ 0.26 & 903 $\pm$ 80 & -3 $\pm$ 32 \\
		\hline
		\multicolumn{6}{c}{$^{(*)}$ We used the nuclear Ly$\alpha$ to define the fiducial systemic velocity of the radio galaxy MRC 0030--219.} \\
		\hline
	\end{tabular}
\end{table*}
\section{Data Analysis}
\label{analysis}
\subsection{Line profile fitting}

We created a {\scshape python} routine to fit the emission and absorption line parameters, with Gaussian and Voigt profiles being used to model the emission and absorption lines, respectively. The routine minimizes the sum of the squares of the difference between the model and data using the {\scshape lmfit} algorithm \citep{lmfit}. The analysis was applied first to a 3\arcsec$\,$ aperture centred on the HzRG, and then applied to the 2D spectrum at each position (pixel) along the slit, where the signal to noise ratio (S/N) of the emission line profile (based on the total flux) is $\geqslant$ 7.

The Ly$\alpha$ profile was parametrized using a single emission (Gaussian) kinematic component for those HzRGs which show no \ion{H}{I} absorption features (i.e. MRC 0030-219 and 4C-00.54), with a single absorption Voigt profile added for the three HzRGs which show clear \ion{H}{I} absorption (i.e. MRC 0406--244, TN J0920--0712 and PKS 1138--262). 

For the emission doublets, two Gaussians were used, with the two double components constrained to have equal FWHM, fixed wavelength separation and a fixed flux ratio (e.g., R$_{\ion{N}{V}}$ =  F$_{\lambda 1239}$ / F$_{\lambda 1243}$ = 2.0 and R$_{\ion{C}{IV}}$ = F$_{\lambda 1548.2}$ / F$_{\lambda 1550.8}$ = 2.0)\footnotemark\footnotetext[1]{ The doublet ratio for both emission lines ranges from 2:1 in the optically thin case to 1:1 in the optically thick case. We adopted the optically thin case, but using the optically thick case did not result in significant changes to the recovered kinematic properties.}. No significant \ion{C}{IV} absorption features were detected in our data, but the S/N and spectral resolution are insufficient to place scientifically useful upper limits on the column density of \ion{C}{IV}.

As a non-resonant recombination line, \ion{He}{II} is expected to provide a more reliable determination of the systemic velocity of a HzRG than Ly$\alpha$, \ion{N}{V} or \ion{C}{IV}, which can be susceptible to absorption and radiative transfer effects that can result in line broadening and velocity shifts. Following \citet{VM1}, we use the nuclear \ion{He}{II} to define the fiducial systemic velocity of each HzRG, with the exception of MRC 0030--219 where we instead use Ly$\alpha$ due to non-detection of \ion{He}{II}. 

The kinematic properties of the extended gas were determined using the fitting routine already mentioned, from where we obtained the FWHM and the velocity offset for the emission and absorption lines, at each spatial position along the slit.

In Table \ref{instru02}, we show the emission lines detected for each radio galaxy along with the parameters from the best fits. In addition, in Table \ref{instru03} we show the best fit parameters for the Ly$\alpha$ absorption.
In Figures \ref{mrc0030kin} to \ref{pks1138kin}, we show (i) the 2-D spectra of the Ly$\alpha$ spectral region, which were smoothed using a Gaussian with kernel = 3.0 pixels (or 0.9\arcsec); (ii) the spatial variation of the flux of the Ly$\alpha$ emission line; (iii) Ly$\alpha$ spatial profile compared with the seeing; (iv) the 1-D spectra of the Ly$\alpha$ profile; (v) the spatial variation of FWHM corrected for the instrumental broadening; (vi) the spatial variation of velocity offset relative to our fiducial systemic velocity; (vii) the FWHM as a function of the velocity offset of Ly$\alpha$; (viii) the spatial variation of the flux of the \ion{He}{II} emission line (the radio galaxy 4C--00.54 was the only one that provided a sufficient S/N  at each spatial position along the slit); (ix) the Ly$\alpha$/\ion{He}{II} flux ratio for the radio galaxy 4C--00.54; (x) the spatial variation of the velocity offset for the extended \ion{H}{I} absorber when detected and (xi) the spatial profile of the \ion{H}{I} column density compared with the seeing. 
\subsection{Spatial extent constraints using the seeing}

In order to investigate whether the Ly$\alpha$ halos detected in our sample are spatially extended along the slit, we have compared the Ly$\alpha$ spatial profile with the seeing profile reconstructed from stars in images taken immediately before or after the observation of the science target (see \citealt{humphrey2015} and \citealt{montse2016} for more details of this methodology). The flux of a non-saturated star was extracted by simulating the slit width used for the spectroscopic observation of the science targets (see Table \ref{observations}). The sky background was then removed from the stellar spatial profile.  

\begin{figure*}
	\centering	\textbf{MRC 0030--219}
	
	\subfloat[Ly$\alpha$ 2-D spectrum]{
		\includegraphics[width=\columnwidth,height=2.10in,keepaspectratio]{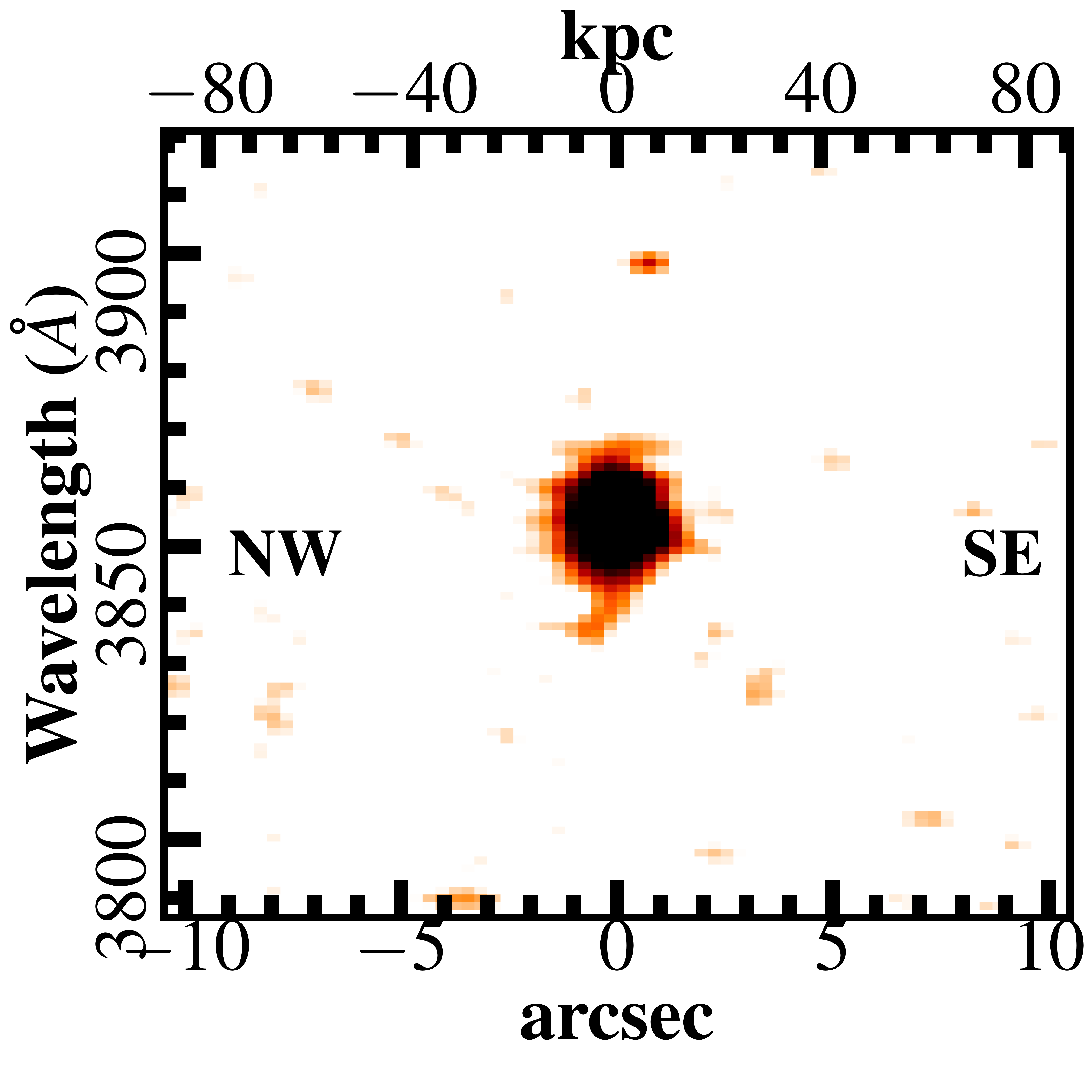}
		\label{2dspec_mrc0030}}
	\subfloat[Ly$\alpha$ Flux]{
		\includegraphics[width=\columnwidth,height=1.90in,keepaspectratio]{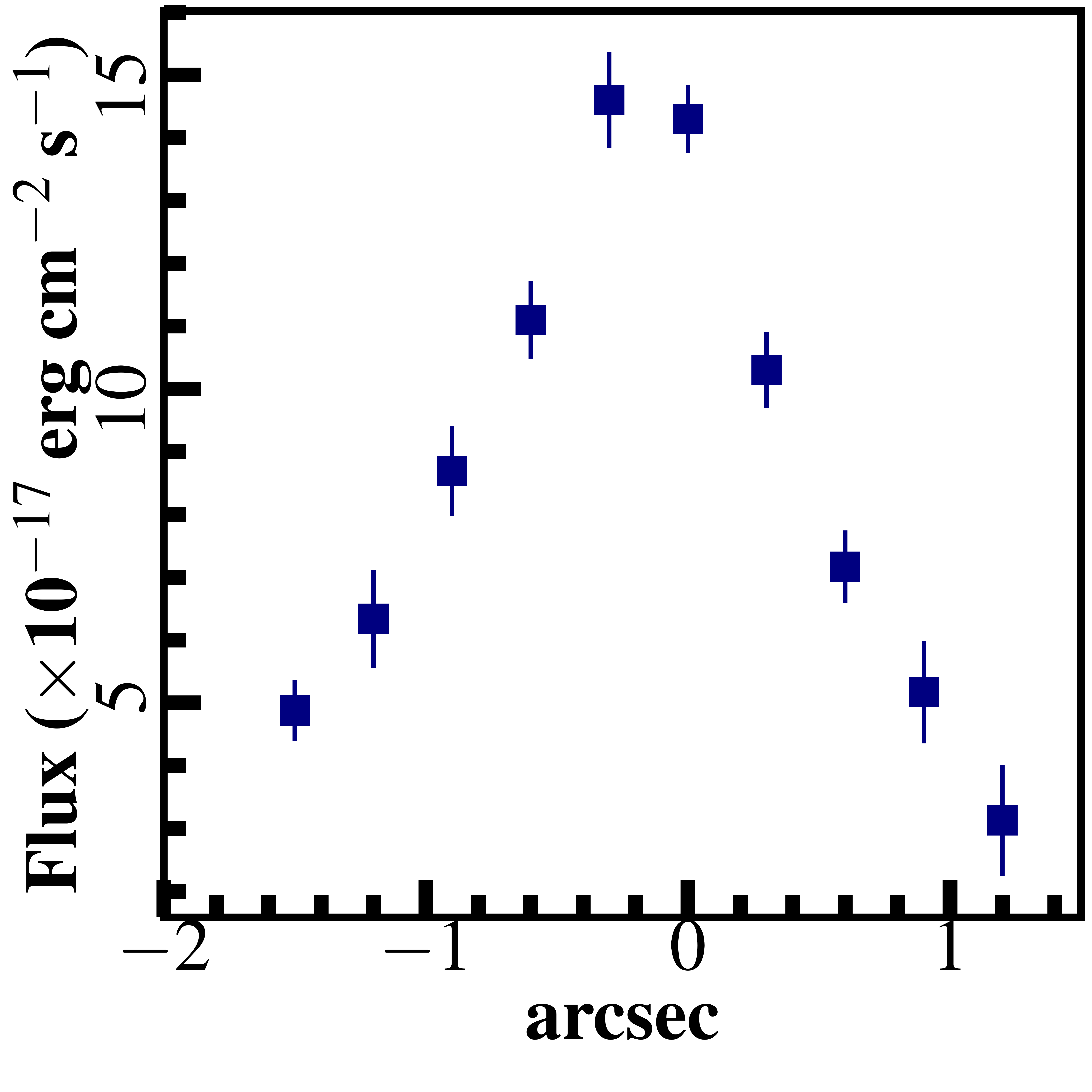}
		\label{ly_flux_mrc0030}}
	\subfloat[Ly$\alpha$ Source \textit{vs.} Seeing]{
		\includegraphics[width=\columnwidth,height=1.90in,keepaspectratio]{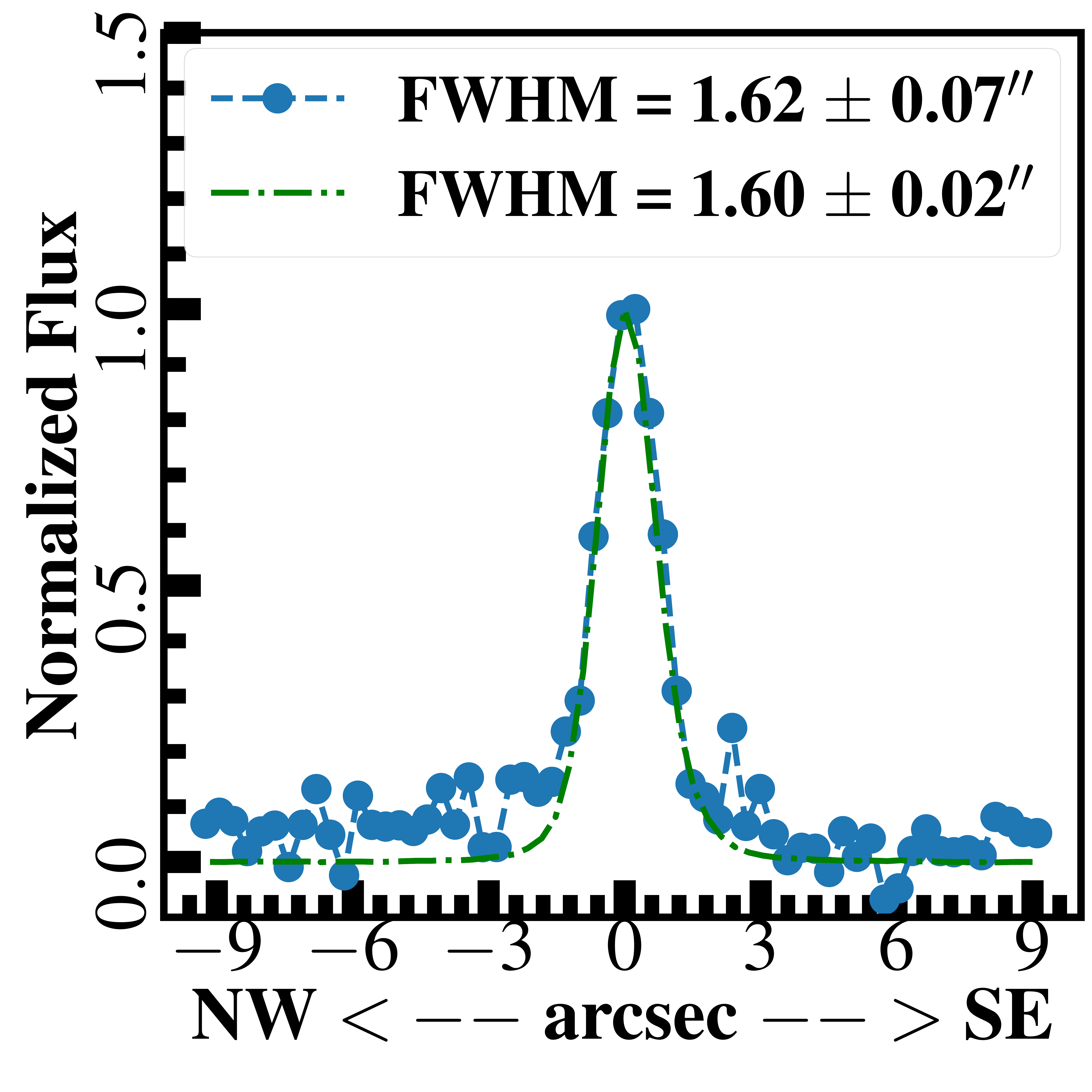}
		\label{mrc0030_seeing}}
	\quad
	\subfloat[Ly$\alpha$ 1-D spectrum]{
		\includegraphics[width=\columnwidth,height=1.93in,keepaspectratio]{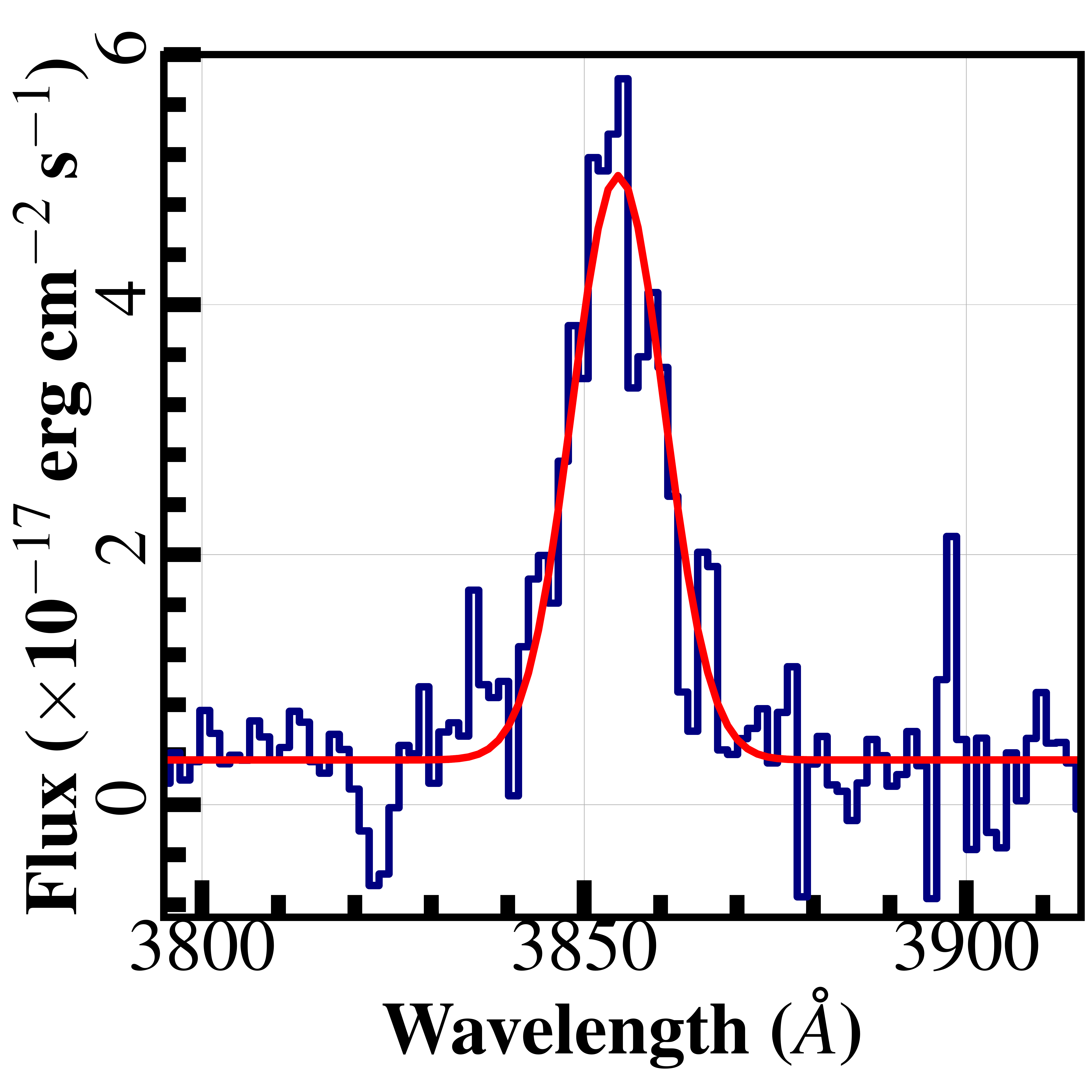}
		\label{1dspec_mrc0030}}
	\subfloat[Ly$\alpha$ FWHM]{
		\includegraphics[width=\columnwidth,height=1.90in,keepaspectratio]{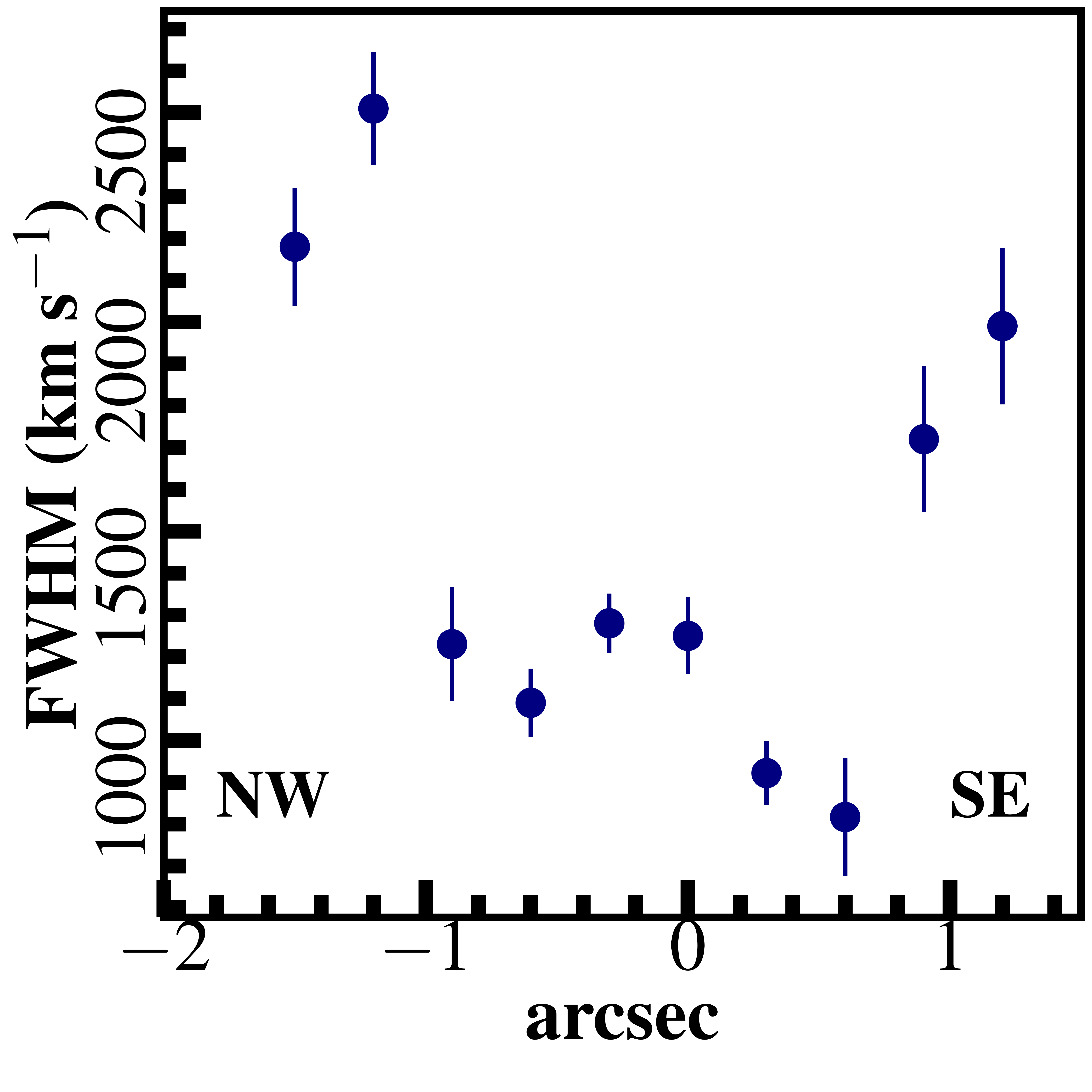}
		\label{fwhm_mrc0030}}
	\subfloat[Ly$\alpha$ Velocity]{
		\includegraphics[width=\columnwidth,height=1.90in,keepaspectratio]{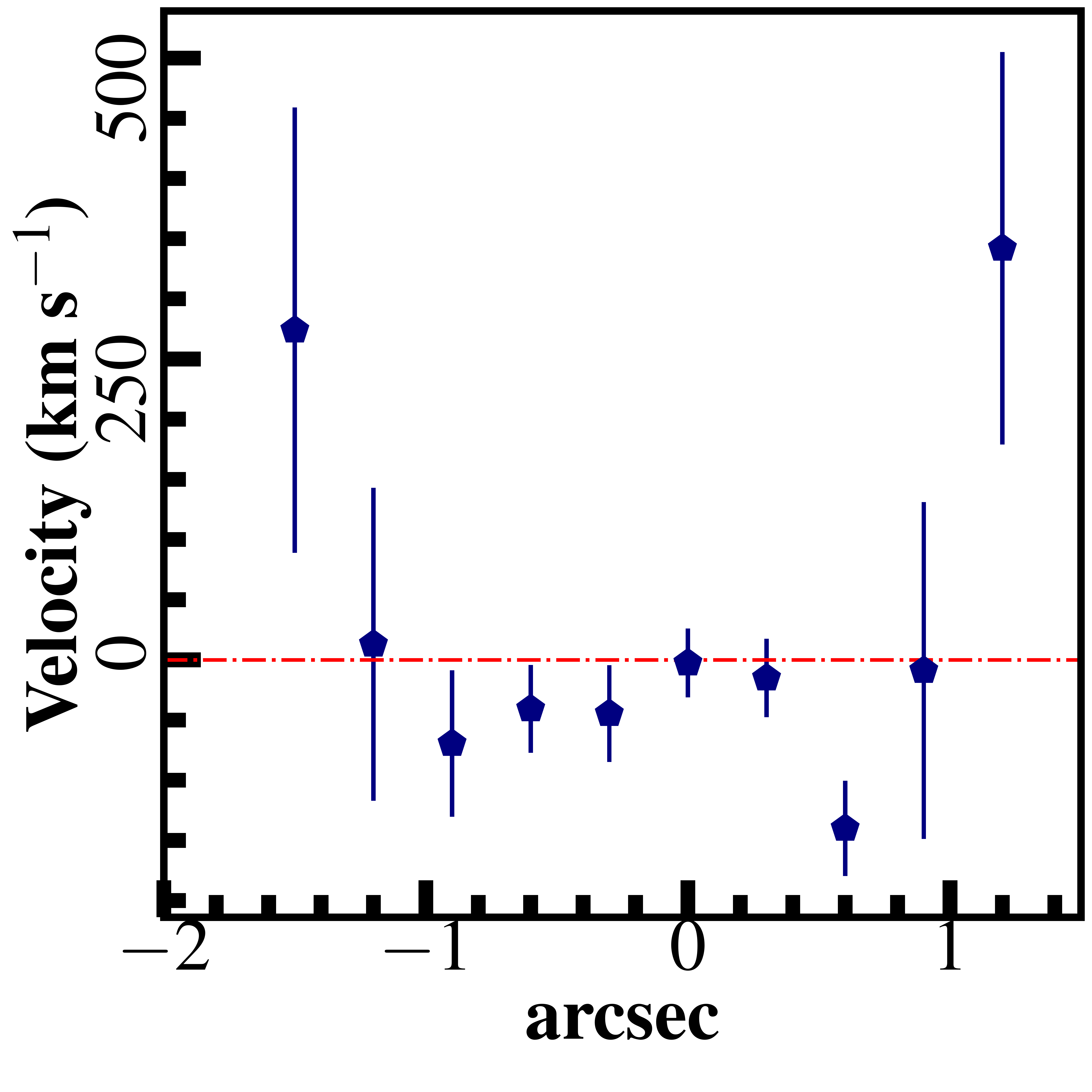}
		\label{velo_mrc0030}}
	\quad
	\subfloat[Ly$\alpha$ FWHM \textit{vs.} Velocity]{
		\includegraphics[width=\columnwidth,height=1.90in,keepaspectratio]{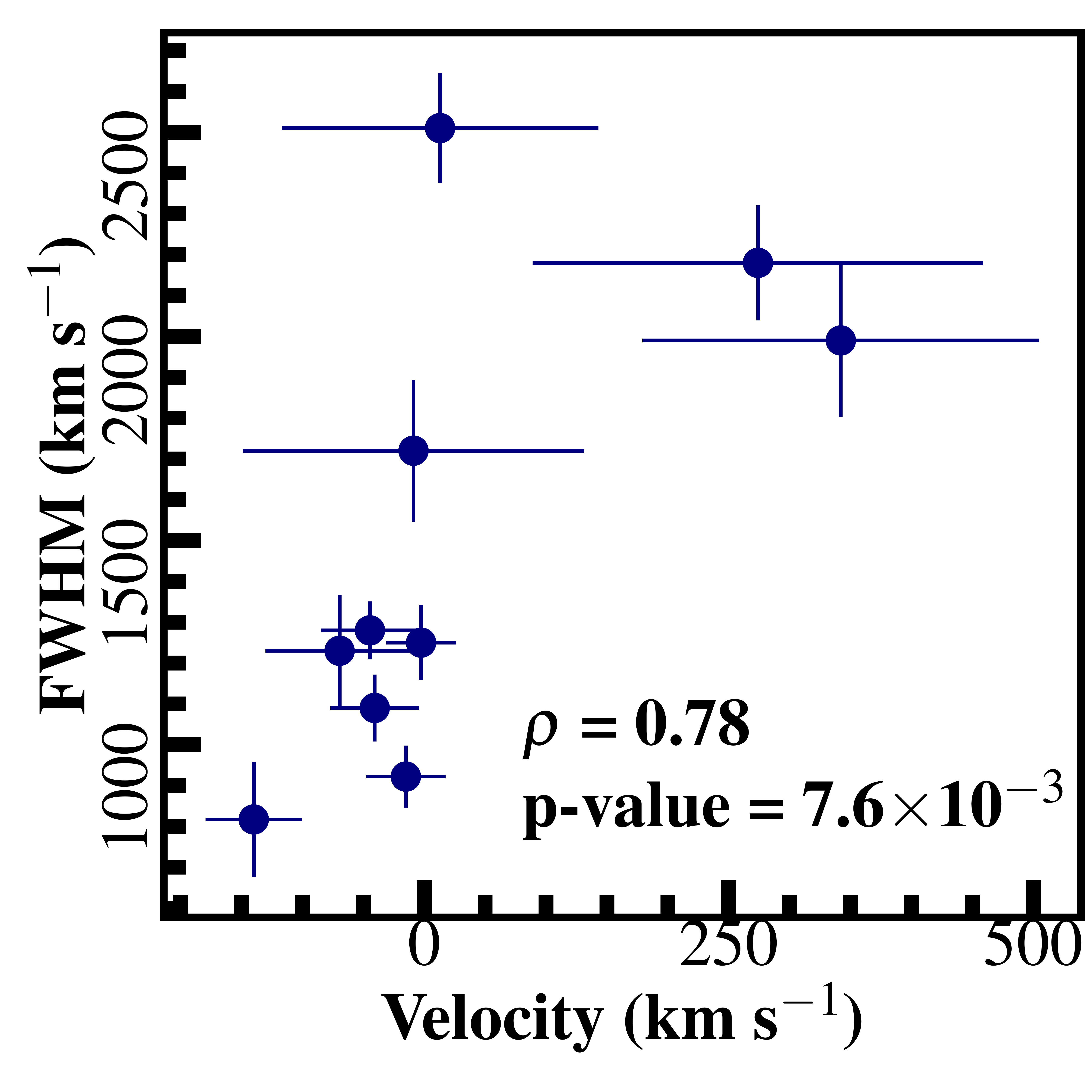}
		\label{corr_mrc0030}}

	\caption{Radio galaxy MRC 0030--219: (a) 2-D spectrum of the Ly$\alpha$ spectral region, (b) Flux of the Ly$\alpha$ emission line, (c) Ly$\alpha$ spatial profile (blue circle with dashed lines) compared with the seeing (green dot dashed lines) and (d) 1-D spectrum of the Ly$\alpha$ spectral region extracted from the SOAR long-slit. The Ly$\alpha$ emission-line was extracted by summing over a 3\arcsec$\,$ region of the slit length. Spatial variations of (e) FWHM, (f) Velocity and (g) Variation of FWHM as a function of the velocity offset of Ly$\alpha$ with $\rho$ and p-value representing the Spearman's rank correlation coefficient and t-distribution, respectively.}
	\label{mrc0030kin}
\end{figure*}

In Figures \ref{mrc0030_seeing} to \ref{pks1138_seeing}, we show the spatial profile of the Ly$\alpha$ emission along with the spatial profile of the seeing. We measured the FWHM (FWHM$_{source}$ $\pm$ $\Delta$FWHM$_{source}$) of the spatial profile of the Ly-alpha line by fitting a single Gaussian. If the seeing has FWHM$^{\prime}$ $\pm$ $\Delta$FWHM$^{\prime}$, we assume that the source is spatially unresolved when
\begin{align}
FWHM_{source} \leq FWHM^{\prime} + \Delta FWHM^{\prime} + \Delta FWHM_{source}.
\end{align}
In this case, the right hand side of this inequality will be used as an upper limit for the intrinsic FWHM. For a resolved source, we estimate the intrinsic FWHM by subtracting the seeing FWHM from the observed FWHM in quadrature. 

\section{Results for individual objects}
\label{prev}

\subsection{MRC 0030--219}

\subsubsection{Previous results}

In the z = 2.17 radio galaxy MRC 0030--219 the radio source consists  of a single compact component with maximum angular size $<$ 0.3\arcsec (or $<$ 2.5 kpc in the adopted cosmology) and has a steep radio spectrum ($\alpha \approx$ -1.0; \citealt{carilli1997}), as revealed by VLA observations. Optical spectroscopy observations from the Cerro Tololo 4 m Telescope revealed strong UV emission lines such as Ly$\alpha$, \ion{C}{IV} and \ion{He}{II} \citep{McCarthy1990}. The Ly$\alpha$ emission line has a rest equivalent width W$^{rest} _{\lambda}$ = 174 \AA$\,$ and luminosity L(Ly$\alpha$) = 10$^{43.66}$ erg s$^{-1}$ \citep{McCarthy1990}.

\subsubsection{Results from SOAR}

\begin{figure*}
	\centering	\textbf{MRC 0406--244}
	
	\subfloat[Ly$\alpha$ 2-D spectrum]{
		\includegraphics[width=\columnwidth,height=2.10in,keepaspectratio]{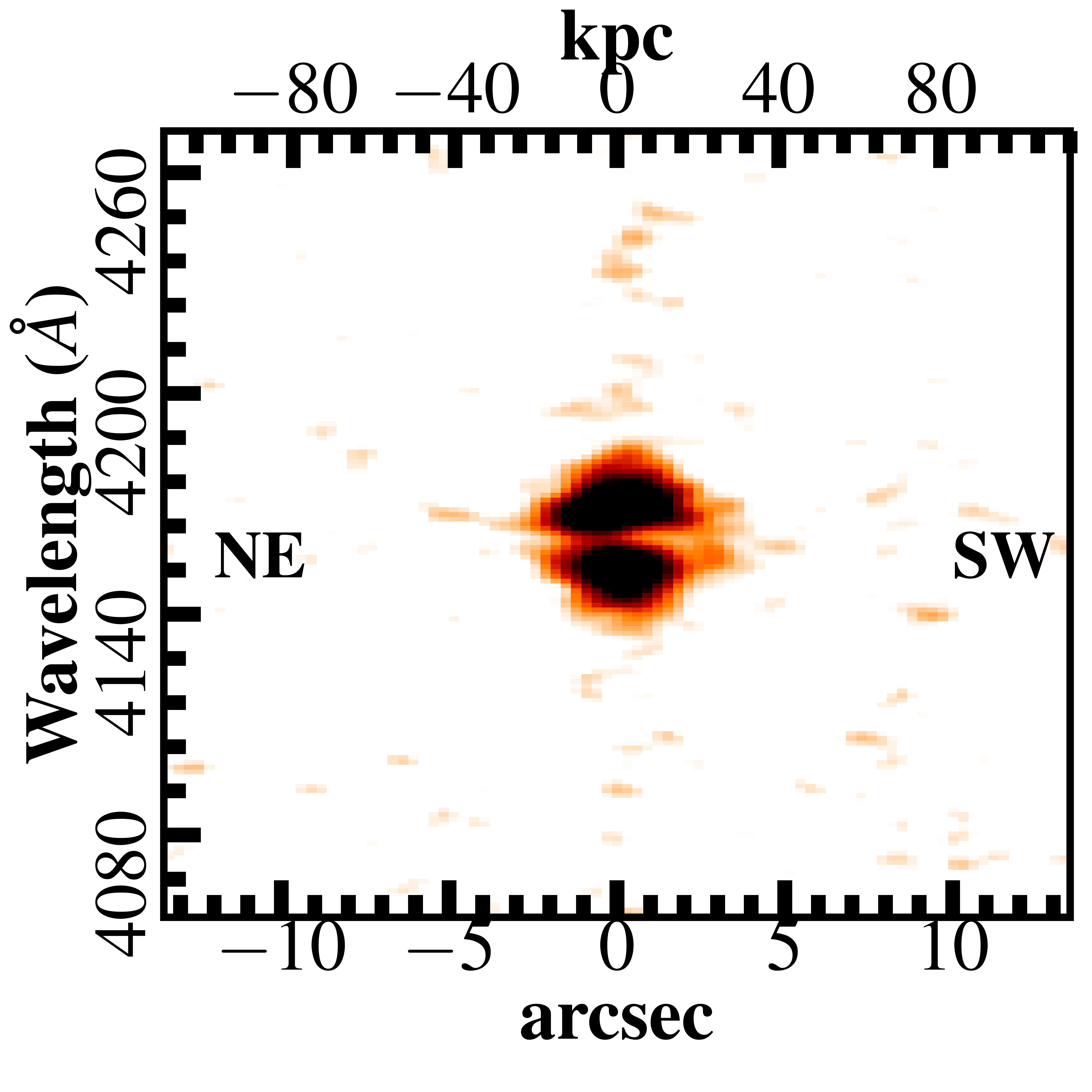}
		\label{2dspec_mrc0406}}
	\subfloat[Ly$\alpha$ Flux]{
		\includegraphics[width=\columnwidth,height=1.9in,keepaspectratio]{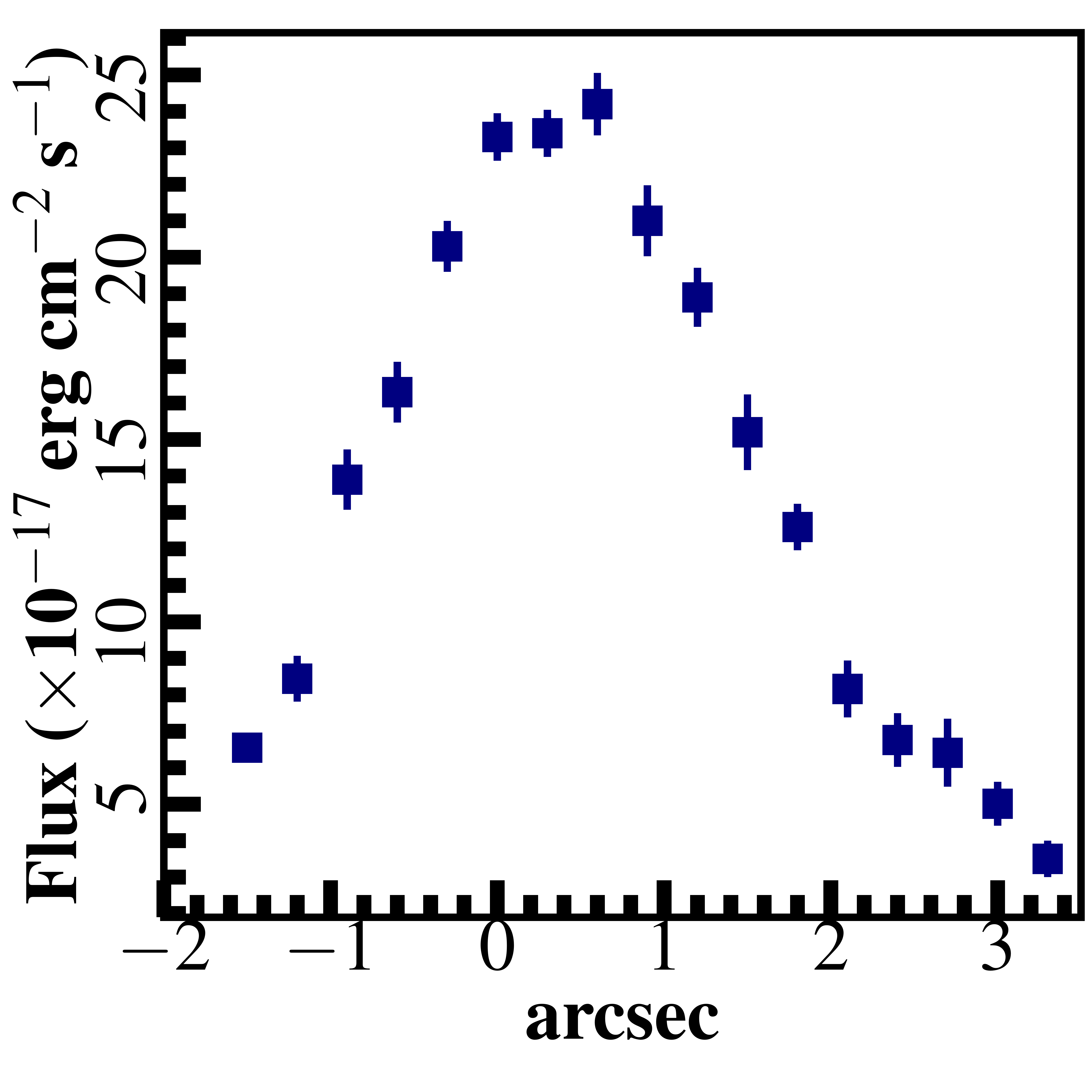}
		\label{ly_flux_mrc0406}}
	\subfloat[Ly$\alpha$ Source \textit{vs.} Seeing]{
		\includegraphics[width=\columnwidth,height=1.9in,keepaspectratio]{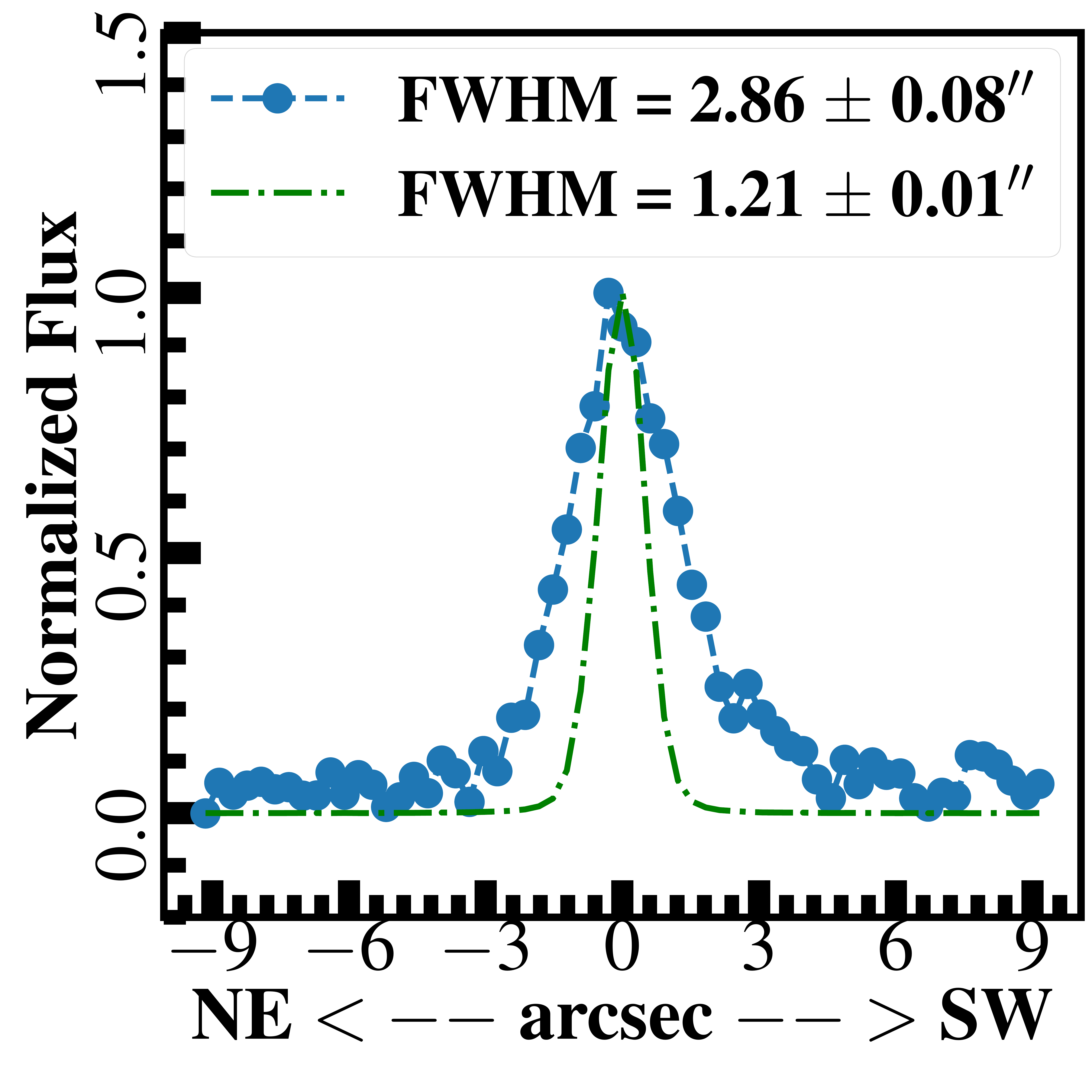}
		\label{mrc0406_seeing}}
	\quad
	\subfloat[Ly$\alpha$ 1-D spectrum]{
		\includegraphics[width=\columnwidth,height=1.9in,keepaspectratio]{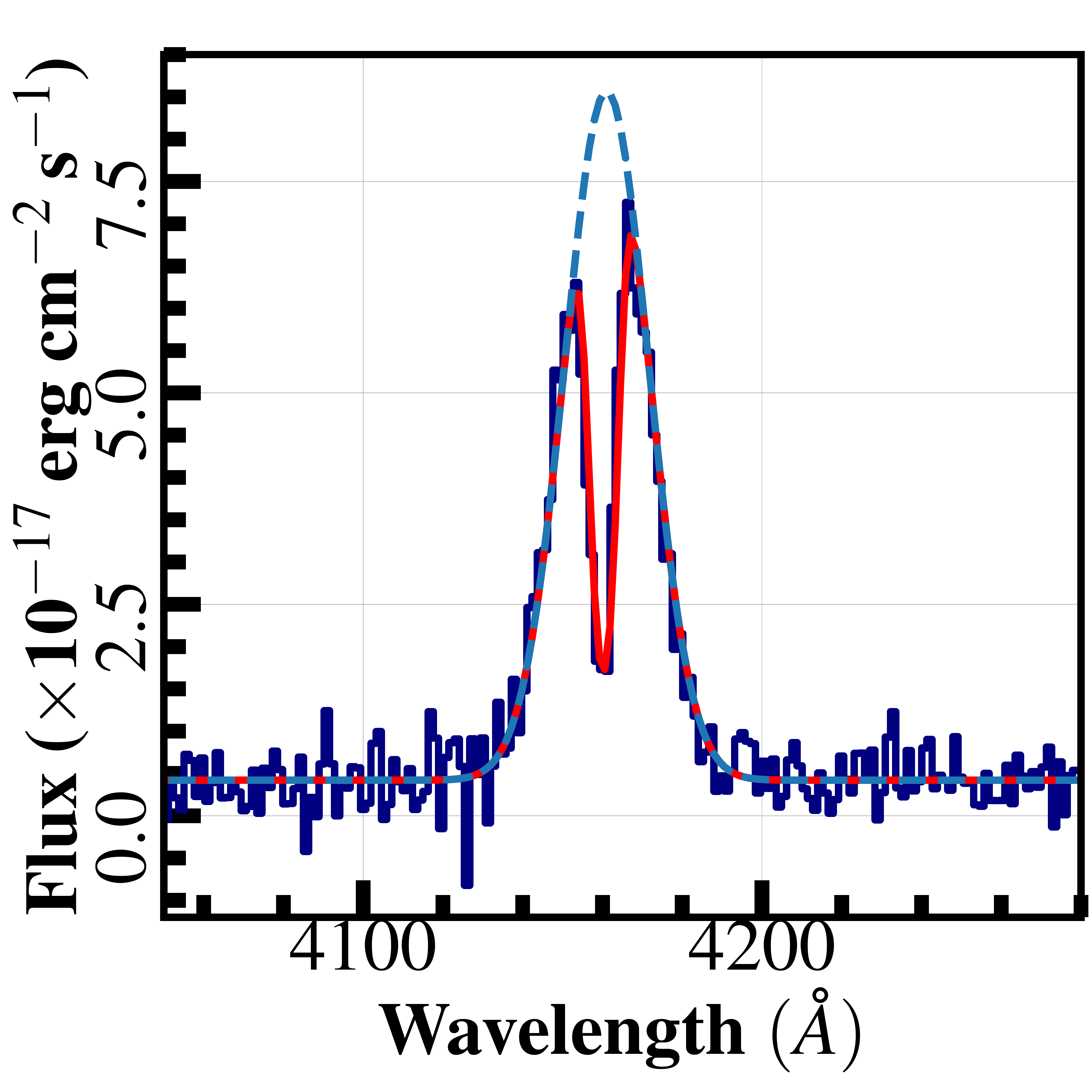}
		\label{1dspec_mrc0406}}
	\subfloat[Ly$\alpha$ FWHM]{
		\includegraphics[width=\columnwidth,height=1.9in,keepaspectratio]{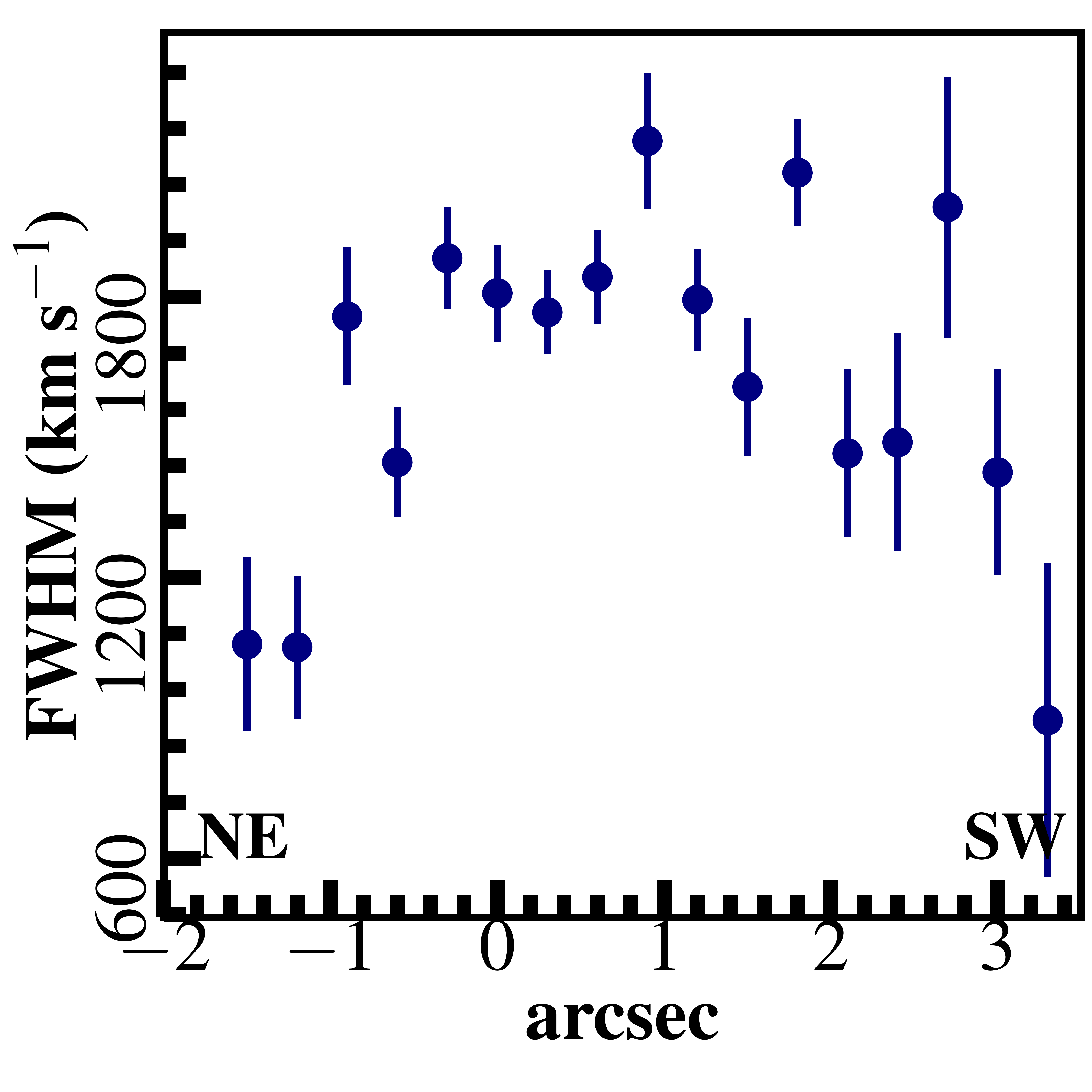}
		\label{fwhm_mrc0406}}
	\subfloat[Ly$\alpha$ Velocity]{
		\includegraphics[width=\columnwidth,height=1.9in,keepaspectratio]{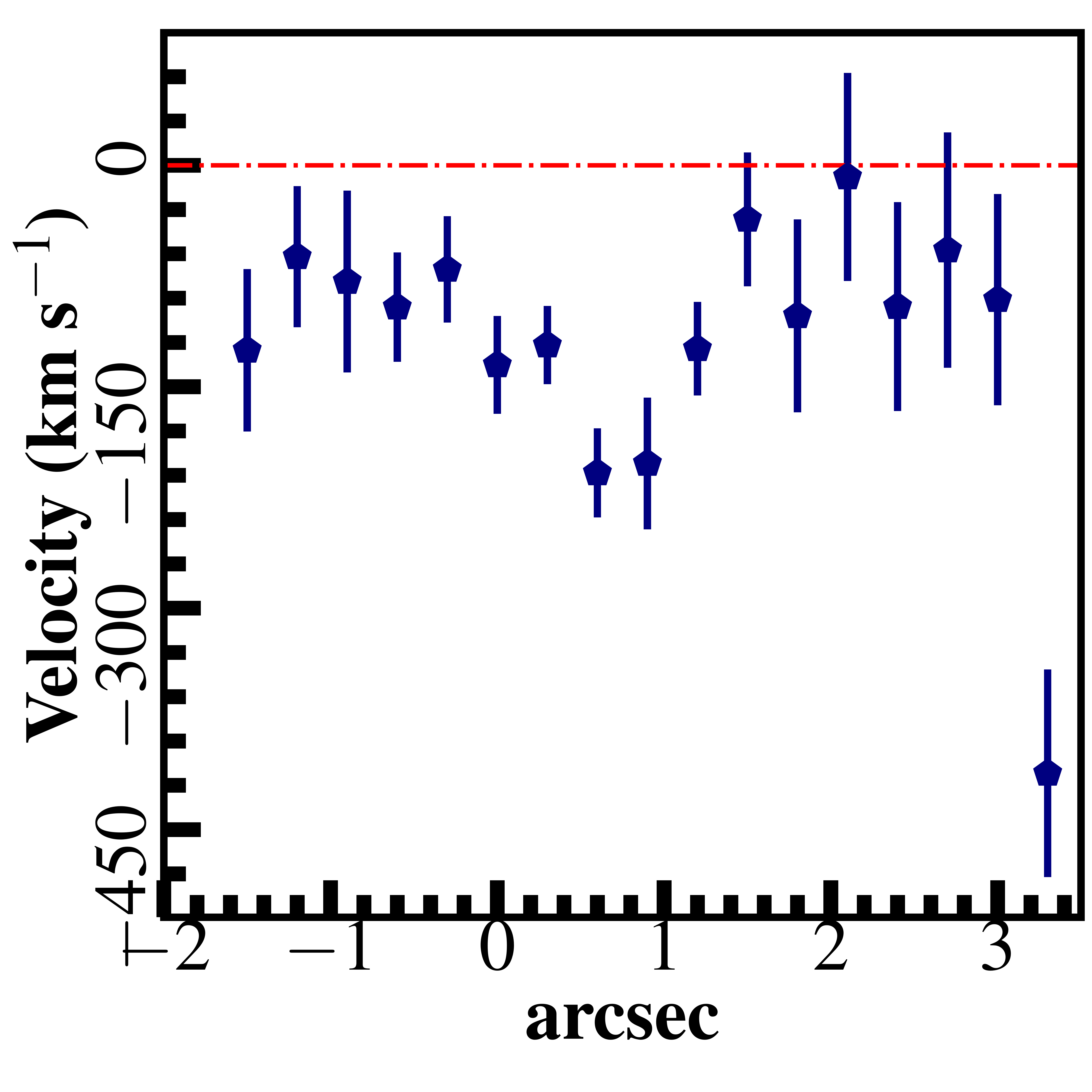}
		\label{velo_mrc0406}}
	\quad
	\subfloat[Ly$\alpha$ FWHM \textit{vs.} Velocity]{
		\includegraphics[width=\columnwidth,height=1.9in,keepaspectratio]{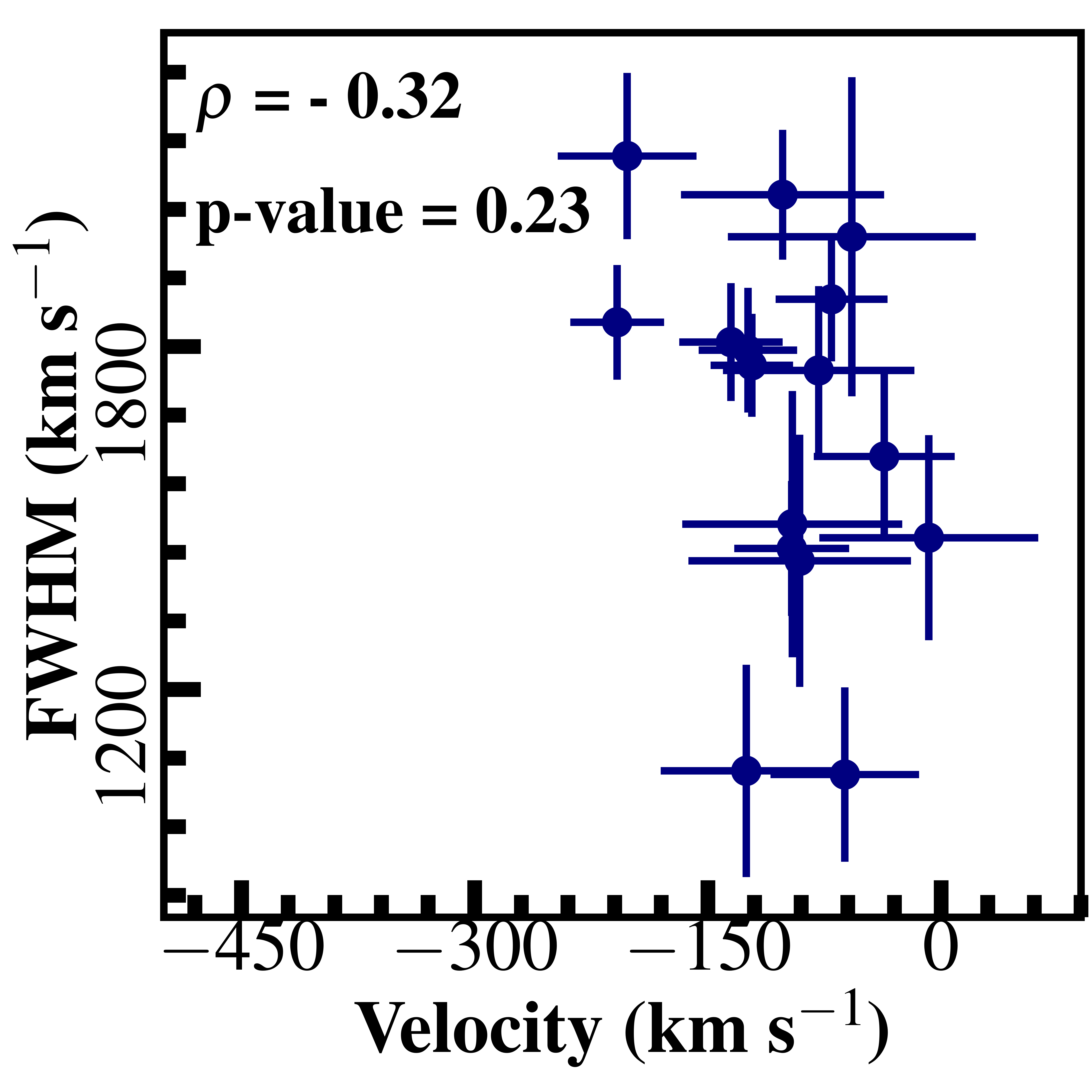}
		\label{corr_mrc0406}}
	\subfloat[Ly$\alpha$ Velocity Abs.]{
		\includegraphics[width=\columnwidth,height=1.9in,keepaspectratio]{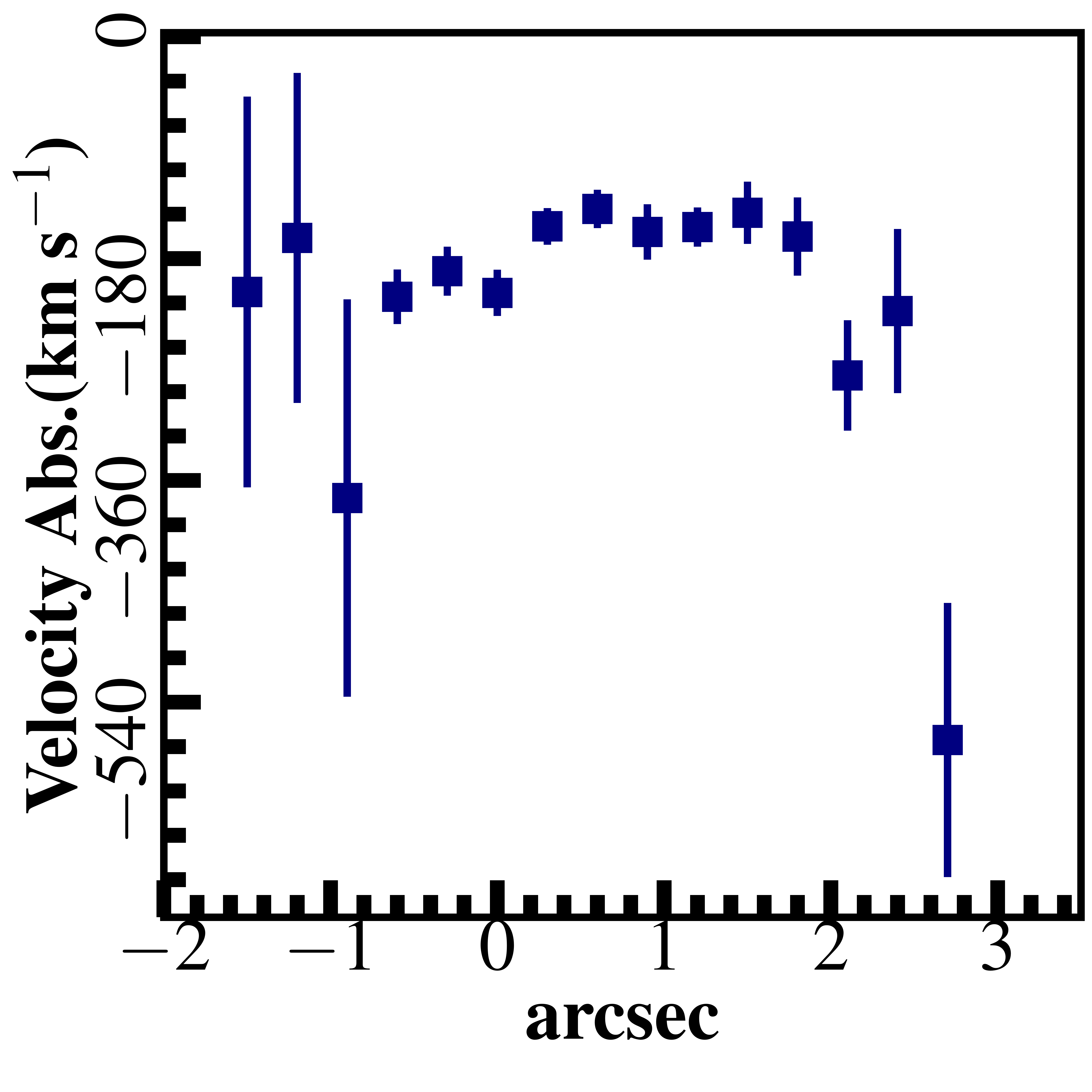}
		\label{vabs_mrc0406}}
	\subfloat[Ly$\alpha$ N(\ion{H}{I}) \textit{vs.} Seeing]{
		\includegraphics[width=\columnwidth,height=1.9in,keepaspectratio]{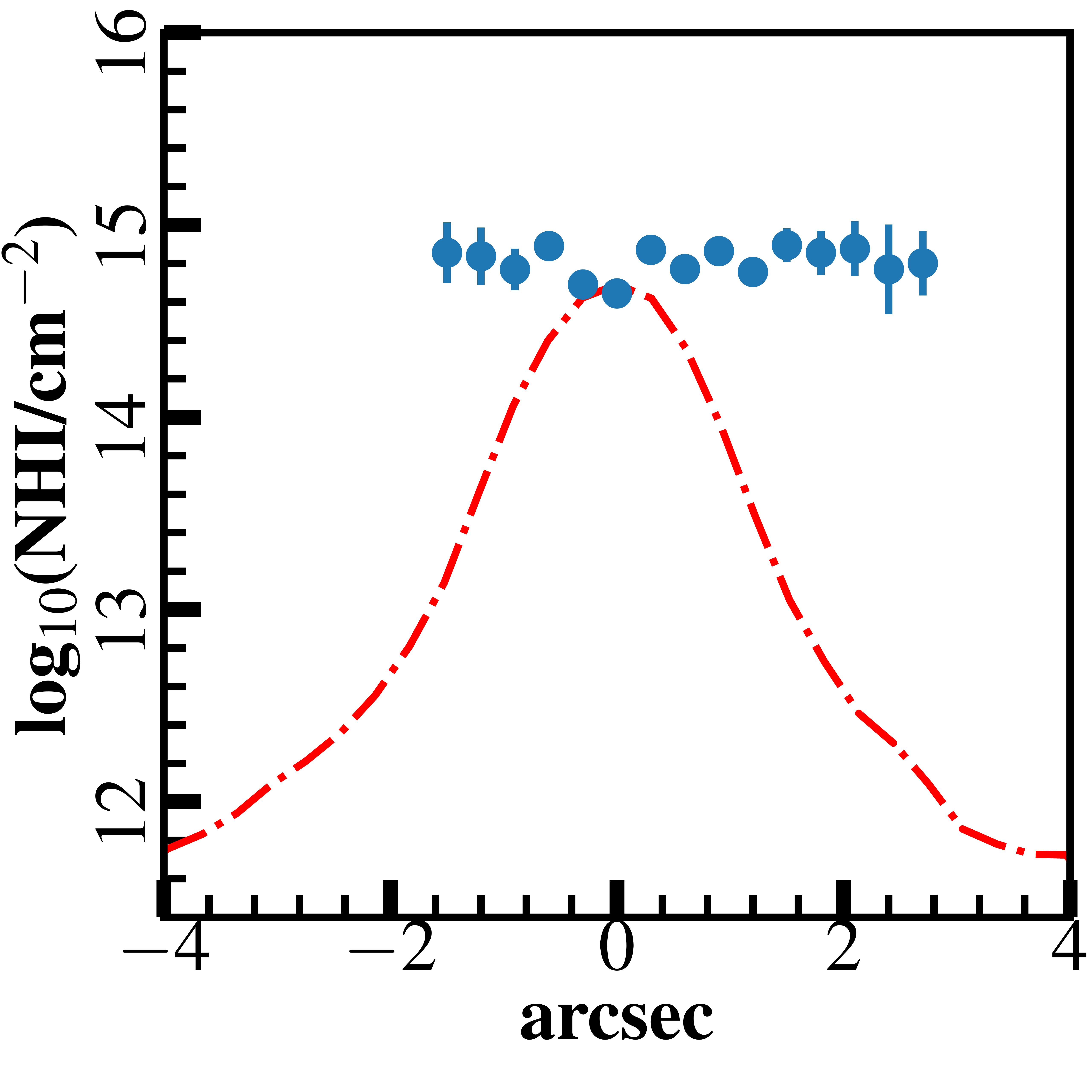}
		\label{NHImrc0406_seeing}}
	
	\caption{Radio galaxy MRC 0406-244: (a) 2-D spectrum of the Ly$\alpha$ spectral region, (b) Spatial variation of the flux of Ly$\alpha$ line, (c) Ly$\alpha$ spatial profile (blue circle with dashed lines) compared with the seeing (green dot dashed lines) and (d) 1-D spectrum of the Ly$\alpha$ spectral region extracted from the SOAR long-slit. The Ly$\alpha$ emission-line was extracted by summing over a 3\arcsec$\,$ region of the slit length. Spatial variations of (e) FWHM, (f) Velocity, (g) Variation of FWHM as a function of the velocity offset of Ly$\alpha$ with $\rho$ and p-value representing the Spearman's rank correlation coefficient and t-distribution, respectively, (h) Velocity of the \ion{H}{I} absorber and (i) Spatial profile of the \ion{H}{I} column density (blue circle points) compared with the seeing (red dot dashed lines), both on a logarithmic scale. In addition, the seeing profile has been normalised and shifted in order to allow the comparison.}
	\label{mrc0406kin}
\end{figure*}

Figure \ref{2dspec_mrc0030} reveals the spatially compact Ly$\alpha$ emission of MRC 0030--219. Figure \ref{ly_flux_mrc0030} shows the spatial variation of the Ly$\alpha$ flux, with this emission being detected further to the NW direction.
In Figure \ref{1dspec_mrc0030}, we also show the integrated 1-D spectrum of the Ly$\alpha$ profile.
Figures \ref{fwhm_mrc0030} and \ref{velo_mrc0030} show the spatial variation of the FWHM and velocity offset of the Ly$\alpha$ emission line. The line width varies in the range FWHM = 800 -- 2500 km s$^{-1}$. Within a radius of $\leq$ 1\arcsec$\,$ of the nucleus, Ly$\alpha$ shows FWHM $<$ 1400 km s$^{-1}$, increasing to 1600 -- 2500 km s$^{-1}$ at radii of $\geq$ 1\arcsec. The spatially integrated Ly$\alpha$ emission line appears blueshifted from the systemic velocity with velocity offset -43 $\pm$ 25 km s$^{-1}$. In order to investigate the possible correlations between the FWHM and the velocity offset (see Fig. \ref{corr_mrc0406}), we use the Spearman correlation ($\rho$) and the t-distribution (p-value), which indicates a positive relationship between the FWHM and velocity curve of the Ly$\alpha$ emission with $\rho$ = 0.78 and p-value = 7.6 $\times$ 10$^{-3}$.
The kinematic properties will be discussed in $\S$\ref{dynamics}.

The Ly$\alpha$ spatial profile with FWHM = 1.62 $\pm$ 0.07\arcsec$\,$ is dominated by a central compact source (see Fig. \ref{mrc0030_seeing}), which appears barely resolved in the central regions compared with the seeing (1.60 $\pm$ 0.02\arcsec). Correcting for seeing broadening in quadrature, we infer FWHM$_{obs}$ =  0.54 $\pm$ 0.09\arcsec$\,$ or 4.6 $\pm$ 0.7 kpc. In addition, the variation seen in the kinematic properties of the radio galaxy  (Figures \ref{fwhm_mrc0030} and \ref{velo_mrc0030}), in the outer parts of the Ly$\alpha$ profile at $\geq$ 1\arcsec$\,$ from the centroid also suggests that it may be barely resolved. None of the other UV emission lines are found to be extended in this spectrum.
%
%%\newpage
\subsection{MRC 0406--244}

\subsubsection{Previous results}

This object consists of a massive host galaxy (M$_{\star}$ $\sim$ 10$^{11}$ M$_{\odot}$; \citealt{seymour2007,Hatch2013}) with a high star-formation rate (790 $\pm$ 75 M$_{\odot}$ yr$^{-1}$; \citealt{Hatch2013}). 
Hubble Space Telescope (HST) images show spatially resolved continuum emission, with several connecting bright clumps in a figure of eight morphology elongated along the radio source of the radio galaxy \citep[e.g.][]{rush1997,Pe2001,Hatch2013}. 
\cite{rush1997} concluded that the complex morphology of the spatially resolved continuum in MRC 0406--244 could be a consequence of a recent merger. 
On the other hand, \cite{Tani2001} and \cite{humphrey2009} argued that this morphology might be a consequence of AGN-driven winds (AGN feedback) or a superwind from a starburst event which swept up superbubbles from the ambient ISM. \cite{Hatch2013}, however argued that the continuum emission is most likely to be due to young stars or dust-scattered light from the AGN.
Using VLT/SINFONI imaging spectroscopy of the rest-frame optical emission lines, \cite{nesvadba2008,nesvadba17,nesvadba2017} find extended emission line regions with large velocity offset in the range -600 to +600 km s$^{-1}$ and line widths in the range 500 to 1500 km s$^{-1}$ consistent with very turbulent outflowing gas. They conclude that the radio jets are the main driver of the gas kinematics.
\cite{Tani2001} and \cite{humphrey2009} studied the emission line ratios of the extended gas and concluded that photoionization by the AGN is the most probable excitation mechanism of this gas. A Ly$\alpha$ image taken using the 2.5 m du Pont Telescope reveals line emission with an extent of 3\arcsec $\times$ 5\arcsec$\,$ (or 24.9 kpc $\times$ 41.5 kpc at the adopted cosmology) in which the long axis is about 130$^{\circ}$ east of north aligned with the radio source \citep{rush1997}. In addition, using data from the 3.58 m ESO New Technology Telescope (NTT), \citet{Pe2001} identified a strong, extended \ion{H}{I} absorption feature superimposed on the bright Ly$\alpha$ emission line.

\subsubsection{Results from SOAR}

In our SOAR data, three emission lines were detected in the spectrum of MRC 0406--244 (see Table \ref{instru02}). Strong Ly$\alpha$ emission is detected in the direction perpendicular to the radio axis of the galaxy (see Fig. \ref{2dspec_mrc0406}). The Ly$\alpha$ emission shows an asymmetric spatial distribution, which is detected further in the SW direction (see Fig. \ref{ly_flux_mrc0406}). \ion{C}{IV} and \ion{He}{II} emission lines are also detected, however they are spatially compact and also detected more in the SW direction. In Figure \ref{1dspec_mrc0406}, we show the integrated 1-D spectrum of the Ly$\alpha$ profile.
Figure \ref{fwhm_mrc0406} shows the spatial variations of the FWHM of the Ly$\alpha$ nebula (1000 -- 2300 km s$^{-1}$). The line width is relatively high across the full extent of the emission line, decreasing to 1000 km s$^{-1}$ in the outermost regions.
The Ly$\alpha$ velocity shift relative to fiducial velocity shows the most blueshifted gas ($\sim$ 200 km s$^{-1}$) around the nuclear region of the nebula (within 1\arcsec). Towards to the outermost regions of the nebula we find velocity offset varying from -135 km s$^{-1}$ to -8 km s$^{-1}$ (see Fig. \ref{velo_mrc0406}). Investigating the possible correlations between the FWHM and the velocity offset (see Fig. \ref{corr_mrc0406}), we find a weak negative relationship with $\rho$ = - 0.32 and p-value = 0.23.

The Ly$\alpha$ spatial profile with FWHM = 2.86 $\pm$ 0.08\arcsec$\,$ is clearly spatially resolved compared with the seeing FWHM = 1.21 $\pm$ 0.01 (see Fig. \ref{mrc0406_seeing}), indicative of a extended emission line. Correcting for seeing broadening, the intrinsic FWHM is 2.59 $\pm$ 0.09\arcsec (or 22 $\pm$ 1 kpc). None of the other UV emission lines are found to be extended in this spectrum.

We detected a Ly$\alpha$ absorption feature in the spectrum of MRC 0406--244 (see Fig. \ref{2dspec_mrc0406}).
The best fit to the Ly$\alpha$ profile is shown in Fig. \ref{1dspec_mrc0406}. Table \ref{instru03} lists the parameters of the best fitting model together with the diameter of the absorber, and the maximum detected radius. The absorber has column density log N(\ion{H}{I}/cm$^{-2}$) = 14.81 $\pm$ 0.03 with Doppler parameter $\mathit{b}$ = 205 $\pm$ 11 km s$^{-1}$. This structure is detected across the full spatial extent of the Ly$\alpha$ emission where S/N in the line is sufficient to detect an absorber with that column density. In Figure \ref{vabs_mrc0406}, we show the line of sight velocity of the \ion{H}{I} absorber measured from the SOAR spectrum across its detected spatial extent. The absorbing gas appears blueshifted from the systemic velocity with line of sight velocity -184 $\pm$ 8 km s$^{-1}$. 
We find that the absorber is detected extending over across 4.2\arcsec$\,$ (or 35 kpc) in the direction perpendicular to the radio axis of the galaxy. In addition, the Ly$\alpha$ absorption feature detected in the spectrum of MRC 0406--244 shows a constant N(\ion{H}{I}) along the slit, which suggests to be spatially extended when comparing the spatial profile of the \ion{H}{I} column density with the seeing profile (see Fig. \ref{NHImrc0406_seeing}). If the Ly$\alpha$ emission is extended and the absorber is not extended, we should expect a radial decline in the strength of the absorber (e.g. N(\ion{H}{I})).

Assuming that this structure is a spherically symmetric shell, its \ion{H}{I} mass is given by
\begin{align}
M_{\ion{H}{I}} = 4 \pi R^{2} N(\ion{H}{I}) m_H
\end{align} 
which simplifies to

\begin{align}
\label{m_HI}
M_{\ion{H}{I}} \gtrsim 5.3\times10^{3}(R/23\,kpc)^{2}(N(\ion{H}{I})/10^{14} cm^{-2}) M_{\sun},  
\end{align} 
where $R$ is the radius of the absorption system in kpc, and N(\ion{H}{I}) is the \ion{H}{I} column density in cm$^{-2}$, and $m_H$ is the mass of a hydrogen atom. We estimate the mass of the absorbing shell of gas to be log ($M_{\ion{H}{I}}/M_{\sun}$) $\gtrsim$ 4.5. If the absorbing gas is partly ionized, then its total mass (i.e. M(\ion{H}{I}) + M(\ion{H}{II})) could be substantially higher.
\subsection{4C--00.54}
\label{previous_4c}

\begin{figure*}
	\centering	\textbf{4C--00.54}
	
	\subfloat[Ly$\alpha$ 2-D spectrum]{
		\includegraphics[width=\columnwidth,height=2.10in,keepaspectratio]{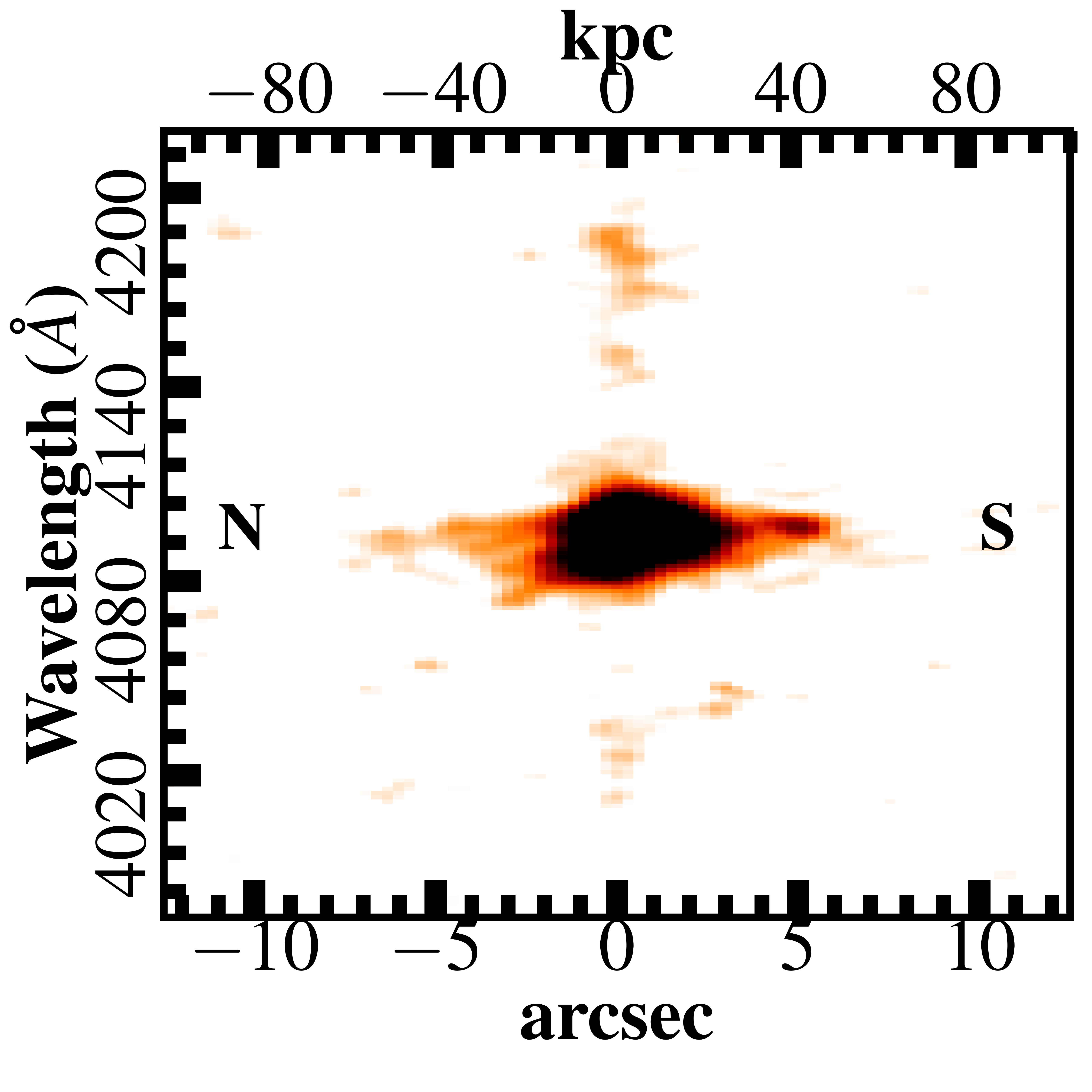}
		\label{2dspec_4c}}
	\subfloat[Ly$\alpha$ Flux]{
		\includegraphics[width=\columnwidth,height=1.9in,keepaspectratio]{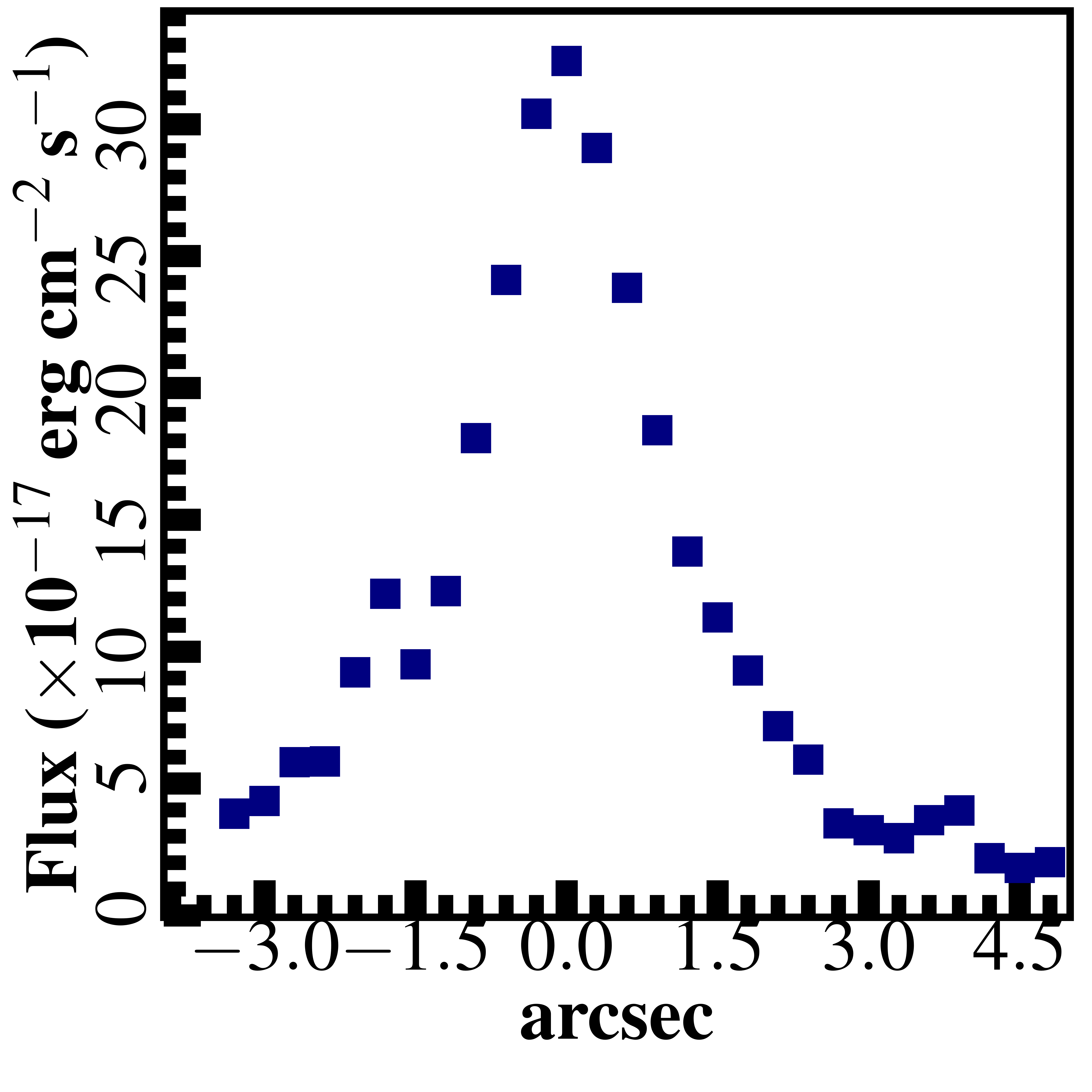}
		\label{ly_flux_4c}}
	\subfloat[Ly$\alpha$ Source \textit{vs.} Seeing]{
		\includegraphics[width=\columnwidth,height=1.90in,keepaspectratio]{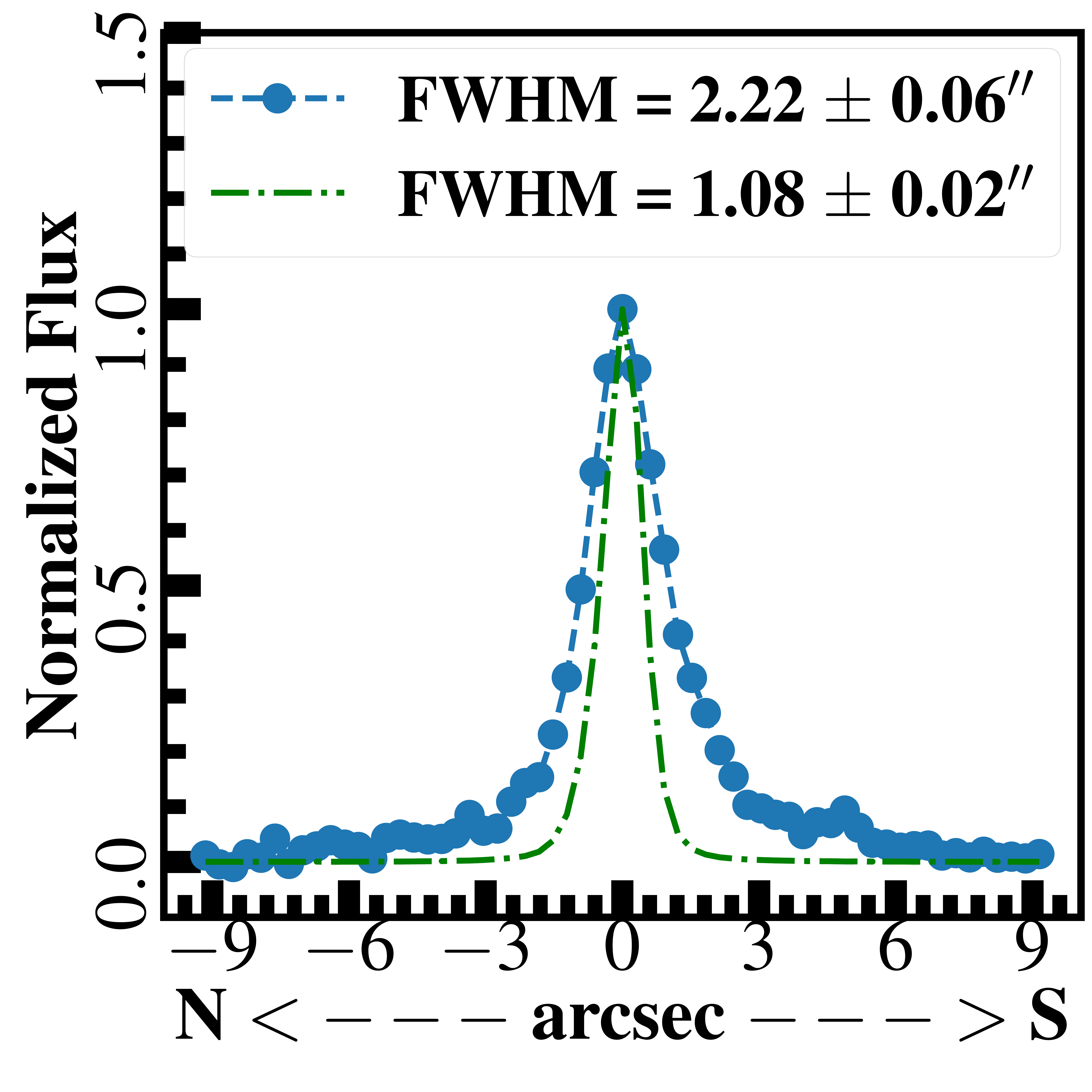}
		\label{4c0054_seeing}}
	\quad
	\subfloat[\ion{He}{II} Flux]{
		\includegraphics[width=\columnwidth,height=1.9in,keepaspectratio]{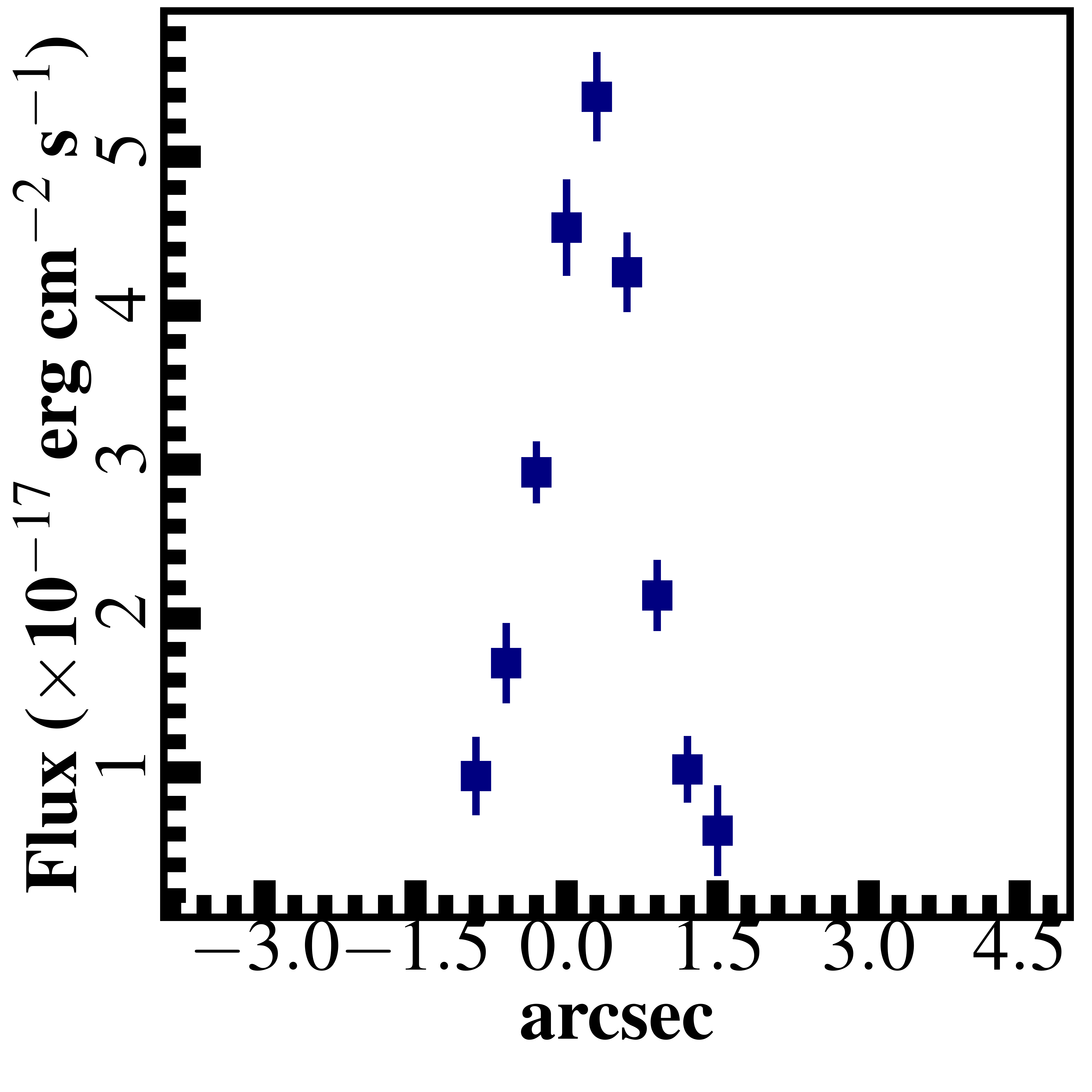}
		\label{heii_flux_4c}}
	\subfloat[Ly$\alpha$/\ion{He}{II}]{
		\includegraphics[width=\columnwidth,height=1.9in,keepaspectratio]{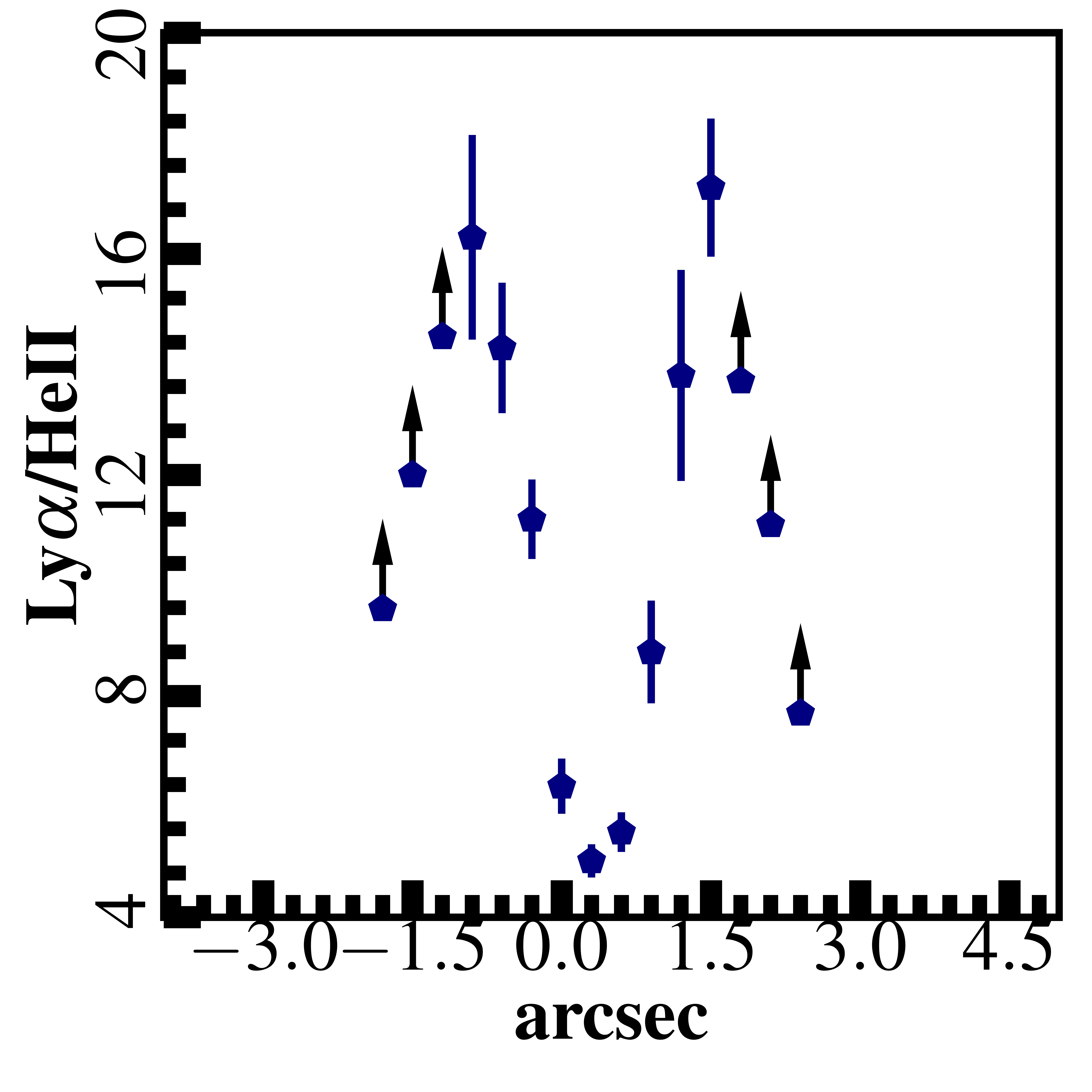}
		\label{lnr_4c}}
	\subfloat[Ly$\alpha$ 1-D spectrum]{
		\includegraphics[width=\columnwidth,height=1.9in,keepaspectratio]{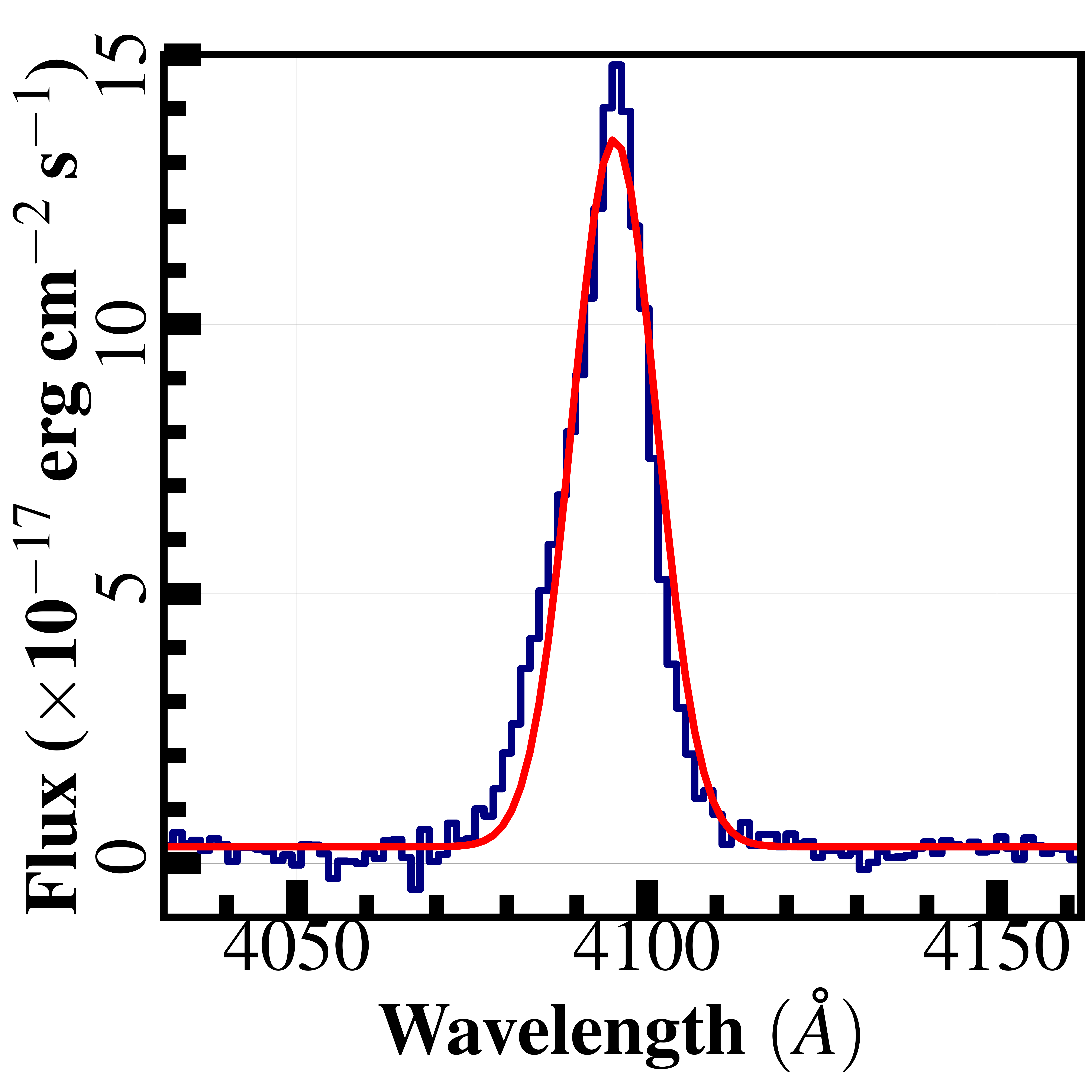}
		\label{1dspec_4c}}
	\quad
	\subfloat[Ly$\alpha$ FWHM]{
		\includegraphics[width=\columnwidth,height=1.9in,keepaspectratio]{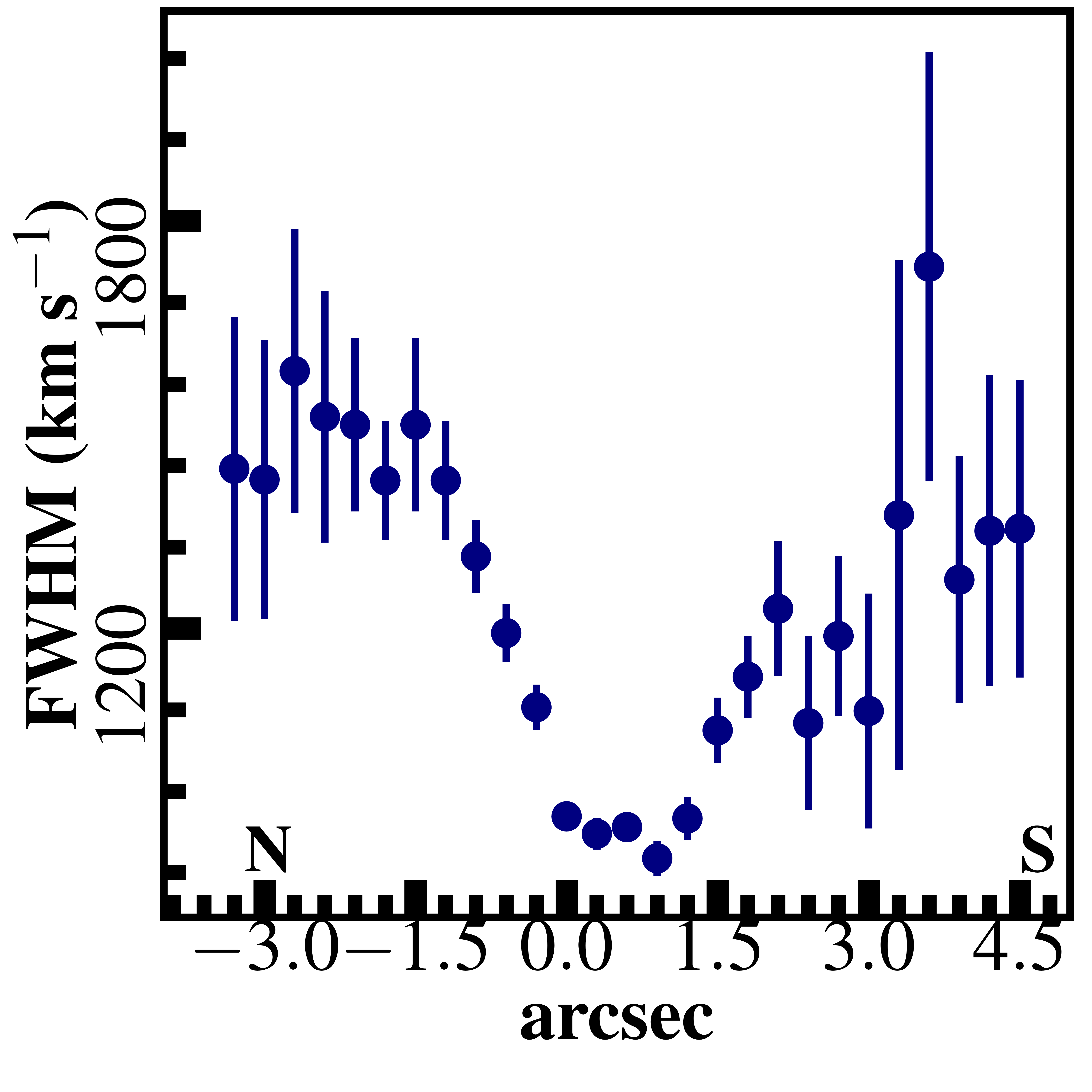}
		\label{fwhm_4c}}
	\subfloat[Ly$\alpha$ Velocity]{
		\includegraphics[width=\columnwidth,height=1.9in,keepaspectratio]{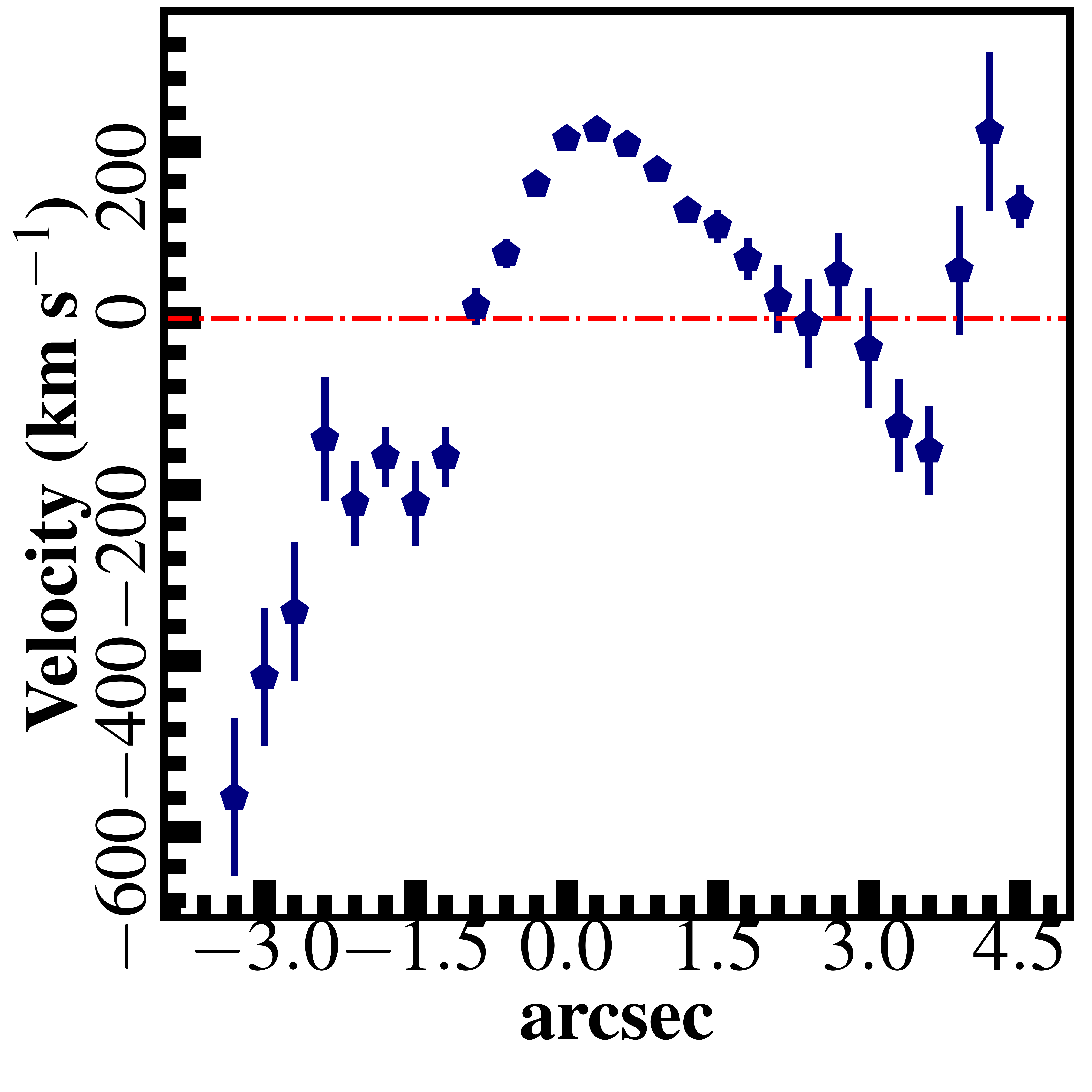}
		\label{velo_4c}}
	\subfloat[Ly$\alpha$ FWHM \textit{vs.} Velocity]{
		\includegraphics[width=\columnwidth,height=1.93in,keepaspectratio]{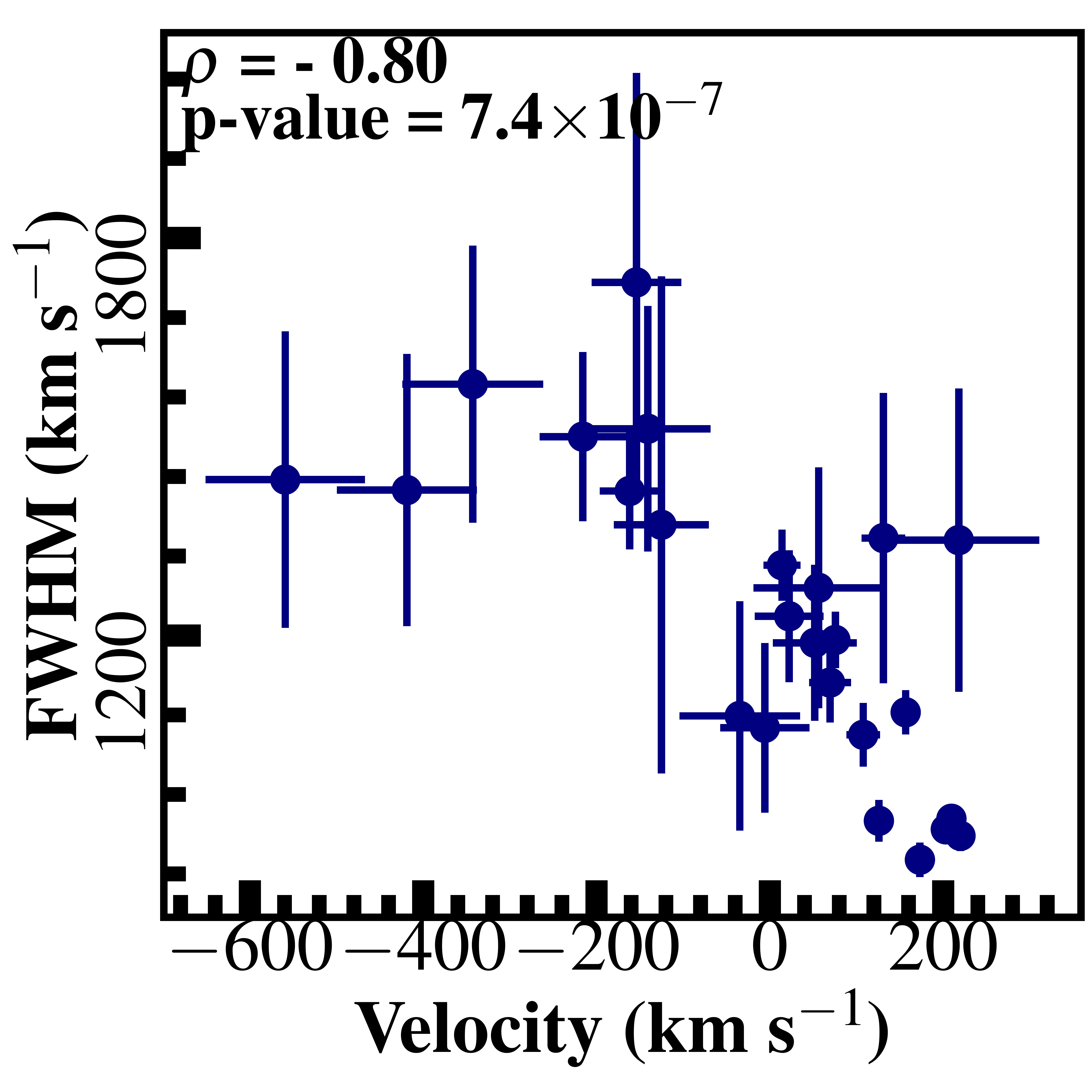}
		\label{corr_4c}}
	
	\caption{Radio galaxy 4C--00.54: (a) 2-D spectrum of the Ly$\alpha$ spectral region, (b) Spatial variation of the flux of Ly$\alpha$ line, (c) Ly$\alpha$ spatial profile (blue circle with dashed lines) compared with the seeing (green dot dashed lines), (d) Spatial variation of the flux of \ion{He}{II} line, (e) the Ly$\alpha$/\ion{He}{II} line ratio and (f) 1-D spectrum of the Ly$\alpha$ spectral region extracted from the SOAR long-slit. The Ly$\alpha$ emission-line was extracted by summing over a 3\arcsec$\,$ region of the slit length. Spatial variations of (g) FWHM, (h) Velocity, (i) Variation of FWHM as a function of the velocity offset of Ly$\alpha$ with $\rho$ and p-value representing the Spearman's rank correlation coefficient and t-distribution, respectively. The black arrows seen in the Ly$\alpha$/\ion{He}{II} line ratio diagram represent the 3$\sigma$ lower limit of the line ratios.}
	\label{4c0054kin}
\end{figure*}
\subsubsection{Previous results}

The z = 2.36 radio galaxy 4C--00.54 (also known as USS 1410--001) consists of a very optically elongated host galaxy \citep[e.g.][]{Pe1999,Pe2001} for the UV/optical morphology. With M$_{\star}$ $\sim$ 10$^{11.4}$ M$_{\odot}$ \citep[e.g.][]{seymour2007} and a star formation rate $\sim$ 460$^{+480} _{-270}$ M$_{\odot}$ yr$^{-1}$ inferred from 7.7 $\mu$m  polycyclic aromatic hydrocarbon luminosity \citep[e.g.][]{rawlings2013}. The radio source has a double lobe morphology with total extent of 24\arcsec$\,$ (or 196 kpc at the adopted cosmology) and it has a misalignment of $\sim$ 45$^{\circ}$ relative to the UV-optical continuum emission in the central few tens of kpc of the galaxy  \citep[e.g.][]{Pe1999,Pe2001}. Spectroscopic studies with the ESO-NTT have revealed extended Ly$\alpha$ emission ($\sim$ 80 kpc) without any sign of \ion{H}{I} absorption \citep[e.g.][]{Vo}. Long slit spectropolarimetry using Keck II has shown that the UV continuum emission along the radio axis is significantly polarized (P$_{\%}$ = 11.7 $\pm$ 2.7) with contribution from the scattered AGN continuum (47 -- 88 $\%$), young stellar population (41 -- 0 $\%$) and nebular continuum (12 $\%$) \citep[e.g.][]{Ve2001}. Using the same data, \cite{VM1} and \cite{Hu2}  concluded that the extended gas along the radio axis is part of a giant quiescent halo (FWHM $\leqslant$ 472 -- 800 km s$^{-1}$) without any evidence of jet-gas interactions. Using VLT/SINFONI imaging spectroscopy of the rest-frame optical emission lines of HzRGs,   \cite{nesvadba17,nesvadba2017} find line emission extending over an area of 2\arcsec$\times$5.6\arcsec (or 14 kpc $\times$ 41 kpc) with the major axis going from south to north. They find line widths in the range FWHM = 400 -- 1300 km s$^{-1}$ in which a small region to the north-east of the nucleus shows the highest FWHM. In addition, with velocity offsets in the range -400 -- +400 km s$^{-1}$, they conclude that the radio jets are the main driver of the turbulent outflowing gas.
Also using rest-frame optical emission lines obtained with the VLT/ISAAC, \cite{Hu4} concluded that AGN photoionization is the dominant ionization mechanism operating in the extended emission line region possibly with a fractional contribution from shocks.
%\newpage
\subsubsection{Results from SOAR}

The Ly$\alpha$ emission detected in the direction perpendicular to the UV-optical emission of the galaxy (see Fig. \ref{2dspec_4c}) shows an asymmetric spatial distribution, which is slightly more extended in the S direction (see also Fig. \ref{ly_flux_4c}). In addition to Ly$\alpha$ (see the 1-D spectrum in Fig. \ref{1dspec_4c}), we could also detect \ion{N}{V}, \ion{C}{IV} and \ion{He}{II} emission lines (see Table \ref{instru02}).
In Figures \ref{fwhm_4c} and \ref{velo_4c}, we show the spatial variations of the FWHM and velocity of the Ly$\alpha$ emission line, respectively. We find that the extended emission line halo shows a central region of kinematically quiescent gas\footnote{We define kinematically `quiescent' to mean FWHM$<$1000 km s$^{-1}$ and kinematically `perturbed' to mean FWHM$\ge$1000 km s$^{-1}$. This definition, although somewhat arbitrary, is motivated by some previous studies which found ionized gas with FWHM$\ge$1000 km s$^{-1}$ associated with the radio structures of HzRGs, and gas FWHM$<$1000 km s$^{-1}$ present across the full spatial extent of the extended nebulae of the HzRGs \citep[e.g.][]{VM1}, suggesting that the `quiescent' gas is a common feature of the nebulae of HzRGs and represents gas that has not been disturbed by radio mode feedback.} with FWHM $<$ 1000 km s$^{-1}$. From the nucleus towards the outermost regions we find a gradual increase in FWHM, with values reaching up to 1600 km s$^{-1}$. The velocity offset of the extended gas varies from -578 km s$^{-1}$ to 216 km s$^{-1}$. In Figure \ref{corr_4c}, we show the correlation between the FWHM and the velocity offset of Ly$\alpha$. The diagram suggests a strong negative relationship between the parameters with $\rho$ = -0.60 and p-value = 7.4 $\times$ 10$^{-7}$.  We note that a similar anticorrelation between velocity shift and FWHM is also present in Fig. 4 of \citet{VM1}, albeit along a different slit PA to ours.

With the slit position angle set to 90$^{\circ}$, which is perpendicular to the UV-optical emission in the HST images, we note that the slit is at an angle of 45$^{\circ}$ to the radio axis as defined by the line running through the 2 hotspots and the radio core \citep[see][]{Pe1999,Pe2001}. However, as \citet{Pe1999} suggest, the AGN axis (or the radio jet axis) might be precessing, in which case the radio axis near the nucleus might be different to what the radio axis was when the material in the hotspots was ejected from the nucleus.
It does seem plausible that due to jet precession, the slit position angle used might intersect gas that has been disturbed by the radio jets. However, it seems odd since that at the nucleus and along the UV-optical axis the kinematics are more quiescent. We should expect the jets going through and perturbing gas in the nuclear region as well.

The Ly$\alpha$ spatial profile of the radio galaxy 4C--00.54 with FWHM = 2.22 $\pm$ 0.06\arcsec$\,$ shows that the emission line is spatially extended compared with the seeing FWHM = 1.08 $\pm$ 0.02\arcsec$\,$ (see Fig. \ref{4c0054_seeing}).  In addition, the spatial profile shows an excess above the seeing disk especially obvious towards the South. Correcting for seeing broadening in quadrature, we infer the intrinsic FWHM as 1.94 $\pm$ 0.07\arcsec$\,$ or 16 $\pm$ 1 kpc. None of the other UV emission lines are found to be extended in this spectrum.

In Figure \ref{lnr_4c}, we show the spatial variation of the Ly$\alpha$/\ion{He}{II} flux ratio along the slit. 
We find a significant variation in Ly$\alpha$/\ion{He}{II}, with a radial increase from 5.0 $\pm$ 0.3 near the spatial zero position, up to 16.3 $\pm$ 1.9 (r = - 0.9\arcsec) and 17.2 $\pm$ 1.3 (r = + 1.5\arcsec). The spatial variation of the Ly$\alpha$/\ion{He}{II} shows flux ratios that are consistent with the standard photoionization models, which predict Ly$\alpha$/\ion{He}{II} in the range 15 -- 20 for HzRGs \cite[e.g.][]{VM3}. The possible reasons for such variation will be discussed in $\S$\ref{Ly_heii_ratio}.

\subsection{TN J0920--0712}

\begin{figure*}
	\centering	\textbf{TN J0920--0712}
	
	\subfloat[Ly$\alpha$ 2-D spectrum]{
		\includegraphics[width=\columnwidth,height=2.10in,keepaspectratio]{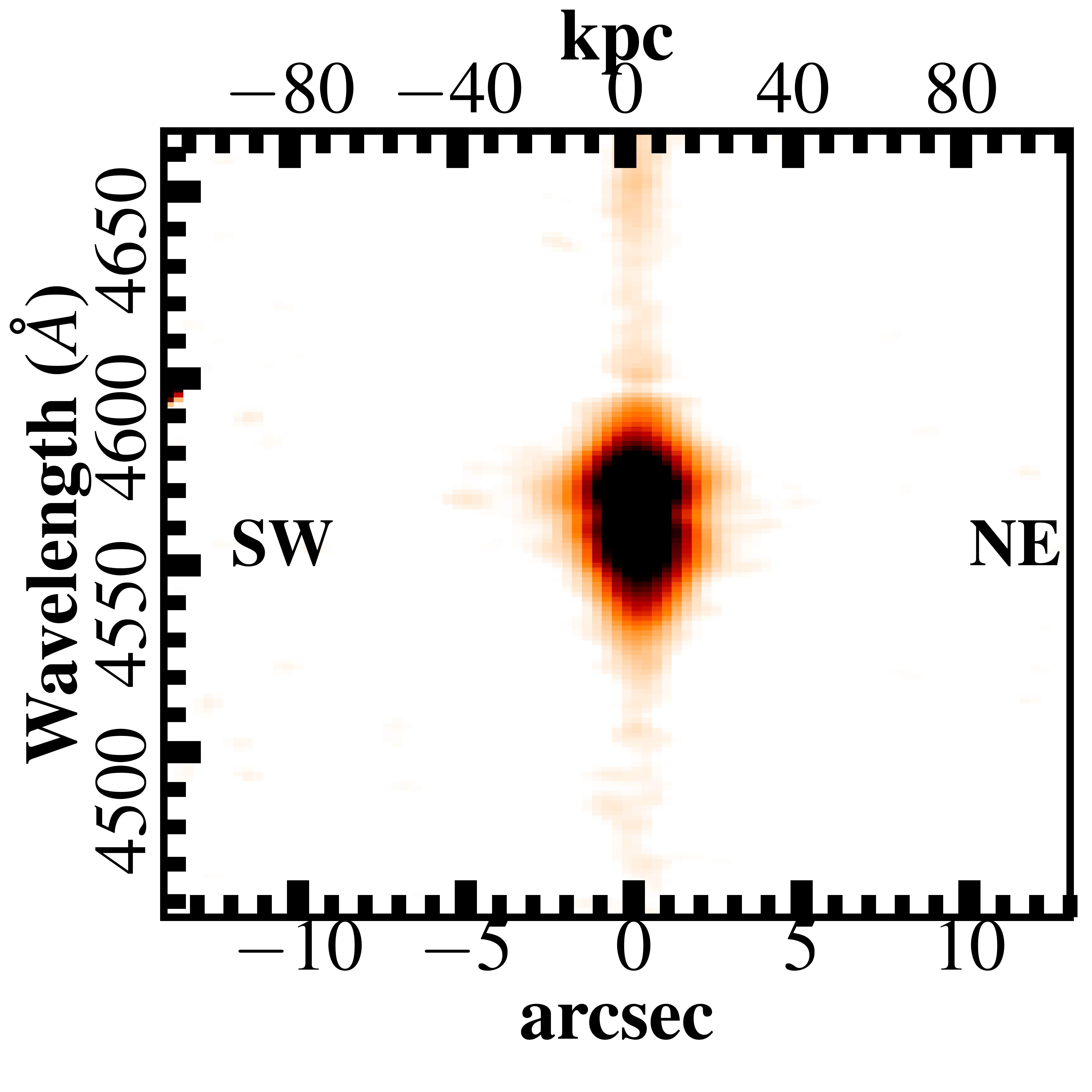}
		\label{2dspec_tnj0920}}
	\subfloat[Ly$\alpha$ Flux]{
		\includegraphics[width=\columnwidth,height=1.9in,keepaspectratio]{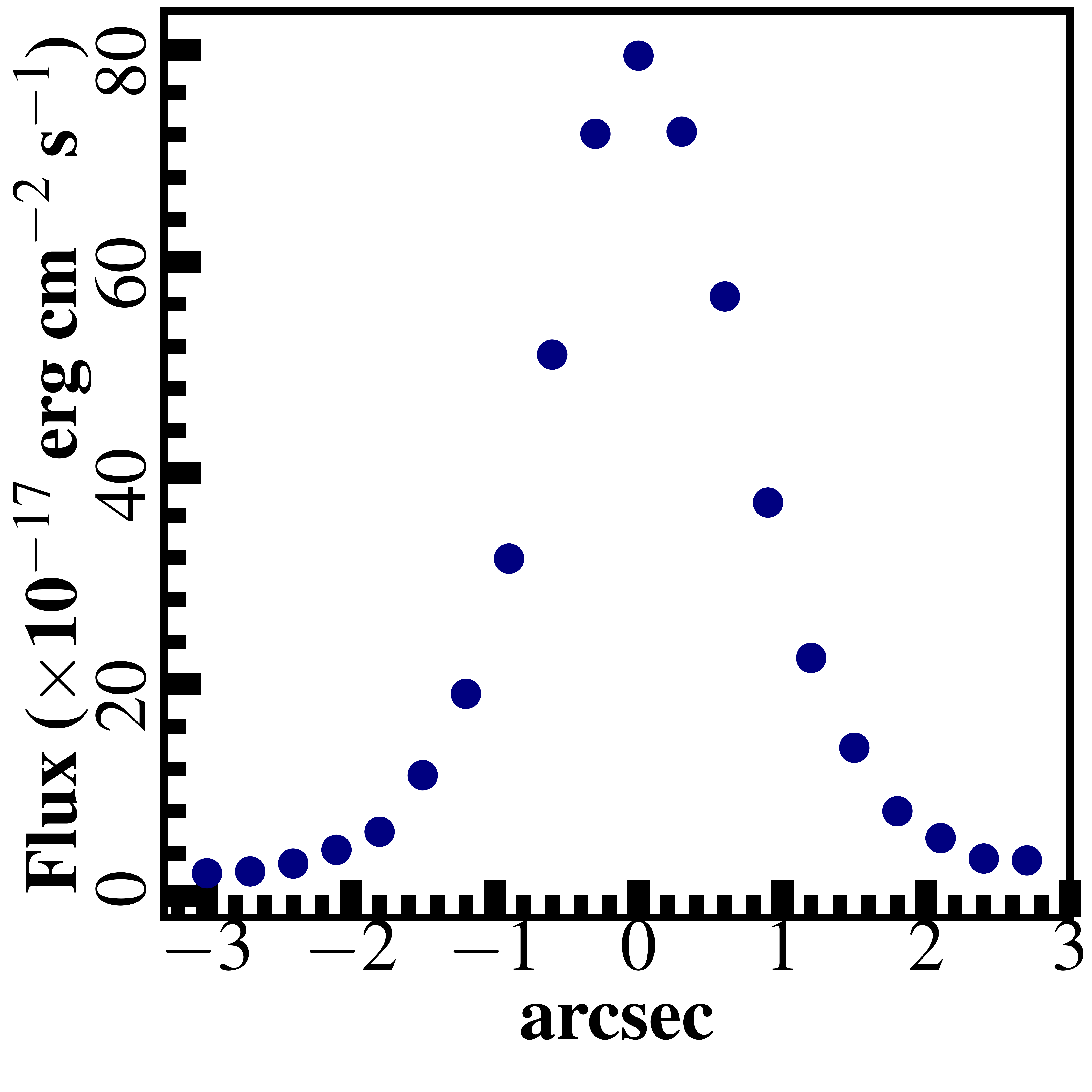}
		\label{ly_flux_tnj0920}}
	\subfloat[Ly$\alpha$ Source \textit{vs.} Seeing]{
		\includegraphics[width=\columnwidth,height=1.9in,keepaspectratio]{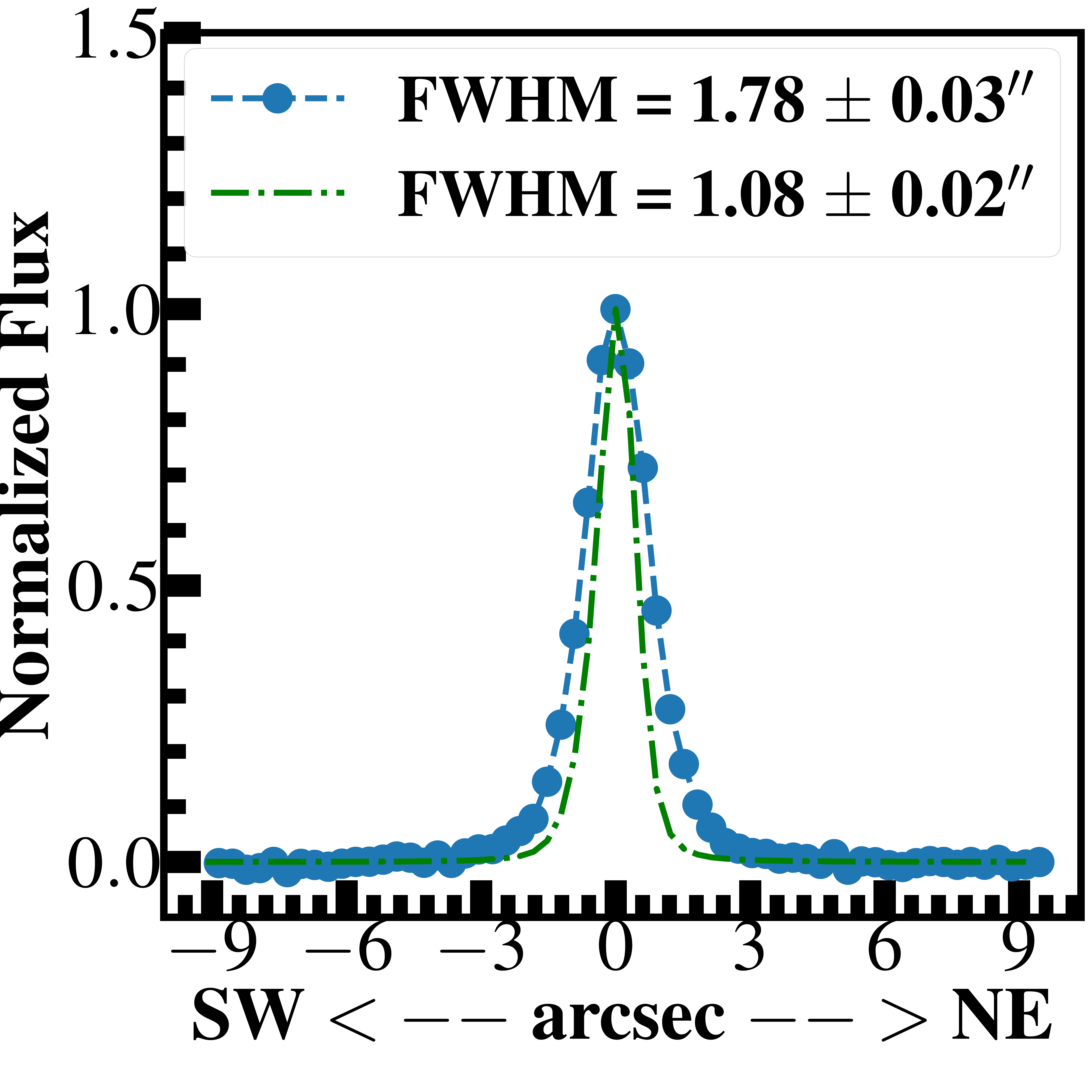}
		\label{tnj0920_seeing}}
	\quad
	\subfloat[Ly$\alpha$ 1-D spectrum]{
		\includegraphics[width=\columnwidth,height=1.96in,keepaspectratio]{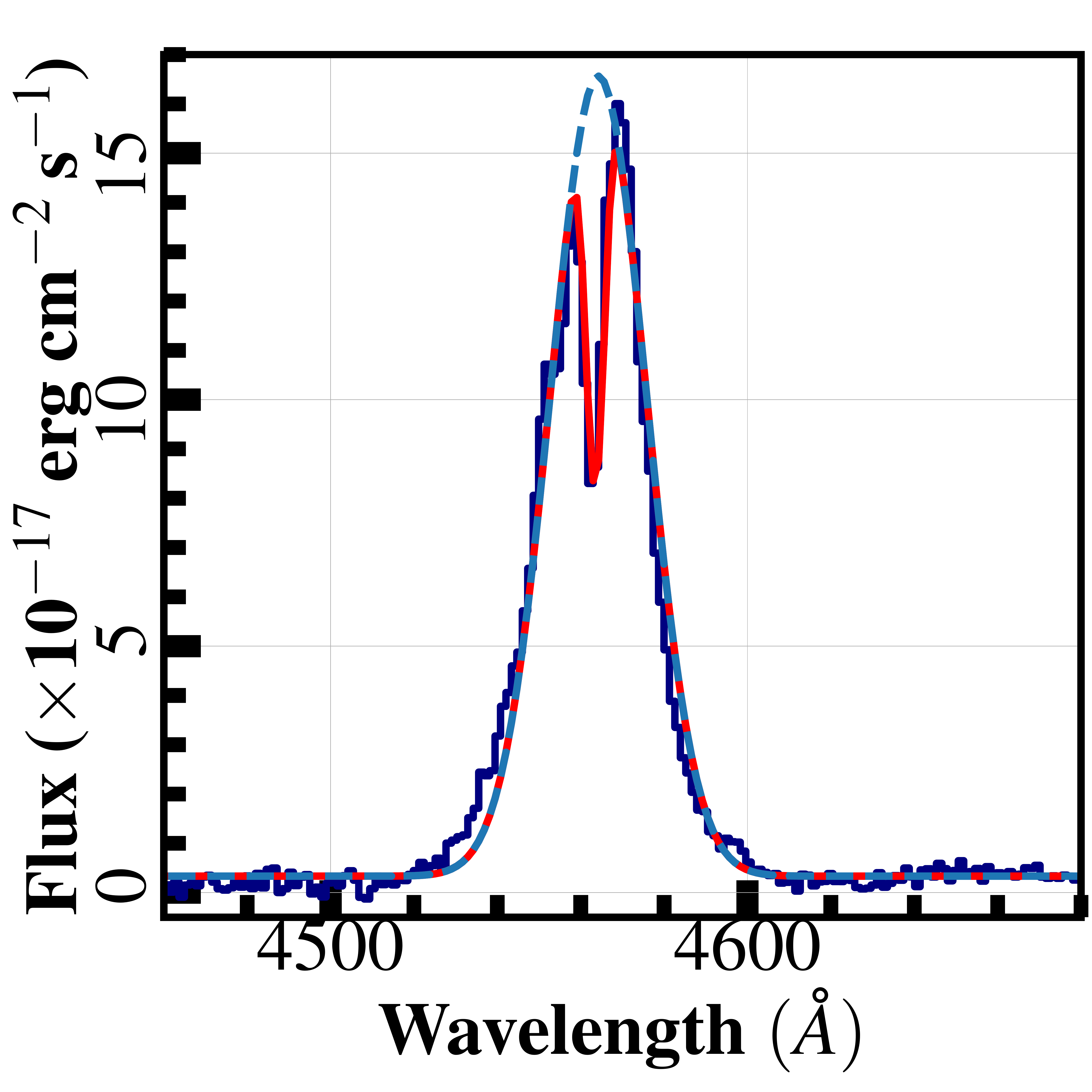}
		\label{1dspec_tnj0920}}
	\subfloat[Ly$\alpha$ FWHM]{
		\includegraphics[width=\columnwidth,height=1.9in,keepaspectratio]{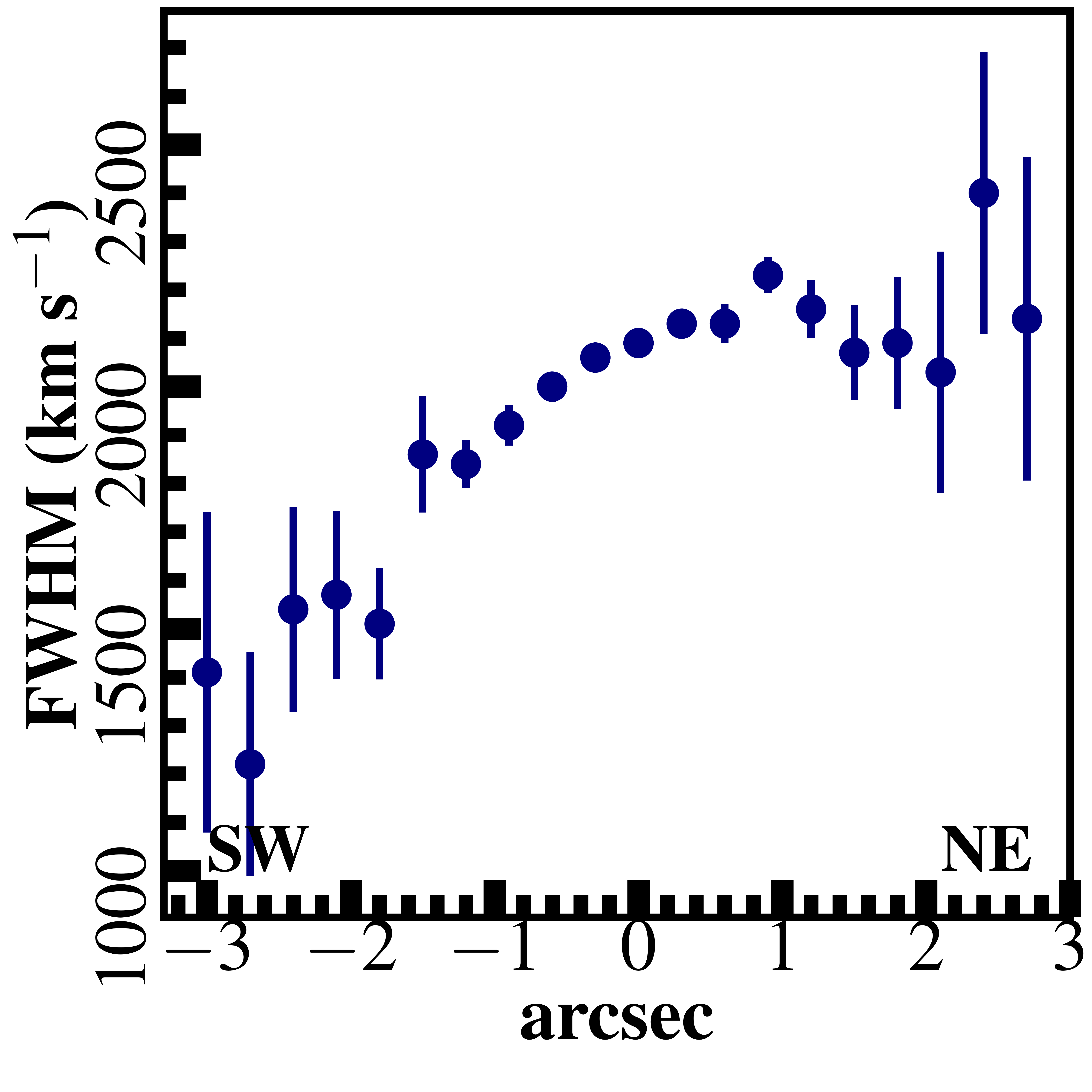}
		\label{fwhm_tnj0920}}
	\subfloat[Ly$\alpha$ Velocity]{
		\includegraphics[width=\columnwidth,height=1.9in,keepaspectratio]{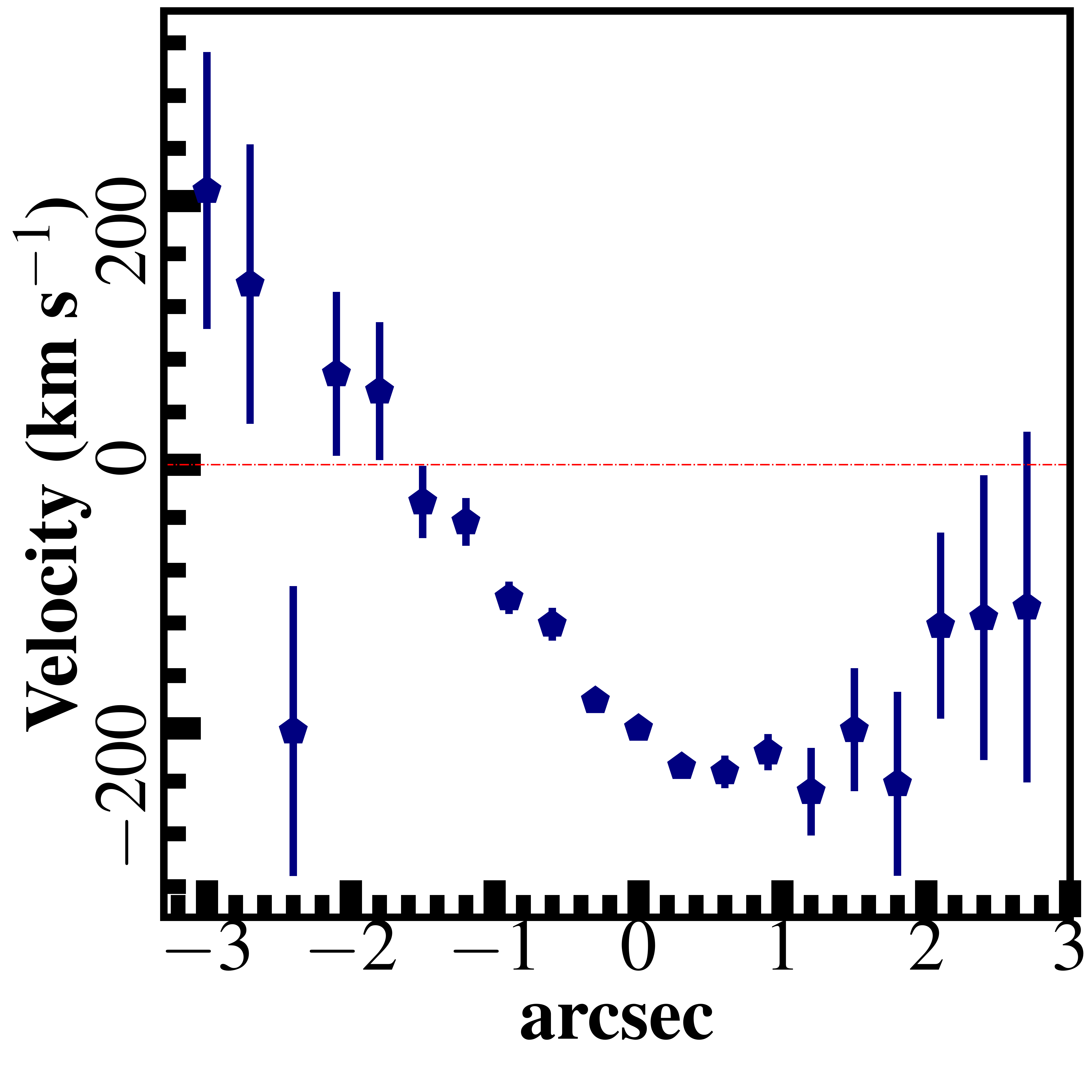}
		\label{velo_tnj0920}}
	\quad
	\subfloat[Ly$\alpha$ FWHM \textit{vs.} Velocity]{
		\includegraphics[width=\columnwidth,height=1.93in,keepaspectratio]{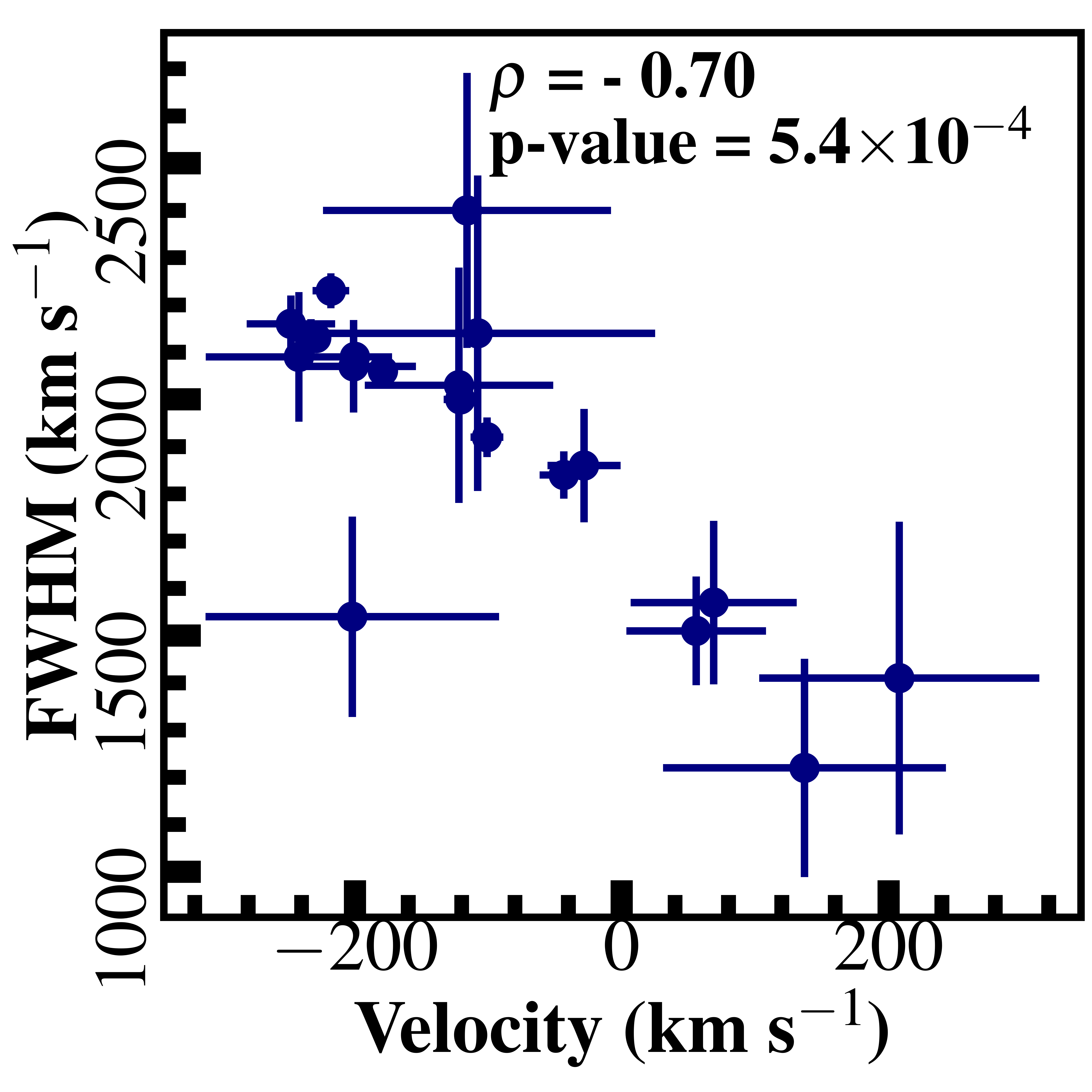}
		\label{corr_tnj0920}}
	\subfloat[Ly$\alpha$ Velocity Abs.]{
		\includegraphics[width=\columnwidth,height=1.9in,keepaspectratio]{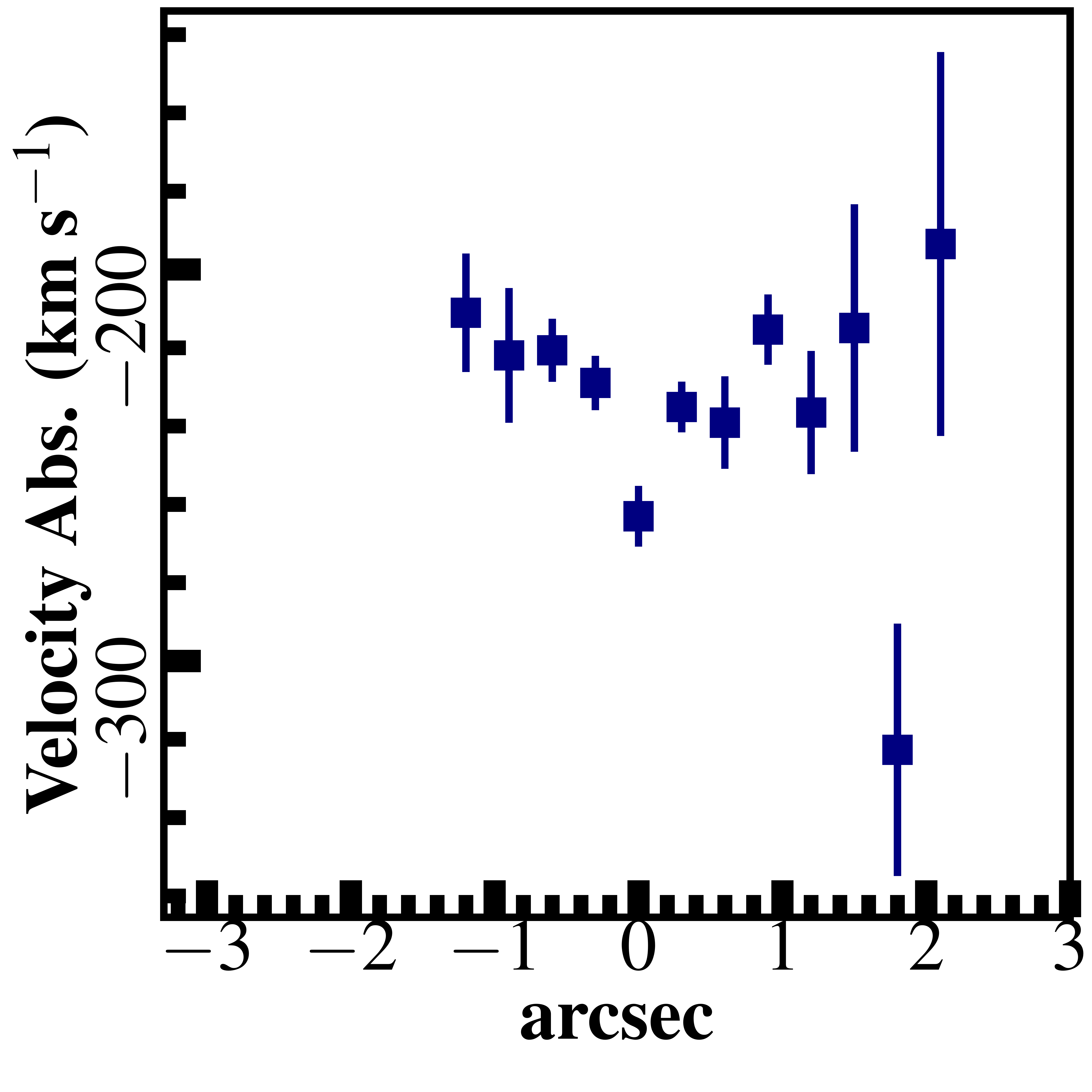}
		\label{vabs_tnj0920}}
	\subfloat[Ly$\alpha$ N(\ion{H}{I}) \textit{vs.} Seeing]{
		\includegraphics[width=\columnwidth,height=1.9in,keepaspectratio]{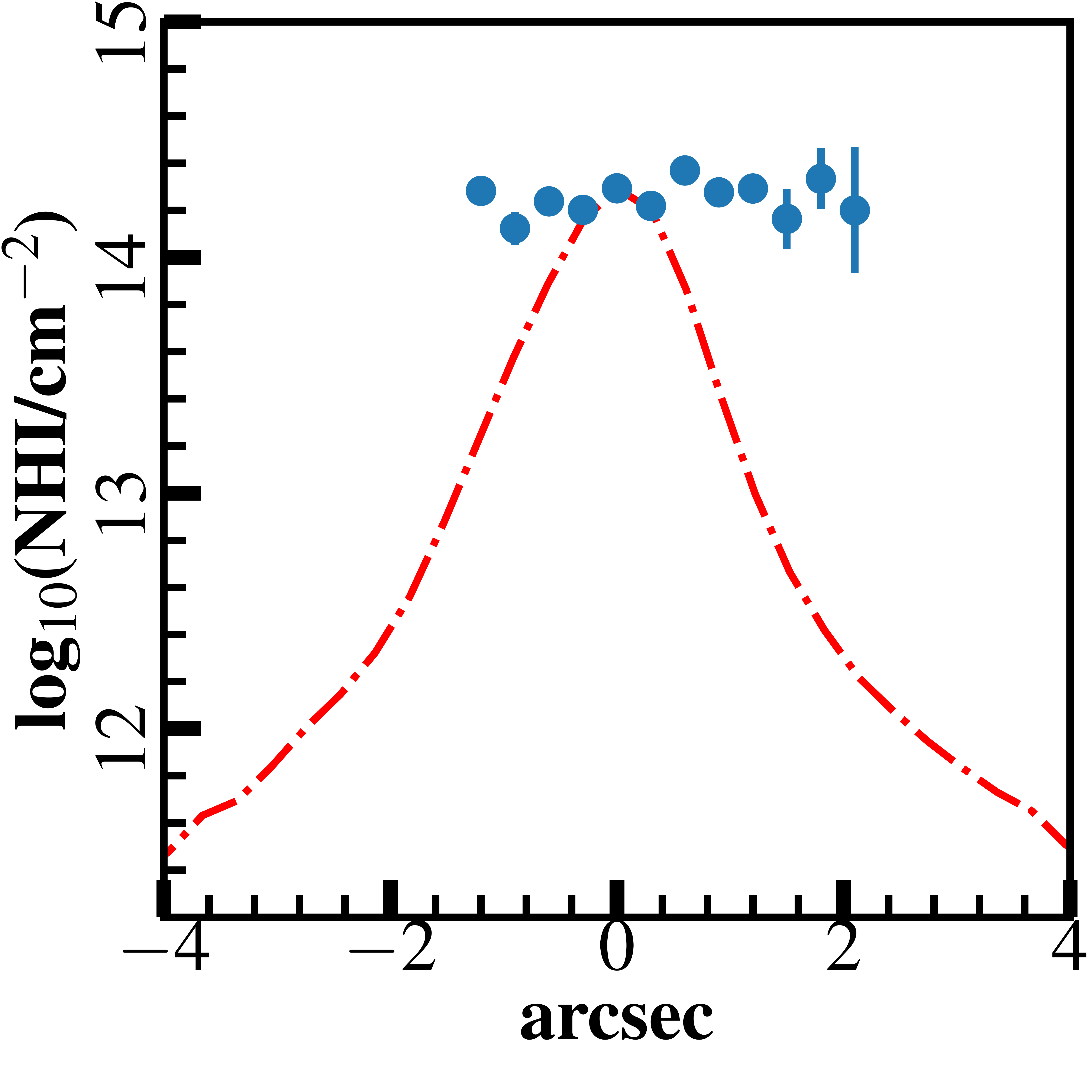}
		\label{NHItnj0920_seeing}}
	
	\caption{Radio galaxy TN J0920--0712: (a) 2-D spectrum of the Ly$\alpha$ spectral region, (b) Spatial variation of the flux of the Ly$\alpha$ line, (c) Ly$\alpha$ spatial profile (blue circle with dashed lines) compared with the seeing (green dot dashed lines) and (d) 1-D spectrum of the Ly$\alpha$ spectral region extracted from the SOAR long-slit. The Ly$\alpha$ emission-line was extracted by summing over a 3\arcsec$\,$ region of the slit length. Spatial variations of (e) FWHM, (f) Velocity, (g) Variation of FWHM as a function of the velocity offset of Ly$\alpha$ with $\rho$ and p-value representing the Spearman's rank correlation coefficient and t-distribution, respectively, (h) Velocity of the \ion{H}{I} absorber and (i) Spatial profile of the \ion{H}{I} column density (blue circle points) compared the seeing (red dot dashed lines), both on a logarithmic scale. In addition, the seeing profile has been normalised and shifted in order to allow the comparison.}
	\label{tnj0920kin}
\end{figure*}
\subsubsection{Previous results}

VLA observations of the HzRG TN J0920--0712 at z = 2.76 revealed a radio source consisting of a single component with a
maximum angular size of 1.4\arcsec$\,$ (11 kpc in the adopted cosmology), and spectral index $\alpha$ = -1.51 \citep[e.g.][]{Bre2000b,Bre2001}. Using the Faint Object Spectrograph and Camera 1 (EFOSC1) on the ESO 3.6 m telescope at La Silla, \citet{Bre2001} found extended Ly$\alpha$ emission with FWHM = 2050 $\pm$ 150 km s$^{-1}$ and a large equivalent width (W$_{\lambda} ^{rest}$ = 350 $\pm$ 60 \AA). Additionally, they also detected \ion{Si}{IV} + \ion{O}{IV}, \ion{C}{IV} and \ion{He}{II} emission lines. They also reported the detection of an extended \ion{H}{I} absorber on the blue side of the Ly$\alpha$ line althoulgh it has not been explored. \citet{Bre2000}  compared the UV line ratios of this HzRG, and found that photoionization by the central AGN offers a plausible explanation for the excitation of the line emitting gas, with potential for a significant additional contribution from shock ionization.

\subsubsection{Results from SOAR}

The Ly$\alpha$ nebula shows an elongated and symmetric spatial distribution (see the 2-D spectrum of the Ly$\alpha$ spectral region in Fig. \ref{2dspec_tnj0920}, and also Fig. \ref{ly_flux_tnj0920}).
In Figure \ref{1dspec_tnj0920}, we show the integrated 1-D spectrum of the Ly$\alpha$ extracted from the SOAR long-slit, with the Gaussian emission component plus absorption model overlaid. Together with Ly$\alpha$, we also detected the emission lines such as \ion{O}{VI} + \ion{C}{II}, \ion{N}{V}, \ion{C}{IV} and \ion{He}{II} (see Table \ref{instru02}). The spatial variation of the FWHM shows a smooth transition from SW to the NE during which the line width varies from 1220 km s$^{-1}$ to 2400 km s$^{-1}$ across the entire nebula (see Fig. \ref{fwhm_tnj0920}). The velocity shift varies from - 242 km s$^{-1}$ to 208 km s$^{-1}$ (see Fig. \ref{velo_tnj0920}). We find a correlation between the FWHM and the velocity offset of Ly$\alpha$, such that regions with higher FWHM also tend to have a higher blueshift (see Fig. \ref{corr_tnj0920}). The result indicates that there is a relationship between the parameters with $\rho$ = - 0.70 and p-value = 5.4 $\times$ 10$^{-4}$.

The Ly$\alpha$ spatial profile with FWHM = 1.78 $\pm$ 0.03\arcsec$\,$ shows that the emission line is spatially extended compared with the seeing FWHM = 1.08 $\pm$ 0.02\arcsec$\,$ (see Fig. \ref{tnj0920_seeing}). In addition, the spatial profile shows excess of emission above the seeing wings at both sides of the central source showing that the Ly$\alpha$ halo is extended. We infer the intrinsic FWHM to be 1.42 $\pm$ 0.04\arcsec$\,$ (or 11.4 $\pm$ 0.3 kpc). None of the other UV emission lines are found to be extended in this spectrum.

Like \citet{Bre2001}, we also detect a spatially extended Ly$\alpha$ absorption feature on the blue side of the Ly$\alpha$ emission in the spectrum of TN J0920--0712 (see Fig. \ref{2dspec_tnj0920}). The best fit to the Ly$\alpha$ is shown in Fig. \ref{1dspec_tnj0920}. Table \ref{instru03} lists the parameters of the best fitting model together with the diameter of the absorber, and the maximum detected radius. We obtain a column density log N(\ion{H}{I}/cm$^{-2}$) = 14.23 $\pm$ 0.03 with Doppler parameter $\mathit{b}$ = 185 $\pm$ 10 km s$^{-1}$, and  velocity offset of the absorber -232 $\pm$ 6 km s$^{-1}$. In Figure \ref{vabs_tnj0920}, we show the line of sight velocity of the \ion{H}{I} absorber measured from the SOAR spectrum across its detected spatial extent, which shows that the absorbing gas appears blueshifted from the systemic velocity. This absorber is detected across the full spatial extent of the Ly$\alpha$ emission where the S/N of the line is sufficient to detect an absorber with that column density, i.e., 3.3\arcsec$\,$ or 27 kpc. This absorption is clearly spatially extended compared to the the seeing profile (see Fig. \ref{NHItnj0920_seeing}).
Assuming that the absorbing gas is a spherically symmetric shell, using the expression \ref{m_HI} we estimate log ($M_{\ion{H}{I}}/M_{\sun}$) $\gtrsim$ 3.7.
\subsection{PKS 1138--262}

\begin{figure*}
	\centering	\textbf{PKS 1138--262}
	
	\subfloat[Ly$\alpha$ 2-D spectrum]{
		\includegraphics[width=\columnwidth,height=2.10in,keepaspectratio]{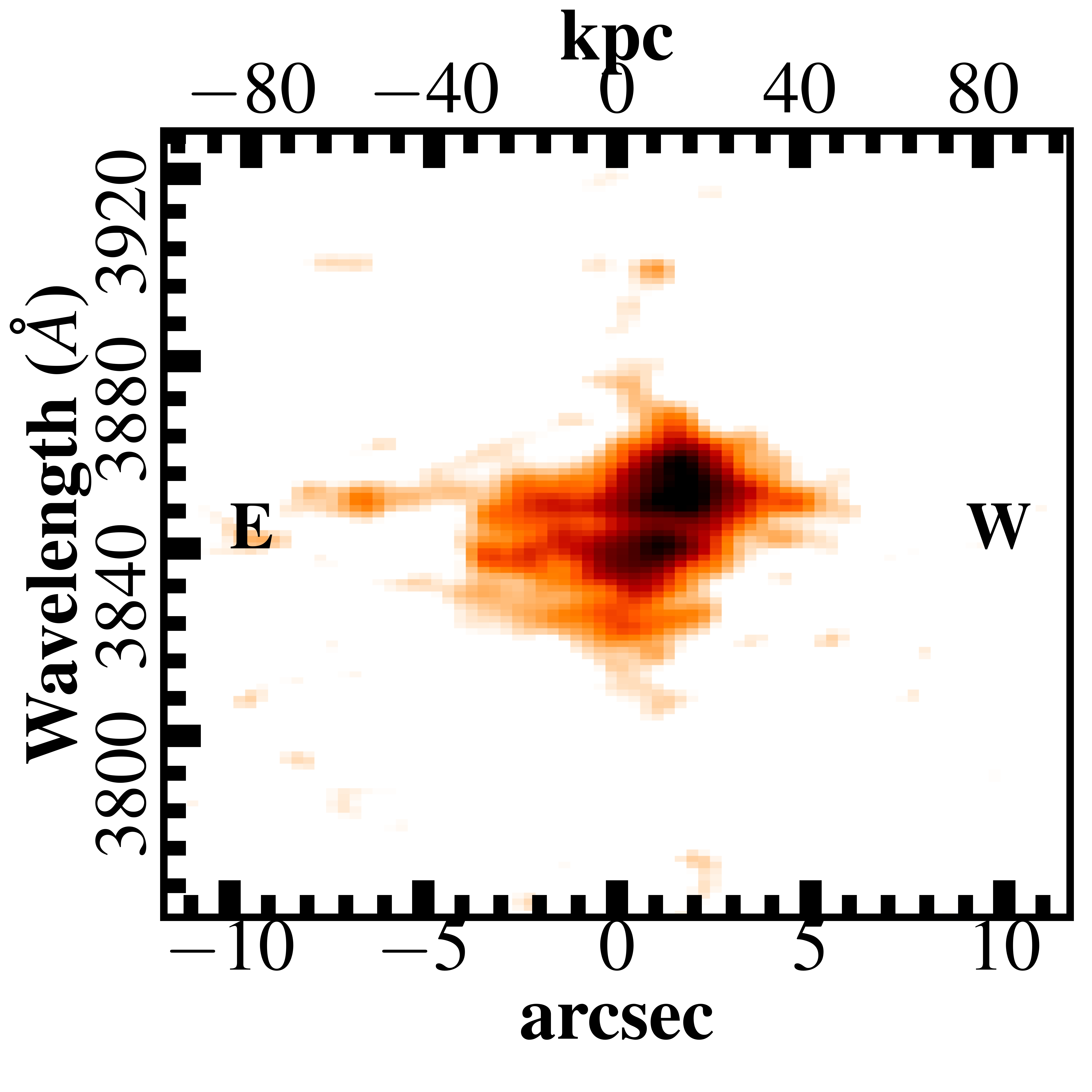}
		\label{2dspec_pks1138}}
	\subfloat[Ly$\alpha$ Flux]{
		\includegraphics[width=\columnwidth,height=1.9in,keepaspectratio]{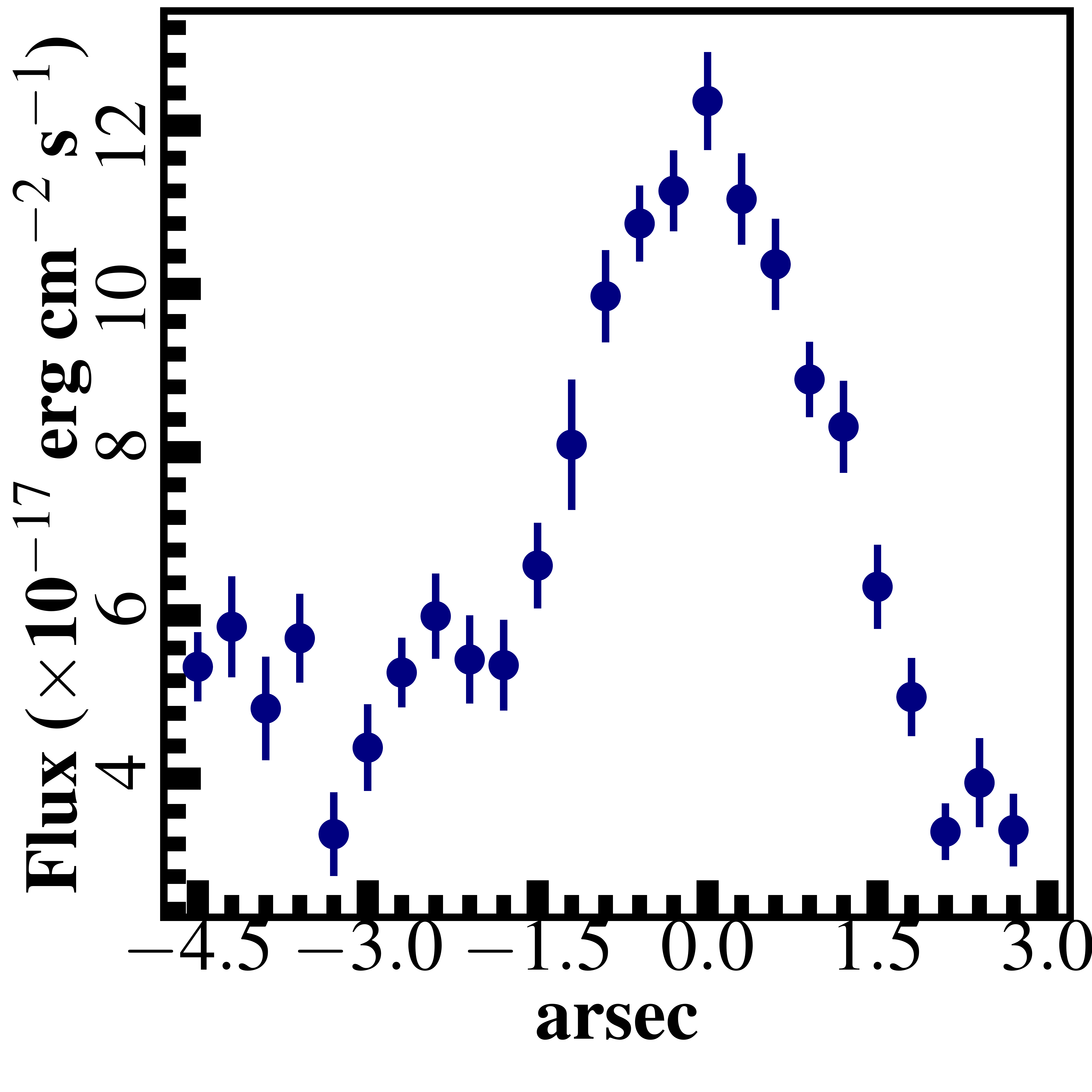}
		\label{ly_flux_pks1138}}
	\subfloat[Ly$\alpha$ Source \textit{vs.} Seeing]{
		\includegraphics[width=\columnwidth,height=1.9in,keepaspectratio]{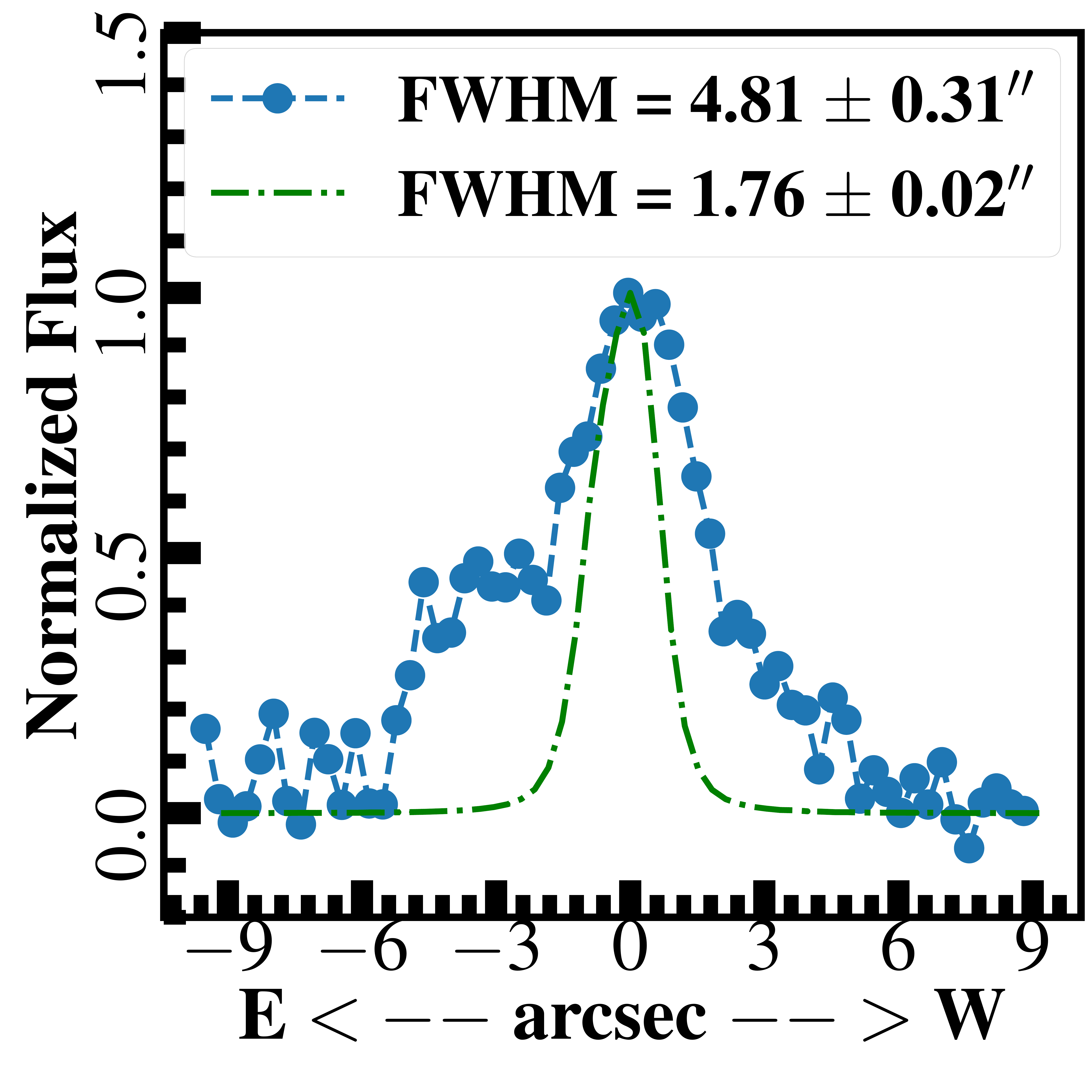}
		\label{pks1138_seeing}}
	\quad
	\subfloat[Ly$\alpha$ 1-D spectrum]{
		\includegraphics[width=\columnwidth,height=1.96in,keepaspectratio]{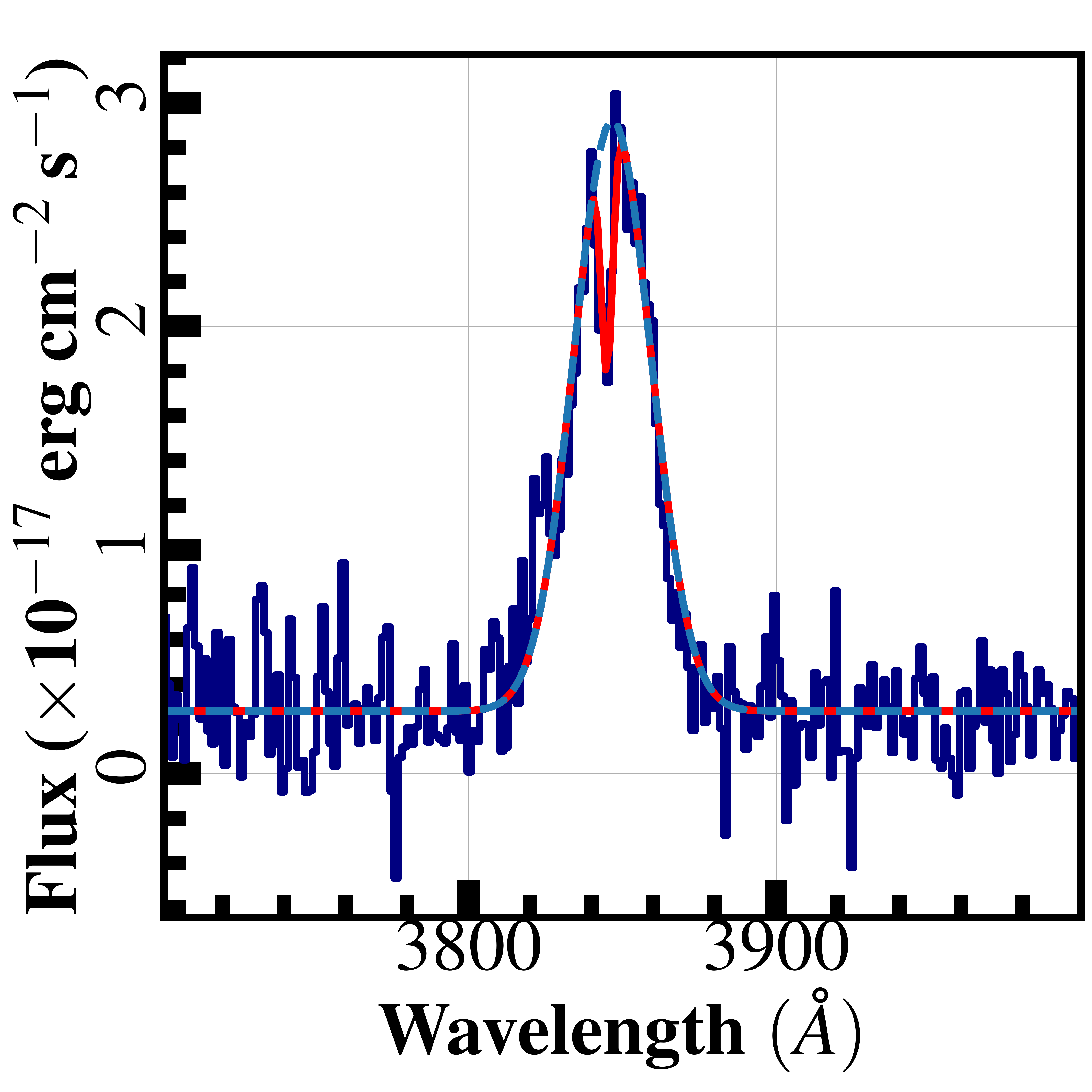}
		\label{1dspec_pks1138}}
	\subfloat[Ly$\alpha$ FWHM]{
		\includegraphics[width=\columnwidth,height=1.9in,keepaspectratio]{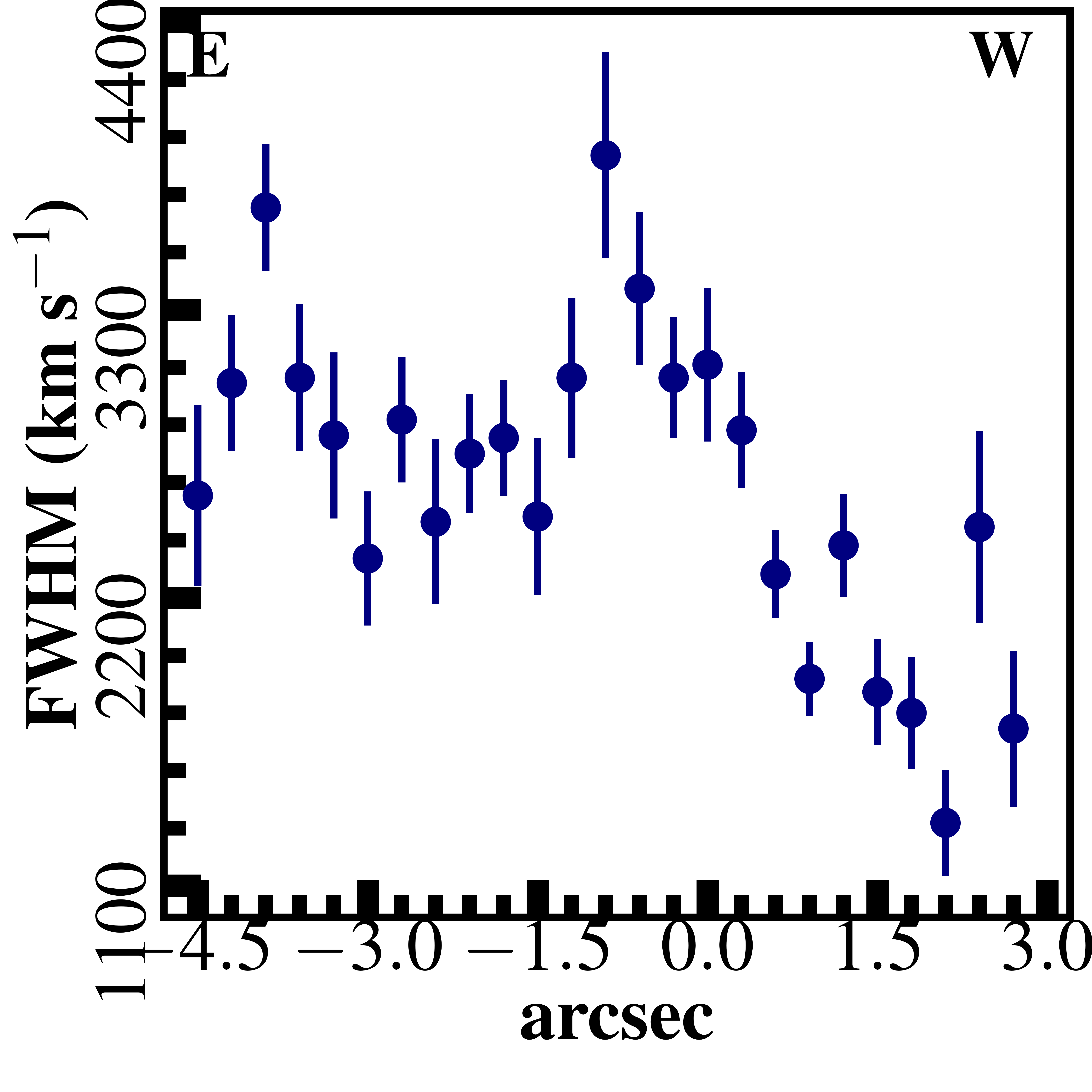}
		\label{fwhm_pks1138}}
	\subfloat[Ly$\alpha$ Velocity]{
		\includegraphics[width=\columnwidth,height=1.9in,keepaspectratio]{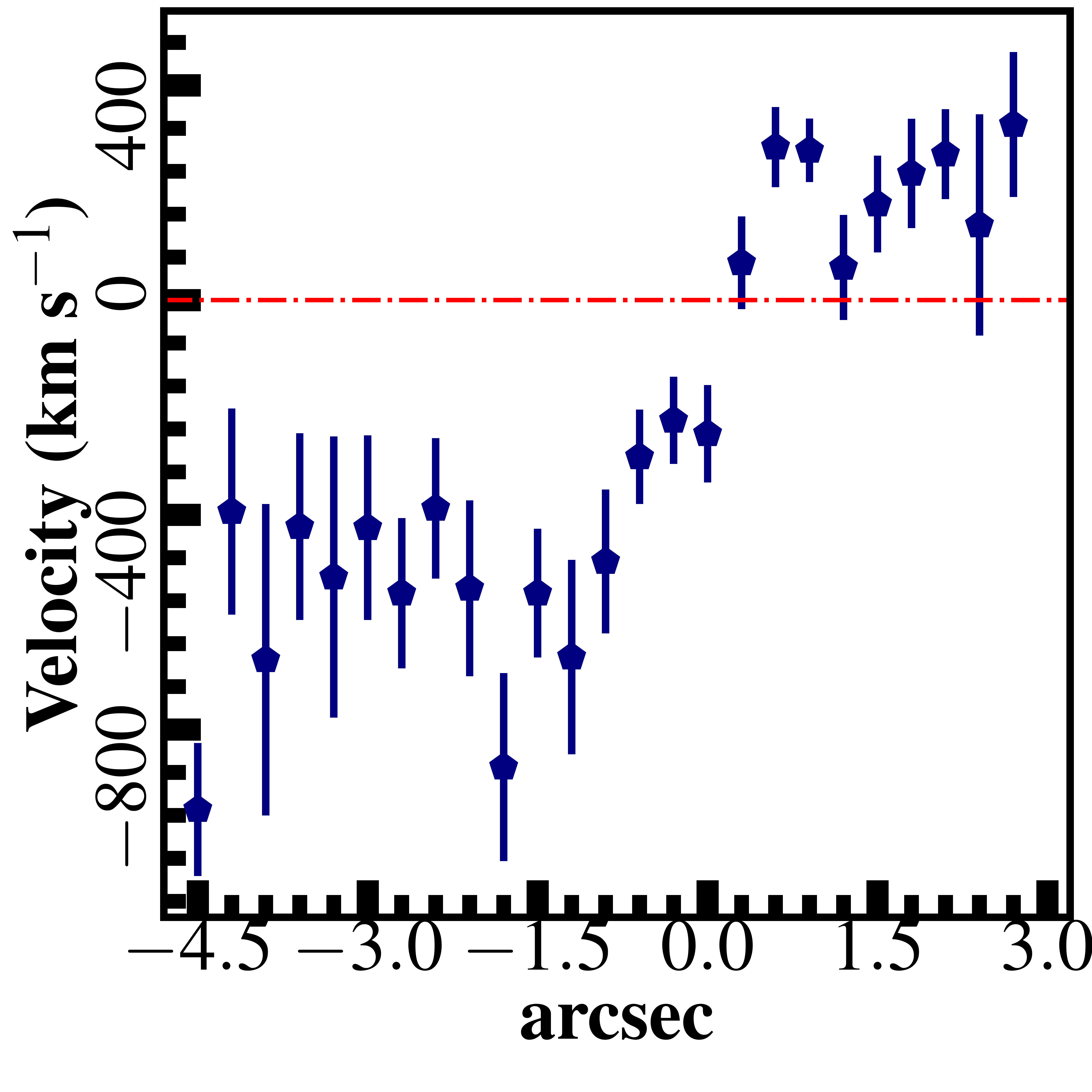}
		\label{velo_pks1138}}
	\quad
	\subfloat[Ly$\alpha$ FWHM \textit{vs.} Velocity]{
		\includegraphics[width=\columnwidth,height=1.93in,keepaspectratio]{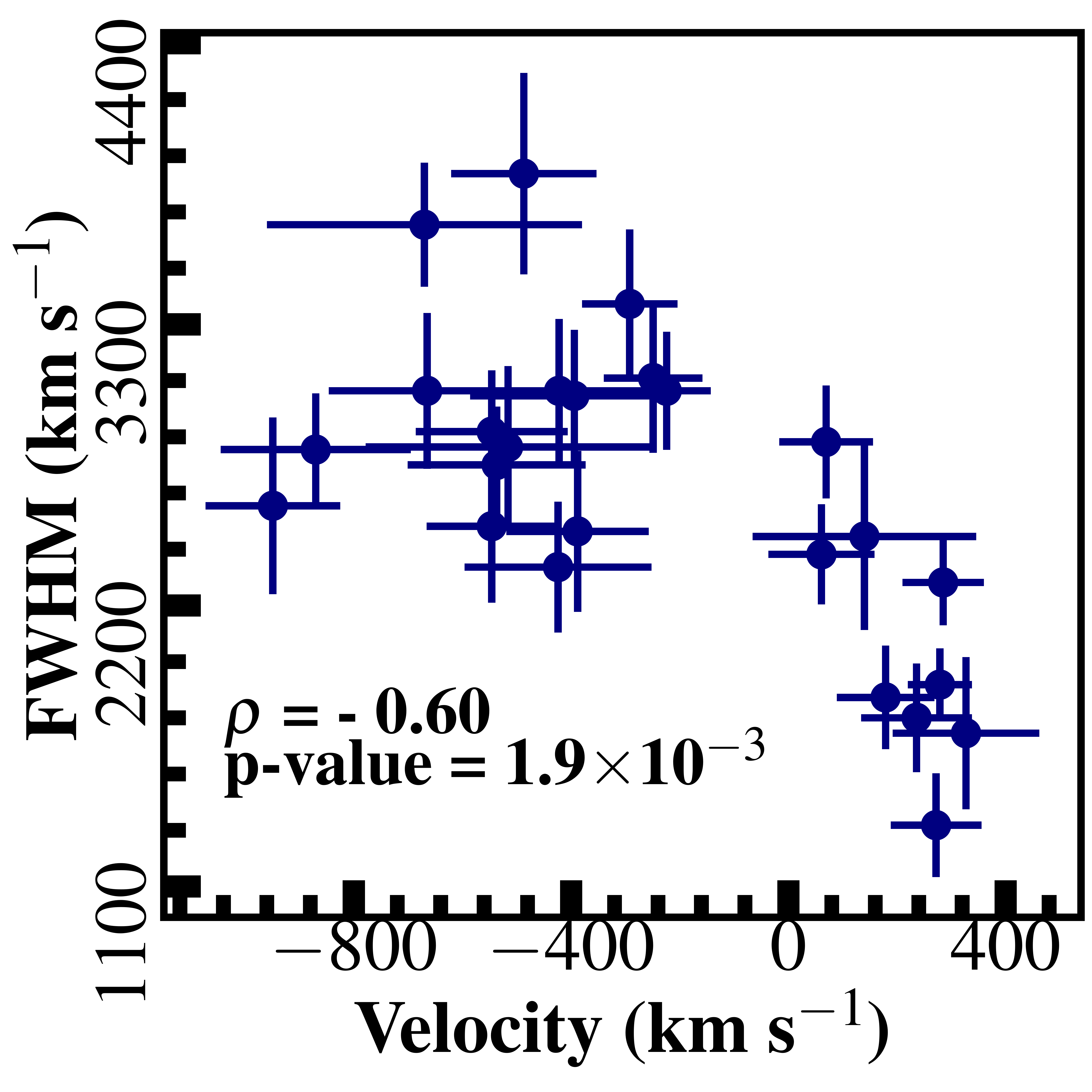}
		\label{corr_pks1138}}
	\subfloat[Ly$\alpha$ Velocity Abs.]{
		\includegraphics[width=1.9in,height=1.9in]{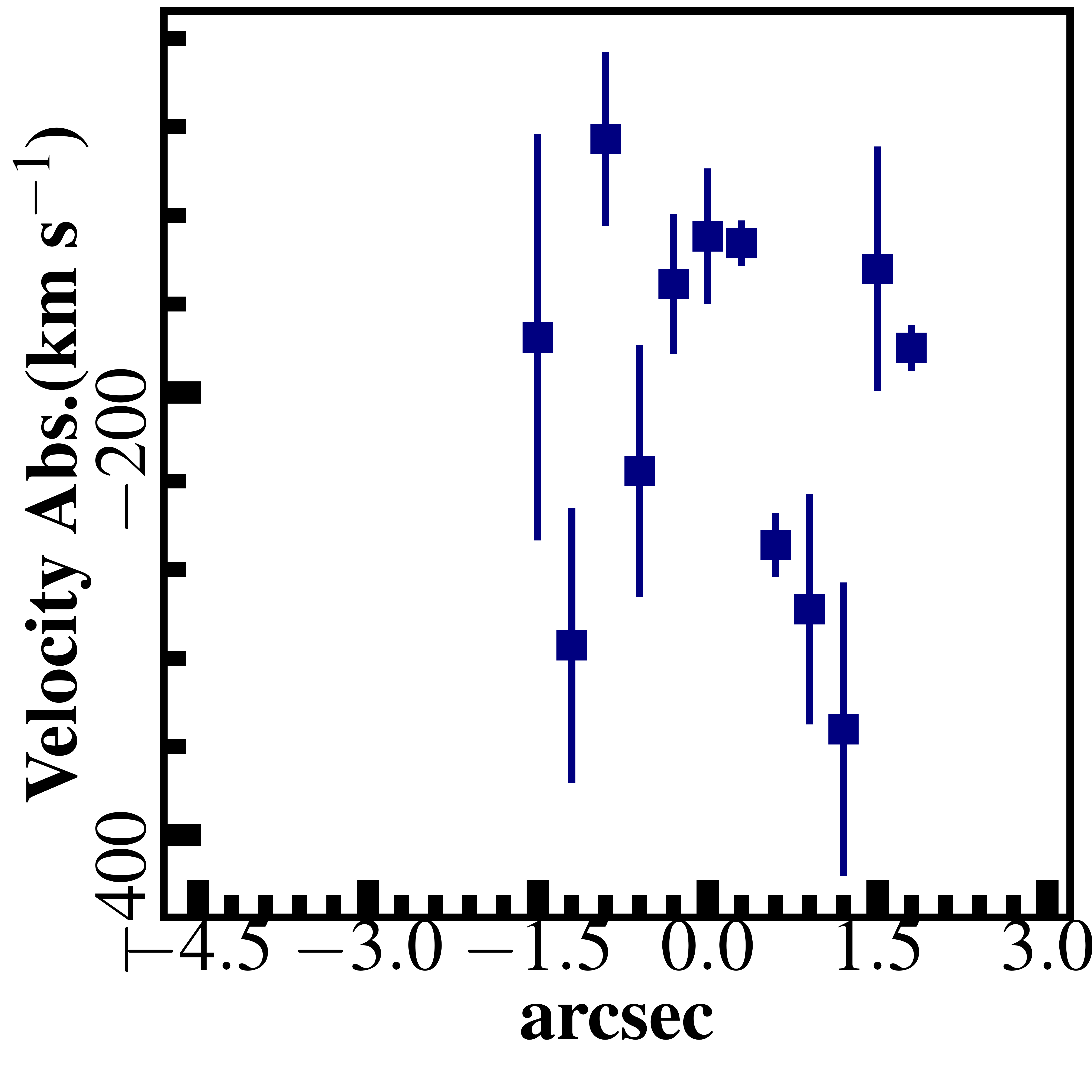}
		\label{vabs_pks1138}}
	\subfloat[Ly$\alpha$ N(\ion{H}{I}) \textit{vs.} Seeing]{
		\includegraphics[width=\columnwidth,height=1.9in,keepaspectratio]{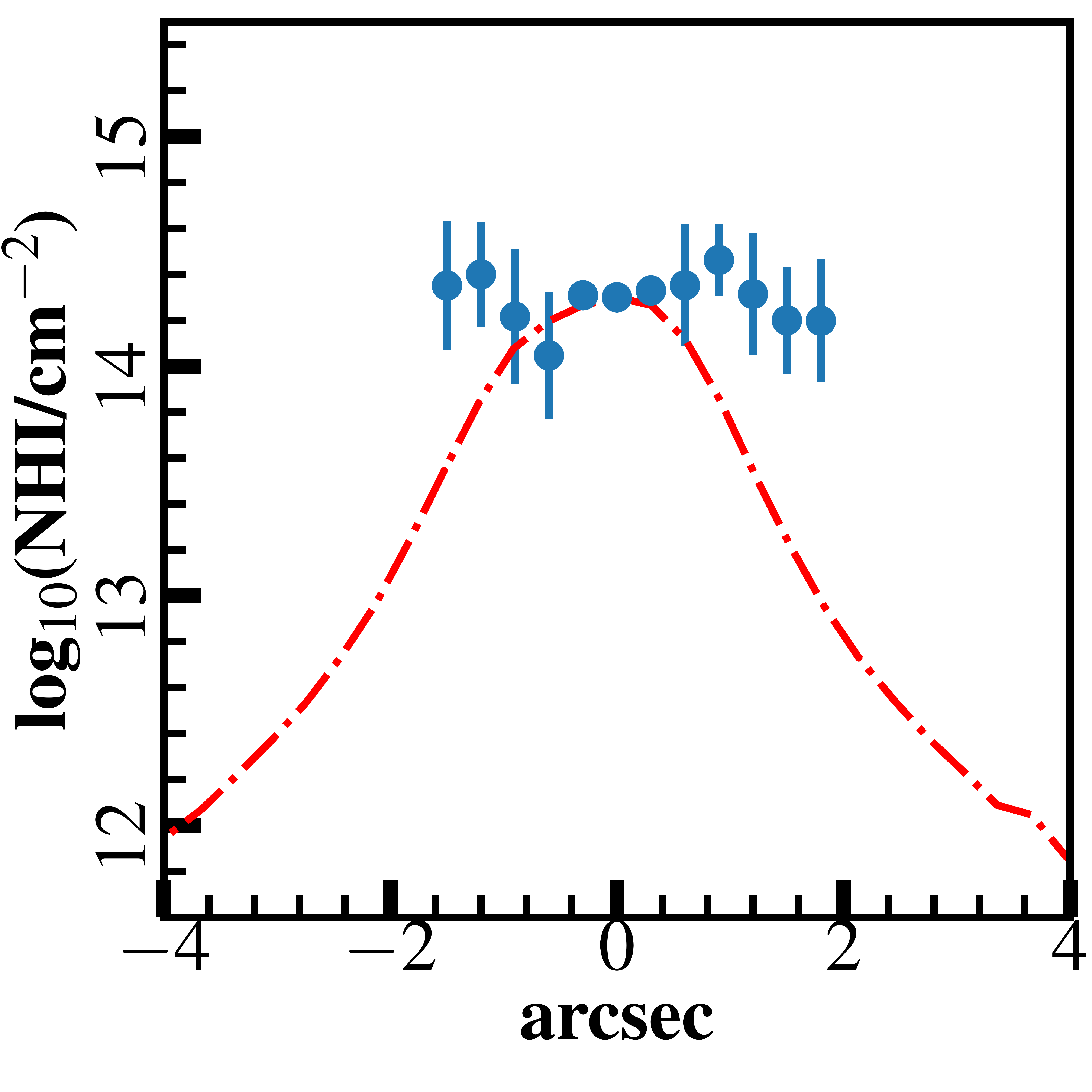}
		\label{NHIpks1138_seeing}}
	
	\caption{Radio galaxy PKS 1138--262: (a) 2-D spectrum of the Ly$\alpha$ spectral region, (b) Spatial variation of the flux of the Ly$\alpha$ line, (c) Ly$\alpha$ spatial profile (blue circle with dashed lines) compared with the seeing (green dot dashed lines) and (d) 1-D spectrum of the Ly$\alpha$ spectral region extracted from the SOAR long-slit. The Ly$\alpha$ emission-line was extracted by summing over a 3\arcsec$\,$ region of the slit length. Spatial variations of (e) FWHM, (f) Velocity, (g) Variation of FWHM as a function of the velocity offset of Ly$\alpha$ with $\rho$ and p-value representing the Spearman's rank correlation coefficient and t-distribution, respectively, (h) Velocity of the \ion{H}{I} absorber and (i) Spatial profile of the \ion{H}{I} column density (blue circle points) compared the seeing (red dot dashed lines), both on a logarithmic scale. In addition, the seeing profile has been normalised and shifted in order to allow the comparison.}
	\label{pks1138kin}
\end{figure*}
\subsubsection{Previous results}

VLA observations of the z = 2.16 radio galaxy PKS 1138--262 (Spiderweb galaxy) shows that the radio source consists of multiple hotspots in the eastern lobe, which shows a double bend towards south \citep[e.g.][]{carilli1997,pentericci1997}. 
The radio source has a maximum angular size of 11.4\arcsec$\,$ (97 kpc in the adopted cosmology) and spectral index $\alpha$ = -1.3 \citep[e.g.][]{carilli1997,pentericci1997}. 
The rest-frame near-infrared (near-IR) light and the K-band luminosity indicate that the host galaxy is very massive  (e.g. M$_{\star}$ $\sim$ 10$^{12}$ M$_{\odot}$; \citealt{seymour2007,hatch2009}) and infrared luminosity implies a star formation rate of 1390 $\pm$ 150 M$_{\odot}$ yr$^{-1}$ \citep[see][]{seymour2012}.
The radio galaxy presents a diffuse halo of young stars \citep{hatch2008} and observations with the Australia Telescope Compact Array (ATCA) have allowed the detection of CO(1-0) emission from cold molecular gas \citep{emonts2013}. 
Chandra X-ray imaging of the galaxy shows extended X-ray emission aligned with the radio structure \citep[e.g.][]{carilli2002}, and observations with the HST indicate that the optical emission is clumpy and disturbed \citep[e.g.][]{Pe98,kuiper2011} suggestive of a merging structure \citep{miley2006,kuiper2011}. 
With the major axis oriented along the radio axis, the enormous Ly$\alpha$ halo extends by at least 25\arcsec$\,$ (214 kpc in the adopted cosmology) surrounding the host galaxy, which indicates a large reservoir of atomic gas \citep{pentericci1997,miley2006}. 
Using VLT/SPIFFI IFU (SPectrometer for Infrared FaintField Imaging) of the rest-frame optical emission line, \citet{nesvadba2006} conclude that the kinematic properties and energy arguments favor the AGN as the plausible mechanism that efficiently power the outflowing gas.
Furthermore, using VLT/SINFONI imaging spectroscopy of the rest-frame optical emission line of the warm ionized gas, \citet{nesvadba17} find that the kinetic energy and momentum injection rates from the radio jets seem to be the most efficient mechanism powering the outflowing gas.

Investigating the Ly$\alpha$ emission line using medium resolution spectroscopy (instrumental FWHM = 2.8 \AA) with the slit approximately along the radio axis, \citet{pentericci1997} report the detection of several \ion{H}{I} absorbers at a similar redshift to the HzRG. They find column densities of the order log N(\ion{H}{I}/cm$^{-2}$) $\backsimeq$ 15 -- 16 and Doppler parameters between 40 and 100 km s$^{-1}$. In addition, studying optical images of the extended Ly$\alpha$ halo obtained with FORS1 at the 8.2m VLT Antu, \citet{kurk2003} identified the presence of \ion{H}{I} absorption associated with the Ly$\alpha$ emission line. Simulating 3 slits placed in different regions of the Ly$\alpha$ halo, \citeauthor{kurk2003} could identify up to two absorption troughs in the Ly$\alpha$ line with column density in the range log N(\ion{H}{I}/cm$^{-2}$) $\backsimeq$ 14.2 -- 14.8 and Doppler parameters between 140 and 430 km s$^{-1}$.

\subsubsection{Results from SOAR}

In the direction perpendicular to the radio axis, we detect the Ly$\alpha$ emission showing an asymmetric spatial distribution, being more extended towards the E (see the 2-D spectrum of the Ly$\alpha$ spectral region in Fig. \ref{2dspec_pks1138}, see also the spatial variation of the Ly$\alpha$ flux in Fig. \ref{ly_flux_pks1138}).
In Figure \ref{1dspec_pks1138}, we show the integrated 1-D spectrum of the Ly$\alpha$ profile. In addition to Ly$\alpha$, we also detected the \ion{C}{IV} and \ion{He}{II} emission lines (see Table \ref{instru02}).
In Figures \ref{fwhm_pks1138} and \ref{velo_pks1138}, we show the spatial variations of FWHM and velocity of the extended Ly$\alpha$ nebula emission. The extended emission line shows line widths in the range FWHM = 1700 -- 3890 km s$^{-1}$ while the velocity offset is ranging from -670 km s$^{-1}$ to 292 km s$^{-1}$. As in the previous objects, we find a correlation between the FWHM and the velocity offset of Ly$\alpha$, such that regions with higher FWHM also tend to have a larger blueshift (see Fig. \ref{corr_pks1138}). The result indicates that there is a negative relationship between the parameters with $\rho$ = - 0.60 and p-value = 1.9 $\times$ 10$^{-3}$.

The spatial profile of the radio galaxy PKS 1138--262 (FWHM = 4.81 $\pm$ 0.31\arcsec) is spatially resolved compared with the seeing disk FWHM = 1.76 $\pm$ 0.02. The emission line shows an excess above the seeing disk at $\lesssim$ 5.4\arcsec$\,$, which is particularly obvious towards the East (see Fig. \ref{pks1138_seeing}), confirming that Ly$\alpha$ is dominated by extended emission. The intrisic FWHM for Ly$\alpha$ will be 4.48 $\pm$ 0.33\arcsec or 38 $\pm$ 3 kpc. None of the other UV emission lines are found to be extended in this spectrum.

Unlike \citet{pentericci1997}, only a single absorption feature is detected in the spectrum of PKS 1138--262 (see Fig. \ref{2dspec_pks1138}), which is likely due to the lower resolution of our spectrum. Our best fit to the Ly$\alpha$ is shown in Fig. \ref{1dspec_pks1138}. Table \ref{instru03} lists the parameters of the best fitting model together with the diameter of the absorber, and the maximum detected radius. The absorber has a column density log N(\ion{H}{I}/cm$^{-2}$) = 14.08 $\pm$ 0.10 with Doppler parameter $\mathit{b}$ = 190 $\pm$ 43 km s$^{-1}$, and velocity offset -234 $\pm$ 28 km s$^{-1}$. In Figure \ref{vabs_pks1138}, we show the line of sight velocity of the \ion{H}{I} absorber measured from the SOAR spectrum across its detected spatial extent. The absorbing gas appears blueshifted from the systemic velocity. The absorber is detected extending over across 3.3\arcsec$\,$ (or 28 kpc) in the direction perpendicular to the radio axis of the galaxy. In addition, the Ly$\alpha$ absorption feature is clearly spatially extended compared to the seeing profile (see Fig. \ref{NHIpks1138_seeing}).
Using the column density and extent of the absorber, we estimate the neutral mass to be log ($M_{\ion{H}{I}}/M_{\sun}$) $\gtrsim$ 3.4.

\begin{table*}
	\centering
	\caption{Ly$\alpha$ absorption features best fit parameters. Column (1) gives the object name. Column (2) gives the redshift for the Ly$\alpha$ emission Gaussian. Column (3) gives the redshift for the Ly$\alpha$ absorption. Column (4) gives the column density (N(\ion{H}{I})). Column (5) gives the Doppler width b. Column (6) gives the velocity shift of the absorber with respect to \ion{He}{II} emission line. Column (7) gives the maximum diameter detected of the absorbers. Column (8) gives the maximum radius of the absorbers. We note that the radius over which we detect the absorber only gives us a lower limit to the radius of the shell because the absorption feature can not be detected where the background Ly$\alpha$ $\it emission$ is weak or absent. In addition, without information on the ionization fraction of hydrogen, the \ion{H}{I} column density only gives a lower limit on the total hydrogen column density. Moreover, because of the degeneracy discussed by \citet{marckelson2018} we also note that our column density estimate is likely a lower limit of the true column. Thus, our mass estimates are lower limits.}
	\label{instru03}
	\begin{tabular}{lcccccccr} % four columns, alignment for each
		\hline
		Object & Ly$\alpha$ emission redshift & Absorption redshift & Column Density & Doppler Parameter &  $\Delta$v  &   D   &   R$_{max}$  \\
		& ($z_{em}$)  		    &      ($z_{abs}$)	     &	  (cm$^{-2}$)      & (km s$^{-1}$) & (km s$^{-1}$) &  (kpc)    &  (kpc)  \\
		\hline
                (1) & (2) & (3) & (4) & (5) & (6) & (7) & (8) \\
		\hline
		MRC 0406--244 & 2.42287 $\pm$ 0.00015 & 2.42219 $\pm$ 0.00009 & (6.46 $\pm$ 0.38)$\times10^{14}$ & 250 $\pm$ 11  &  -184 $\pm$ 8 &  35   &   23 \\
		\hline
		TN J0920--0712 & 2.75450 $\pm$ 0.00010 & 2.75385 $\pm$ 0.00010 & (1.70 $\pm$ 0.10)$\times10^{14}$ & 185 $\pm$ 10  &  -232 $\pm$ 6 &  27   &   17  \\
		\hline
		PKS 1138--262 & 2.16430 $\pm$ 0.00032 & 2.16284 $\pm$ 0.00003 & (1.21 $\pm$ 0.29)$\times10^{14}$ & 190 $\pm$ 43  &  -234 $\pm$ 28 &  28   &   15 \\
		\hline
		
	\end{tabular}
\end{table*}
\section{Discussion}
\label{discussion}

\subsection{Gas Dynamics}
\label{dynamics}

Several scenarios have been suggested in order to explain the observed kinematic properties of the giant nebulae associated with HzRGs, including outflows resulting from jet-gas interactions, AGN or starburst driven super-winds \citep[e.g.][]{villar-martin1999b,villar2000,Ja,Hu2,Hu6,nesvadba2006,nesvadba2008,swinbank,Cai2017,nesvadba17}, gas infalling/inflow \citep[e.g.][]{Hu3,VM2,Hu1,nathan2014,marckelson2018} or rotation \citep[e.g.][]{Vo96,VM1,VM2006,arrigoni2018}. 
Clearly, by studying the gas dynamics of the Ly$\alpha$ nebulae it should be possible to obtain information about their origins. 

For 3 out of 5 radio galaxies (e.g. MRC 0406--244, TN J0920--0712 and PKS 1138--262), the kinematic properties of the Ly$\alpha$ emission line show a resemblance to rotation curves (see Figs. \ref{mrc0406kin}, \ref{tnj0920kin} and \ref{pks1138kin}).
However, gas dynamics dominated by the large-scale gravitational potential of the host galaxy ought to show smooth velocity gradients and rather uniform line widths with FWHMs of a few $\times$100 km s$^{-1}$. Given the large values of FWHM we measure in the extended Ly$\alpha$ emitting gas, it seems unlikely that rotation dominates the gas dynamics in these specific regions.
We note that most of the objects show a correlation between the FWHM and the velocity offset of Ly$\alpha$ (see Section \ref{prev}), suggesting that bulk radial motion, rather than rotation or random motion, dominates the gas dynamics in these regions. Furthermore, the fact that this trend does not appear to correlate with the presence (or absence) of strong Ly$\alpha$ \ion{H}{I} absorption suggests that the trend is not caused by the superposition of an extended absorption system.
Given also the facts that the perturbed gas is spatially extended and detected on both sides of the nucleus, and that the most perturbed regions shows usually a blueshifted gas with respect to the systemic velocity, it would be natural to conclude that outflows dominate the kinematics of this gas. However, this interpretation effectively ignores line radiative transfer effects, which can have a significant effect on the observed kinematics of resonant lines such as Ly$\alpha$. Indeed, if the Ly$\alpha$ kinematics are strongly affected by resonant scattering, the correlation between FWHM and blueshift may also be explained by the {\it infall} of gas \citep[see][]{dijkstra2006b,verhamme2006}. In other words, we argue that radial motion appears to dominate the bulk dynamics of this extended gas, but due to degeneracies inherent in the use of Ly$\alpha$ to study gas kinematics, we are unable to discriminate between the infall and outflow scenarios. 
In the case of 4C--00.54, it seems plausible that due to jet precession \citep[see][]{pentericci1997}, the slit position angle used might intersect gas that has been disturbed by the radio jets. It is, however, unexpected that the ionized gas in the nucleus has lower FWHM than the extended gas outside the ionization cones. We should also expect the jets to interact with and perturb gas in the nuclear region of the galaxy, but this not seen in our data. On the other hand, if the jets are not responsible for the high FWHM, we suggest this might be a result of scattering of Ly$\alpha$ photons that were originally produced along the radio axis or within the ionization cones, by neutral gas in regions that are predominantly neutral due to not being illuminated by the AGN. In this scenario, the emergent velocity profile of the scattered Ly$\alpha$ would have a velocity profile that has been artificially broadened as a result of the kinematic diversity of the L$\alpha$ emitting regions in the radio galaxy's extended gas halo. Another possibility might be that Ly$\alpha$ emitted by highly turbulent gas associated with jet-gas interactions is being preferentially scattered by neutral hydrogen within the extended regions in our slit. However, detailed numerical modeling of Ly$\alpha$ radiative transfer within the gas halo would be needed to explore this possible effect in more detail.
The radio galaxy MRC 0030--219 has a single and compact radio source, which is much less extended than the Ly$\alpha$ emission line gas detected. Given the sizes involved between the radio source and the emission line gas, we suggest that jets seem unlikely to be responsible for such turbulence. However, we are unable to definitely say what kind of motion is dominating the gas (i.e. infall, outflow or rotation) or if the observed properties observed are a consequence of the resonant scattering within the nebula.

Ly$\alpha$ is particularly sensitive to radiative transfer effects. Given this nature, 
the best way to avoid any possible uncertainty on this issue would be to compare the kinematic properties of Ly$\alpha$ with lines that are not affected by radiative transfer effects, such as \ion{He}{II} $\lambda$ 1640 or optical forbidden lines such as \ion{[O}{III]} $\lambda\lambda$ 4959,5007. Although we can not neglect the contribution from radiative transfer effects, without other emission lines we are not able to say whether the high FWHM of Ly$\alpha$ is dominated by these effects or is instead due to extreme motions in the extended gas.

\subsection{The radial gradient of Ly$\alpha$/He{\small II} in 4C--00.54}
\label{Ly_heii_ratio}
Previous investigations using long-slit spectroscopy along the radio axis or perpendicular to the radio axis of other HzRGs have shown that line ratios involving Ly$\alpha$ emission line can vary significantly \citep[e.g.][]{Hu4,sandy2016}.  

We find a significant variation in Ly$\alpha$/\ion{He}{II} flux ratio (see Fig. \ref{lnr_4c}). The ratio shows a radial increase from 5.0 $\pm$ 0.3 near the spatial axis origin, up to 16.3 $\pm$ 1.9 (r = - 0.9\arcsec) and 17.2 $\pm$ 1.3 (r = + 1.5\arcsec). Such a variation suggests evidence for some process in which the Ly$\alpha$ is depressed in the centre with this effect becoming less important when moving far from the centre. As it moves out from the centre, we note that the flux ratio reaches values that are typically seen in other HzRGs (Ly$\alpha$/\ion{He}{II} = 15 -- 20; \citealt{VM3}).

Using only the Ly$\alpha$/\ion{He}{II} flux ratio, it is not easy to discriminate what kind of effects might be responsible for the large variation observed. However, we suggest that it may be a consequence of effects such as (i) absorption of Ly$\alpha$, presumably by \ion{H}{I} or interstellar dust \citep[see][]{Ro97,Vo}, (ii) resonant scattering of Ly$\alpha$ photons \citep[see][]{loeb2008,hayes2011,steidel2011} and (iii) enhanced emission of Ly$\alpha$, which can be produced by a low ionization parameter, a relatively soft ionizing spectral energy distribution (SED) or a low gas metallicity \citep[see][]{VM3,humphrey2018}.

Interestingly, higher FWHM are one of the possible signatures of Ly$\alpha$ resonance scattering. If this is true, then the observed Ly$\alpha$ might be a result of resonant scattering strongly affecting the regions across the nebula.
\subsection{The nature of the H\textbf{\small I} Absorbers}

A new perspective about the gaseous environment of HzRGs was opened up by \citet{Ro95} and \citet{Vo} with the discovery of associated, spatially extended \ion{H}{I} absorbers. With the slit positioned along the radio axis, a number of investigations have shown that the \ion{H}{I} absorbers are typically detected with column densities between log N(\ion{H}{I}/cm$^{-2}$) = 13 -- 19 and Doppler width $\mathit{b}$ = 20 -- 874 km s$^{-1}$ \cite[e.g.][]{Ro95,Vo,Bi1,Ja,Wm}. In addition, previous work has shown that the absorbers are extended over tens or even a few hundred kpc, leading to \ion{H}{I} mass estimates of up to log ($M_{\ion{H}{I}}/M_{\sun}$) $\sim$ 9. In addition, it has been argued that the extended absorbers are (i) in outflow; (ii) partly ionized based on detection of \ion{C}{IV} and other lines in a few cases; (iii) probably located outside the Ly$\alpha$ halos based on covering factor $\sim$ 1 and the fact that the absorbers appear to completely cover the (detected) extended Ly$\alpha$ emission \citep[e.g.][]{Bi1}. Based on these properties, it has been argued that the absorbing gas is part of an expanding shell that surrounds the HzRG and its Ly$\alpha$ halo, produced by feedback activity \citep[e.g.][]{Bi1}. 

However, single-slit observations did not provide information about the two-dimensional distribution of the absorbing gas, leaving open the possibility that the absorbers do not cover the entire Ly$\alpha$ halo, and are only present along the radio/optical axis of the HzRGs. A crucial test of the shell hypothesis is the detection of extended \ion{H}{I} absorbers in the direction perpendicular to the radio/optical axis. 

To address this issue, in recent years three HzRGs with extended \ion{H}{I} absorbers were studied with IFU spectroscopy, and in each case the absorber was found to cover the full spatial extent of the Ly$\alpha$ halo (MRC 2025--218: \citealt{Hu6}; TN J1338--1942: \citealt{swinbank}; MRC 0943-242: \citealt{Gu,marckelson2018}). Furthermore, \citet{marckelson2018} were able to measure the radial velocity curve for the main absorber in front of MRC 0943--242, and found a radial gradient consistent with an expanding shell with a radius of at least several tens of kpc. 

In this paper, we have studied the \ion{H}{I} absorbers of a further 3 HzRGs, placing a long slit at large angles ($>$45$^{\circ}$) to the radio axis, and detecting the \ion{H}{I} absorber across the full spatial extent of the Ly$\alpha$ emission. In all three cases, the absorber is blueshifted relative to the systemic velocity across its full detected extent. 

Taking our new results together with the three IFU studies mentioned above, we find that for all 6 of the extended absorbers for which the test has been undertaken, the absorbing gas is extended perpendicularly to the radio/optical axis of the HzRG. We conclude that when present around an HzRG, this type of extended HI absorbing structure commonly covers the entire Ly$\alpha$ halo. These properties are consistent with the absorbing gas being part of a large-scale, expanding shell surrounding the HzRG and its Ly$\alpha$ halo, as suggested by \citet{Bi1}.
\section{Conclusions}
\label{conclusions}

Making use of the Goodman long-slit on the SOAR telescope, we have studied the properties of the extended Ly$\alpha$ halo and the large-scale \ion{H}{I} absorbers structures of 5 HzRGs, in regions that are expected to have little influence from the radiation field of the active nucleus or the radio jets, to examine the global properties of these two gaseous components.

We find spatially extended gas with high line widths (FWHM = 1000 -- 2500 km s$^{-1}$), suggestive of turbulent motion. In addition, we find a correlation between the FWHM and the velocity offset of Ly$\alpha$, such that regions with higher FWHM are more blueshifted. Based on this result, we conclude that an outflowing gas or an infalling gas are the two scenarios consistent with the kinematic properties of the radio galaxies, depending on the level of resonant scattering of the Ly$\alpha$ photons within the nebula but also on the gas geometry, \ion{H}{I} distribution and column density.

Studying the \ion{H}{I} absorbers of 3 HzRGs with the long slit placed at a large angle to the radio axis ($>$45$^{\circ}$), we have detected the \ion{H}{I} absorber across the full spatial extent of the Ly$\alpha$ emission. In all three cases, the absorber is blueshifted relative to the systemic velocity across its full detected extent. Our new results show that the extended \ion{H}{I} absorbers associated with HzRGs, when present, cover the entire Ly-alpha halo, which is consistent with previous IFU results and also consistent with the absorbing gas being part of a giant, expanding shell of gas enveloping the HzRG and its Ly$\alpha$ halo. 

Finally, we comment that our work shows the potential capability of the Goodman Spectrograph on the SOAR 4.1 m telescope to observe objects at high-redshift (z $>$ 2).
\section*{Acknowledgements}
MS acknowledges support from the National Council of Research and Development (CNPq) under the process of number 248617/2013--3. MS and RG thank David Sanmartim for all the support during the observations and also recognize the observational support provided by the SOAR operators Alberto Pasten, Patricio Ugarte, and Sergio Pizarro. MS thanks Montse Villar-Mart\'in and Luc Binette for helpful and valuable discussions.
AH acknowledges FCT support through fellowship SFRH/BPD/107919/2015. PL acknowledges support by the FCT through the grant SFRH/BPD/72308/2010. TS acknowledges the fellowship SFRH/BPD/103385/2014 funded by FCT (Portugal) and POPH/FSE (EC).
PP is supported by the FCT through Investigador FCT contract IF/01220/2013/CP1191/CT0002. 
SGM acknowledges FCT support in the way of PhD fellowship PD/BD/135228/2017. 
MS, AH, PL and PP acknowledge the support of the European Community Programme (FP7/2007--2013) under grant agreement No. PIRSES--GA--2013--612701 (SELGIFS). This work was also supported by Funda\c{c}\~{a}o para a Ci\^{e}ncia e Tecnologia (FCT) through national funds (PTDC/FIS--AST/3214/2012 and UID/FIS/04434/2013), and by FEDER through COMPETE (FCOMP--01--0124--FEDER-029170) and COMPETE2020 (POCI--01--0145--FEDER-007672).
%
%%%%%%%%%%%%%%%%%%%%%%%%%%%%%%%%%%%%%%%%%%%%%%%%%%%
%
%%%%%%%%%%%%%%%%%%%%% REFERENCES %%%%%%%%%%%%%%%%%%
%
%% The best way to enter references is to use BibTeX:
%
\bibliographystyle{mnras}
\bibliography{example} % if your bibtex file is called example.bib
%
%% Alternatively you could enter them by hand, like this:
%% This method is tedious and prone to error if you have lots of references
%%\begin{thebibliography}{99}
%%\bibitem[\protect\citeauthoryear{Author}{2012}]{Author2012}
%%Author A.~N., 2013, Journal of Improbable Astronomy, 1, 1
%%\bibitem[\protect\citeauthoryear{Others}{2013}]{Others2013}
%%Others S., 2012, Journal of Interesting Stuff, 17, 198
%%\end{thebibliography}
%
%%%%%%%%%%%%%%%%%%%%%%%%%%%%%%%%%%%%%%%%%%%%%%%%%%%
%
%%%%%%%%%%%%%%%%%% APPENDICES %%%%%%%%%%%%%%%%%%%%%
%
%\appendix

%%%%%%%%%%%%%%%%%%%%%%%%%%%%%%%%%%%%%%%%%%%%%%%%%%

% Don't change these lines
\bsp	% typesetting comment
\label{lastpage}
\end{document}